\documentclass[fleqn,usenatbib]{mnras}
 
\usepackage{multirow}
\usepackage{multicol}
\usepackage{natbib}
\usepackage{graphicx}
\usepackage{amsmath}
\usepackage{txfonts}
\usepackage[T1]{fontenc}

\usepackage{color}
\hypersetup{colorlinks=true,allcolors=blue}
\usepackage{paralist}\usepackage{epstopdf}
\usepackage{epsfig}
\usepackage{tablefootnote}
\usepackage{amssymb}
\usepackage{bm}
\usepackage{tabularx}
\usepackage{lmodern}
\usepackage{lineno}
\renewcommand{\linenumbers}{}

    \title[Globular Clusters in the times of the JWST.]{Globular Clusters in the Time of the JWST. I. Survey Design and First Results on Multiple Populations and Beyond}
\author[Milone et al.]{
A.\,P.\,Milone$^{1,2}$,
A.\,F.\,Marino$^{2}$,
G.\,Cordoni$^{3}$\thanks{Corresponding author: giacomo.cordoni@anu.edu.au},
E.\,Dondoglio$^{4}$,
M.\,V.\,Legnardi$^{1}$,
T.\,Ziliotto$^{5}$,
\newauthor
E.\,Bortolan$^{1}$,
F.\,Muratore$^{1}$,
F.\,D'Antona$^{6}$,
A.\,Renzini$^{2}$,
G.\,Girardi$^{1}$,
L.\,Gorza$^{1}$,
\newauthor
A.\,Mastrobuono-Battisti$^{1,2,7}$,
C.\,Ventura$^{6}$,
P.\,Ventura$^{6}$,
V.\,Altomonte$^{1}$,
L.\,Bisigello$^{2}$,
Y.\,Cavecchi$^{8}$,
\newauthor
F.\,Dell'Agli$^{6}$,
A.\,Dotter$^{9}$,
E.\,P.\,Lagioia$^{10}$,
C.\,Li$^{11, 12}$,
S.\,Lionetto$^{1}$,
A.\,Marchuk$^{1}$,
\newauthor
J.\,Qi$^{10}$,
G.\,Rodighiero$^{1}$,
M.\,Tailo$^{13}$,
H.\,Wirth$^{14, 15}$
\\
$^1$  Dipartimento di Fisica e Astronomia ``Galileo Galilei'', Universit\`a  di Padova, Vicolo dell'Osservatorio 3, 35122 Padova, Italy \\
$^{2}$ Istituto Nazionale di Astrofisica - Osservatorio Astronomico di Padova, Vicolo dell'Osservatorio 5, 35122 Padova, Italy\\
$^{3}$ Research School of Astronomy \& Astrophysics, Australian National University, Canberra, ACT 2611, Australia \\
$^{4}$Physics Department, American University of Sharjah, P.O. Box 26666, Sharjah, UAE\\
$^{5}$
Space Telescope Science Institute, 3700 San Martin Drive, Baltimore, MD 21218, USA\\
$^{6}$ Istituto Nazionale di Astrofisica, Osservatorio Astronomico di Roma, Via Frascati 33, 00077 Monte Porzio Catone, Italy\\
$^{7}$  Dipartimento di Tecnica e Gestione dei Sistemi Industriali, Universit\`a degli Studi di Padova, Stradella S. Nicola 3, I-36100 Vicenza, Italy\\
$^{8}$ Departamento de Astrofisica, Universidad de La Laguna, 38206, San Cristobal de La Laguna, Tenerife, Spain\\
$^{9}$ Department of Physics and Astronomy, Dartmouth College, 6127 Wilder Laboratory, Hanover, NH 03755, USA\\
$^{10}$ South-Western Institute for Astronomy Research, Yunnan University, Kunming, 650500 P. R. China \\
$^{11}$ School of Physics and Astronomy, Sun Yat-sen University, Daxue Road, Zhuhai, 519082, P.R. China \\
$^{12}$ CSST Science Center for the Guangdong-Hong Kong-Macau Greater Bay Area, Zhuhai, 519082, P.R. China \\
$^{13}$  Dipartimento di Fisica e Astronomia ``Augusto Righi'', Universit\`a di Bologna, Via Gobetti 93/2, 40129 Bologna, Italy \\
$^{14}$ Charles University, Faculty of Mathematics and Physics, Astronomical Institute, V Holešovi\v{c}k\'ach 2, 180 00 Praha, Czech Republic\\
$^{15}$ Department of Theoretical Physics and Astrophysics, Faculty of Science, Masaryk University, Kotl\'a\v{r}sk\'a~2, Brno~611~37, Czech Republic
}

\date{Accepted 2026 August 11. Received 2026 August 9; in original form 2026 June 19}
\pubyear{2026}

\begin{document}
\label{firstpage}
\pagerange{\pageref{firstpage}--\pageref{lastpage}}
\maketitle

\begin{abstract}
Globular clusters (GCs) host multiple stellar populations with distinct chemical compositions, but their properties among very low-mass stars remain poorly constrained. The James Webb Space Telescope (JWST) enables precise infrared studies that are highly sensitive to abundance variations in cool stars. 
We initiate a homogeneous survey of Galactic GCs, based primarily on deep JWST GO-8960 observations and complemented by archival JWST and Hubble Space Telescope data, to characterize multiple populations across a wide range of cluster properties. In this first paper, we present the survey and initial NIRCam results.
We analyze eleven GCs, deriving high-precision photometry and astrometry to measure proper motions. Multiple populations are detected among low-mass stars in all clusters, with diverse behaviors. We find discrete main sequences in NGC\,288, NGC\,6723, and NGC\,2808, and more continuous distributions in NGC\,104 and the Type\,II clusters NGC\,1851 and NGC\,6656. The bulge clusters NGC\,6528, NGC\,6553, and NGC\,6440 show patterns consistent with varying helium and oxygen abundances that do not scale simply with cluster mass. In Terzan\,5 and Liller\,1, we identify populations spanning different ages and helium variations within the old population of Terzan\,5. We also detect an M-dwarf gap in NGC\,104 at $\sim0.35 M_{\odot}$, consistent with the Jao Gap of field stars and open clusters.
This work establishes the foundation for a homogeneous JWST survey of Galactic GCs and provides a valuable dataset for studies of cluster evolution, Galactic stellar populations, and background extragalactic sources.
\end{abstract}

\begin{keywords}
globular clusters: general, stars: population II, stars: abundances, techniques: photometry.
\end{keywords}

\section {Introduction}
\label{sec:intro}
Over the past decades, it has become clear that most globular clusters (GCs) are not simple stellar populations but host multiple stellar populations characterized by distinct chemical compositions. Spectroscopic and photometric studies have revealed star-to-star variations in light elements (e.g., He, C, N, O, Na, and Al), with high-precision photometry proving particularly powerful in identifying distinct populations along the evolutionary sequences. In particular, ultraviolet observations obtained with the \textit{Hubble Space Telescope} (HST) have enabled the separation of multiple populations among bright stars, from the upper main sequence (MS) to the red giant branch (RGB), thanks to the strong sensitivity of UV filters to abundance variations affecting molecular bands such as OH, CN, NH, and CH \citep[see][for reviews]{kraft1994a, gratton2012a, gratton2019a, bastian2018a, cassisi2020a, milone2022a}. Large photometric surveys with HST and complementary ground-based observations have demonstrated that multiple populations are a ubiquitous property of massive GCs \citep[e.g.][]{milone2017a, jang2022a, lagioia2025a}.

Despite major observational efforts, the origin of multiple populations remains debated \citep[e.g.][]{renzini2015a}. Two main classes of scenarios are currently under consideration. In the multiple-generations framework, multiple populations arise from distinct episodes of star formation, where second-generation (2P) stars form from material processed and ejected by first-generation (1P) stars. Various types of 1P polluters have been proposed, 
 such as asymptotic giant branch (AGB) stars \citep[][]{ ventura2001a, dercole2010a, dantona2016a}, massive binaries \citep{demink2009a, renzini2022a}, fast-rotating massive stars \cite{krause2013a}, and even super-massive stars \citep{denissenkov2014a}.
These scenarios face the well-known mass-budget problem:
existing 1P stars in present-day GCs fall dramatically short to provide enough material 
for the build-up of the 2P stars observed today, unless GCs, or more likely their hosts, 
were initially much more massive and subsequently the majority of their 1P stars were lost.

Alternatively, aiming to alleviate the mass-budget problem, abundance anomalies would arise in a single stellar generation from the accretion of processed material ejected by massive \citep{bastian2013a} or supermassive \citep{gieles2018a} stars  onto low-mass pre-MS stars. However, the accretion efficiency would depend strongly on stellar mass. For instance, Bondi-like accretion rates scale approximately with the square of the stellar mass, implying that abundance differences between 1P and 2P stars should decrease toward the low-mass regime, particularly among M dwarfs. Probing multiple populations over a wide mass range (especially among very low-mass stars) thus provides a powerful way to discriminate between competing
formation scenarios.

Understanding the formation of GCs and their multiple populations is also crucial for interpreting recent discoveries at high redshift enabled by the \textit{James Webb Space Telescope} (JWST). JWST has revealed a population of compact, intensely star-forming galaxies with supersolar nitrogen-to-oxygen ratios \citep[e.g.][]{bunker2023a, cameron2023a, marqueschaves2024a}. These dense systems, characterized by extreme star-formation rates, exhibit chemical abundance patterns  
 that are not observed in HII regions at comparable metallicities in the local
Universe \citep[e.g.][]{izotov2012a, izotov2023a} and are not reproduced by standard chemical-evolution models \citep[e.g.][]{vincenzo2016a}.

Remarkably, their enhanced [N/O] ratios closely resemble those measured in 2P GC stars, suggesting that JWST may be probing environments analogous to those in which GC multiple populations formed \citep[e.g.][]{bunker2023a, renzini2023a, marqueschaves2024a}. Notable examples include GN-z11 ($z = 10.6$) and GS\_3073 ($z \simeq 5.55$). 
Theoretical models have been proposed in which massive, dense proto-cluster systems at high redshift
self-enrich in nitrogen through hot hydrogen burning in either AGB stars \citep{dantona2023a, dantona2025a}  or in supermassive
stars \citep{charbonnel2023a}, reproducing the abundance patterns observed both
in GC 2P stars and in these proto-galaxies. But the GC mass budget problem remains.

Another potential link between GCs and JWST observations is provided by the so-called Little Red Dots (LRDs; \citealt{matthee2024a}), a numerous population of compact red sources at $z > 3$ discovered by JWST \citep{labbe2023a}. Their characteristic spectral shape—a pronounced V-shape with a blue rest-frame UV continuum and a red UV-to-optical slope—has led \cite{chisholm2026a} to propose that they may represent GCs in formation, with the UV emission produced by a very young stellar population and the optical emission powered by a short-lived supermassive star.

In this broader context, constraining the nature and origin of multiple populations in present-day GCs provides a crucial benchmark for interpreting these high-redshift observations.

However, most observational studies have focused on relatively bright stars, while the properties of multiple populations among very low-mass stars remain less explored. This is mainly because obtaining spectroscopy or high-precision UV photometry for such faint objects is challenging. The investigation of multiple populations has nevertheless been extended to lower-mass stars thanks to pioneering infrared observations obtained with the IR channel of the Wide Field Camera~3 on board HST. These studies revealed split or broadened sequences below the MS knee in several GCs \citep{milone2012b, milone2014a, milone2019a, dotter2015a, dondoglio2022a}, a feature interpreted as a consequence of the different oxygen abundances that characterize 1P and 2P stars.

In the cool atmospheres of M dwarfs, oxygen-bearing molecules produce absorption features that strongly affect the infrared spectral region. As a result, stars with similar atmospheric parameters but different oxygen abundances exhibit distinct flux distributions at wavelengths $\lambda \gtrsim 1.3\,\mu$m, leading to the observed photometric separation of the sequences. 

More recently, we initiated a pioneering project based primarily on GO-2560 (PI A.\,F.\,Marino) data to study the GC 47\,Tucanae (NGC\,104) using JWST photometric and spectroscopic observations. These studies have demonstrated the potential of JWST to identify multiple populations among very low-mass stars with unprecedented precision and to place strong constraints on their chemical composition \citep{milone2023b, milone2025b, marino2024a, marino2024b, legnardi2024a, ziliotto2025a}. Additional JWST photometric investigations of multiple populations in low-mass stars have been presented by various authors \citep[e.g.\,][]{ziliotto2023a, ziliotto2026a, cadelano2023a, scalco2024a, scalco2025a, milone2025a}.

Building on these results, we have started a project based on NIRCam and NIRISS GO-8960 (PIs A.\,P.\,Milone, A.\,F.\,Marino) observations, complemented by archival JWST and HST data, to homogeneously investigate multiple populations among very low-mass stars in a sample of eleven GCs spanning a wide range of properties relevant to the multiple-population phenomenon. The exceptional quality of the JWST data also enables a variety of ancillary studies, including investigations of cluster stellar populations, field stars, and background galaxies.

This paper is organized as follows. Section~\ref{sec:targets} describes the target sample. Section~\ref{sec:data} summarizes the NIRCam observations and the data-reduction procedures. Section~\ref{sec:cmds} presents the main photometric diagrams based on NIRCam data, while Section~\ref{sec:MPs} reports the first results on multiple populations. Finally, Section~\ref{sec:potential} discusses other early results together with the main scientific programs enabled by this dataset.

\section{The target GCs}\label{sec:targets}

The properties of multiple stellar populations in GCs display significant cluster-to-cluster variations, both in terms of chemical abundance patterns and population complexity in the chromosome map (ChM) or other photometric diagrams. On the basis of their characteristics, GCs are commonly divided into two principal groups \citep{milone2017a}.

\textit{Type I GCs} are systems in which distinct stellar populations are primarily characterized by variations in light-element abundances (e.g., C, N, O, Na, Mg, and Al), while maintaining a nearly homogeneous heavy-element content. \textit{Type II GCs}, in contrast, exhibit the typical light-element variations observed in Type I clusters but also show star-to-star differences in heavy elements, particularly those produced by the $s$-process or in iron abundance. 

Our target selection has been designed to ensure representative coverage of both GC types. For the \textit{Type I} category, we include NGC\,288, NGC\,2808, and NGC\,6723, which share similar metallicities. 
NGC\,288 hosts two well-defined stellar populations and exhibits moderate light-element abundance variations, making it a relatively simple example of a Type I GC \citep{carretta2009a, piotto2013a}. 
NGC\,2808, by contrast, represents one of the most chemically complex Type I clusters known, with at least five distinct stellar populations and extreme light-element abundance variations \citep{dantona2005a, piotto2007a, milone2015a, carlos2023a}. 
NGC\,6723 displays intermediate properties, both in terms of the number of discrete populations and the amplitude of abundance variations. In addition, the analyzed sample comprises NGC\,104 for which archive data in similar data are available.

For the \textit{Type II} category, our sample includes NGC\,1851 and M\,22 (NGC\,6656). Both clusters exhibit intrinsic variations in $s$-process element abundances as well as in the overall C$+$N$+$O content. In NGC\,1851, the $s$-rich and $s$-poor stellar populations display significant differences in $s$-process elements while sharing a similar {total} metallicity \citep[e.g.][]{yong2008a, marino2014a, marino2019a}. In contrast, NGC\,6656 hosts $s$-rich and $s$-poor populations that not only differ in their $s$-process element abundances, but also in their metallicity, with the $s$-rich stars being enhanced in metallicity relative to the $s$-poor component by $\sim$0.2 dex \citep[e.g.][]{marino2009a, marino2011a, dacosta2009a, mckenzie2022a}.

Our sample further includes two of the most metal-rich GCs in the Milky Way, NGC\,6528 and NGC\,6553, which have near-solar metallicities \citep[e.g.][]{dias2015a}. These Galactic bulge clusters likely formed in an environment that experienced rapid chemical enrichment, at a time when a substantial fraction of the present-day bulge mass (of order $2\times10^{10}\,{\rm M}_{\odot}$) was already in place \citep[e.g.][]{ortolani1995a}. Constraining the properties of multiple populations in such high-metallicity systems is therefore of particular importance for understanding GC formation in the dense, metal-rich conditions of the early bulge.

While the presence of multiple populations is firmly established in metal-poor and intermediate-metallicity GCs formed during the early assembly of the Milky Way halo, extending this characterization to the high-metallicity regime provides a critical test of the universality of the phenomenon. Spectroscopic studies have reported star-to-star variations in C, N, and Na in both NGC\,6528 and NGC\,6553, demonstrating the presence of multiple populations, albeit based on relatively small stellar samples \citep[e.g.][]{schiavon2017a, kader2022a}. A comprehensive photometric characterization of their multiple-population properties is still lacking. Exploring the detailed behavior of multiple populations at near-solar metallicity offers a unique opportunity to place stringent constraints on the nature of the polluters responsible for the chemical enrichment of 2P stars and to assess how their efficiency and nucleosynthetic signatures depend on metallicity.
The investigation of Bulge clusters also comprises NGC\,6440, based on archive data \citep{cadelano2023a}.

Finally, our sample includes the bulge GCs Terzan\,5 and Liller\,1, which exhibit unusually large metallicity spreads. In addition to old ($\sim$13 Gyr) stellar populations with subsolar metallicities, both systems host younger populations ($\lesssim$7 Gyr) characterized by super-solar iron abundances \citep[e.g.,][]{ferraro2009a, ferraro2021a, origlia2013a, massari2014a, zullo2026a}. Such pronounced age and metallicity variations are highly unusual among Galactic GCs and, to date, have been observed only in M\,54 (NGC\,6715), which resides at the center of the Sagittarius dwarf galaxy \citep[e.g.,][]{siegel2007a}.

\begin{table*}[h]
    \centering
        \caption{Summary of the JWST/NIRCam data used in this work. For each cluster, we list the observed filters, the number of images and total exposure time in each filter, the observation date, and the corresponding observing program and Principal Investigator (PI).}
    \setlength{\tabcolsep}{8pt} 
    \renewcommand{\arraystretch}{1.3} 
    \begin{tabular}{c c c c c c}
        \hline
        ID & Filters & N $\times$ Exposure Time [s]& Date & GO program & PIs \\
        \hline
        NGC\,104 & F115W       &  9$\times$236 & Oct 22 2025 & 8816 & A.\,Rest \\
                 & F200W-F444W &  9$\times$236 & Oct 22 2025 & 8816 & A.\,Rest \\
                 & F277W       &  9$\times$236 & Oct 22 2025 & 8816 & A.\,Rest \\
        NGC\,288 & F115W-F444W & 12$\times$257 & Nov 17 2025 & 8960 & A.\,P.\,Milone, A.\,F.\,Marino\\
                 & F200W-F277W & 12$\times$214 & Nov 17 2025 & 8960 & A.\,P.\,Milone, A.\,F.\,Marino\\
     NGC\,1851 & F115W-F444W & 12$\times$300 & Nov 14 2025 & 8960 & A.\,P.\,Milone,  A.\,F.\,Marino\\
                 & F200W-F277W & 12$\times$214 & Nov 14 2025 & 8960 & A.\,P.\,Milone,  A.\,F.\,Marino\\
     NGC\,2808 & F115W-F444W & 12$\times$343 & Mar 19-21 2026 & 8960 & A.\,P.\,Milone,  A.\,F.\,Marino\\
                 & F200W-F277W & 12$\times$257 & Mar 19-21 2026 & 8960 & A.\,P.\,Milone,  A.\,F.\,Marino\\
    NGC\,6440    & F115W-F444W & 20$\times$344 & Aug 17 2022 & 2204 & P.\,C.\,Freire\\
                 & F200W-F277W & 20$\times$193 & Aug 17 2022 & 2204 & P.\,C.\,Freire\\
    NGC\,6528    & F115W-F444W & 12$\times$387 & Aug 16 2025 & 8960 & A.\,P.\,Milone,  A.\,F.\,Marino\\
                 & F200W-F277W & 12$\times$301 & Aug 16 2025 & 8960 & A.\,P.\,Milone,  A.\,F.\,Marino\\
    NGC\,6553    & F115W-F444W & 12$\times$343 & Sep 16-18 2025 & 8960 & A.\,P.\,Milone,  A.\,F.\,Marino\\
                 & F200W-F277W & 12$\times$275 & Sep 16-18 2025 & 8960 & A.\,P.\,Milone,  A.\,F.\,Marino\\
    NGC\,6656    & F115W-F444W & 12$\times$129 & Sep 20-25 2025 & 8960 & A.\,P.\,Milone,  A.\,F.\,Marino\\
                 & F200W-F277W & 12$\times$107 & Sep 20-25 2025 & 8960 & A.\,P.\,Milone,  A.\,F.\,Marino\\
    NGC\,6723    & F115W-F444W & 12$\times$268 & Sep 04 2025 & 8960 & A.\,P.\,Milone,  A.\,F.\,Marino\\
                 & F200W-F277W & 12$\times$225 & Sep 04 2025 & 8960 & A.\,P.\,Milone,  A.\,F.\,Marino\\
    Liller\,1    & F200W       & 24$\times$1932 & Apr 22 2025 & 5381 & K.\,Burdge\\
    Terzan\,5    & F115W       & 8$\times$21$+$8$\times$966 & Sep 20-25 2024 & 5502 & F.\,R.\,Ferraro\\
                 & F200W       & 8$\times$21$+$8$\times$751 & Sep 20-25 2024 & 5502 & F.\,R.\,Ferraro\\
                 & F115W       & 8$\times$966 & Apr 04 2025 & 5502 & F.\,R.\,Ferraro\\
                 & F200W       & 24$\times$1932 & Apr 2-21 2025 & 5381 & K.\,Burdge\\
     
       \hline
    \end{tabular}
    \label{tab:data}
\end{table*}

\section{The dataset}\label{sec:data}

In this section, we describe the primary dataset used in this work, which consists of JWST/NIRCam images acquired through the F115W and F200W filters of the short-wavelength channel. For Liller\,1, for which F115W observations are not available, we used F814W images obtained with the Wide Field Channel of the Advanced Camera for Surveys (ACS/WFC) on board HST (GO\,15231, PI.\,F.\,Ferraro)\footnote{The dataset consists of eight images acquired on 17 August 2019, with individual exposure times between 836 and 855\,s.} We also incorporated F277W and F444W images from the NIRCam long-wavelength channel, available for all clusters except Liller\,1 and Terzan\,5. The main characteristics of the dataset are summarized in Table~\ref{tab:data}.

To derive stellar proper motions and extend the color baseline, we additionally exploited all suitable JWST archival images, together with HST observations obtained with ACS/WFC and with both the ultraviolet-visible (UVIS) and infrared (IR) channels of the Wide Field Camera 3 (WFC3). The ACS/WFC and WFC3/UVIS data were retrieved from the archive in their charge-transfer efficiency (CTE)-corrected form \citep{anderson2010a}.
 The data used in this paper are available in the Mikulski Archive for Space Telescopes
(MAST) \footnote{
NGC\,104: \url{https://doi.org/10.17909/k0nz-k320  }\\
and  \url{https://doi.org/10.17909/vha2-8275;} \citep{ziliotto2025a},
NGC\,288: \url{https://doi.org/10.17909/cqhd-jk81},\\
NGC\,1851: \url{https://doi.org/10.17909/2q7r-4y77},\\
NGC\,2808: \url{https://doi.org/10.17909/ztzd-kn19}, \\
NGC\,6528: \url{https://doi.org/10.17909/y828-c21}, \\ 
NGC\,6440: \url{https://doi.org/10.17909/vk1a-ne77  }, \\
NGC\,6553: \url{https://doi.org/10.17909/2w6v-gm71},\\
NGC\,6656: \url{https://doi.org/10.17909/r395-2w73},\\
NGC\,6723: \url{https://doi.org/10.17909/40f8-b520}, \\
Liller\,1: \url{https://doi.org/10.17909/g316-0192},  \\
Terzan\,5: \url{https://doi.org/10.17909/kvs2-x684}. \\}.

This section is organized as follows. In the next subsection, we discuss why the adopted filter combinations are particularly effective for identifying multiple stellar populations in GCs. We then describe the observational datasets and the data-reduction procedures in Section~\ref{subsec:data}.

\subsection{Filters}\label{subsec:filters}
The F115W and F200W filters employed in the GO-8960 observations provide the most efficient separation between 1P and 2P M-dwarfs in NIRCam’s short-wavelength channel \citep{milone2023a}.\footnote{
At fixed signal-to-noise ratio, the F090W$-$F200W color yields a larger separation between 1P and 2P M-dwarfs at comparable luminosity; however, this advantage comes at the cost of significantly longer exposure times, as indicated by exposure-time-calculator estimates.}
Color-magnitude diagrams (CMDs) constructed using long-wavelength filters such as F277W in combination with F115W can, in principle, produce an even larger separation between 1P and 2P stars than F115W$-$F200W. However, these filters are less effective in crowded fields due to increased blending and reduced spatial resolution. Nevertheless, complementary observations in F277W and F444W provide valuable constraints on the average oxygen abundance, enabling broad wavelength coverage and a robust characterisation of O variations across stellar populations.

\begin{figure}
    \centering
    \includegraphics[width=1.0\linewidth]{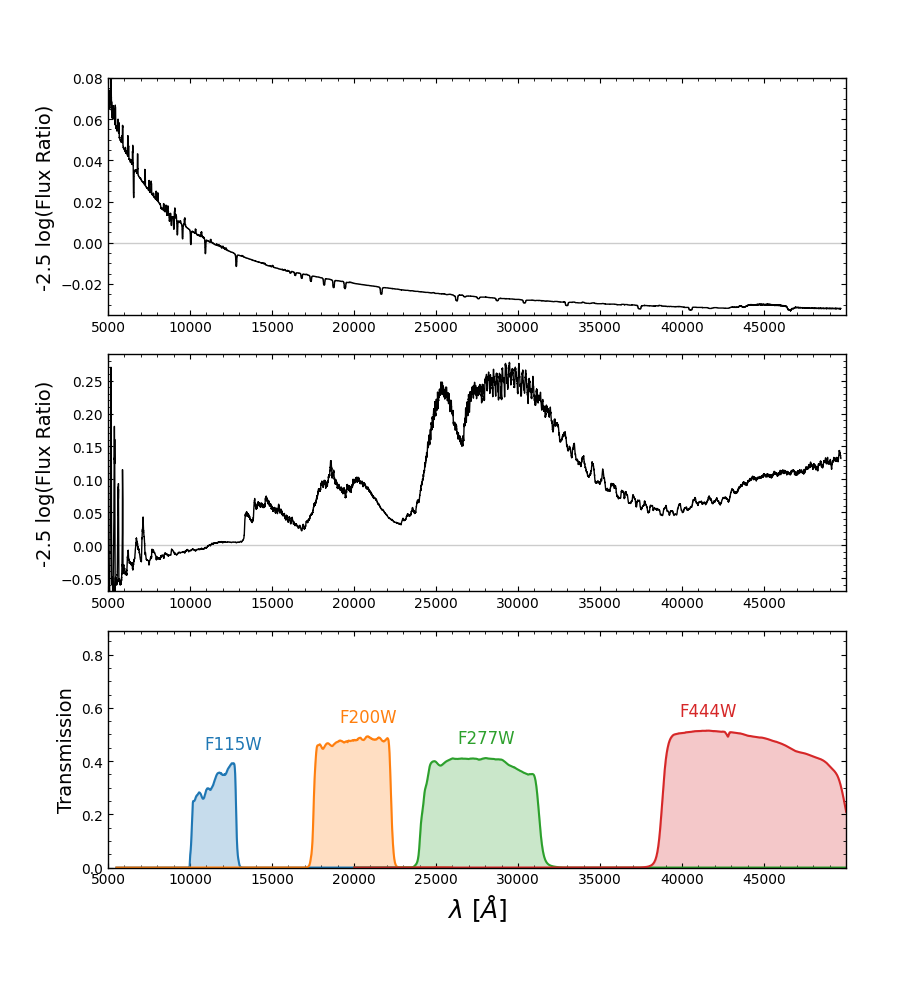}
\caption{\textit{Top and middle panels:} 
{Ratio of the synthetic spectra for stars with [Fe/H]$=-1.5$, computed as the flux of the 1P spectrum divided by that of the 2P spectrum (1P/2P), and expressed as the corresponding magnitude difference}
\citep{milone2023b}.
The 2P models are characterized by enhanced He and N abundances and depleted C and O abundances relative to their 1P counterparts. The top panel refers to bright MS stars, whereas the middle panel shows stars fainter than the MS knee. \textit{Bottom panel:} Transmission curves of the NIRCam filters used in this work.}
    \label{fig:filters}
\end{figure}

Figure\,\ref{fig:filters} compares the transmission curves of the filters adopted in this work (bottom panels) with the flux ratio between synthetic spectra of 1P and 2P stars matched in F115W flux. The spectra, computed by \cite{milone2023b}, assume [Fe/H]$=-1.5$ and [$\alpha$/Fe]$=0.4$. The 2P models are characterised by carbon and oxygen depletions of 0.5 and 0.9\,dex, respectively, and a nitrogen enhancement of 1.2\,dex relative to 1P stars. In addition, we adopt different helium mass fractions, namely $Y=0.247$ for 1P and $Y=0.33$ for 2P stars.

The upper panel shows results for bright MS stars above the knee, with $(T_{\rm eff}, \log g) = (5901\,\mathrm{K}, 4.50)$ for 1P and $(6048\,\mathrm{K}, 4.48)$ for 2P stars. The lower panel refers to M-dwarfs, for which both populations are assumed to have $\log g = 5.03$, with effective temperatures of $T_{\rm eff} = 3762$\,K (1P) and $3808$\,K (2P), respectively. The effective temperature and surface gravity adopted for the synthetic spectra were extracted from the  isochrones of \cite{dotter2008a}
 \citep[see the section 3.4 from][for details]{milone2023b}.

In M-dwarfs, the spectral morphology is primarily governed by oxygen abundance variations. Molecular absorption bands of TiO, CO, OH, and especially H$_2$O strongly shape the spectral regions probed by F200W, F277W, and F444W, while having a comparatively minor impact on the F115W band \citep{milone2012b, marino2024b}.

By contrast, the flux differences among bright MS stars are dominated by effective temperature variations. An enhanced helium abundance increases the mean molecular weight of the stellar interior, leading to higher core temperatures and more efficient nuclear burning. Consequently, helium-rich stars reach higher effective temperatures at fixed luminosity, producing systematically bluer colours compared to helium-normal stars.

\begin{figure}
    \centering
    \includegraphics[width=1.0\linewidth]{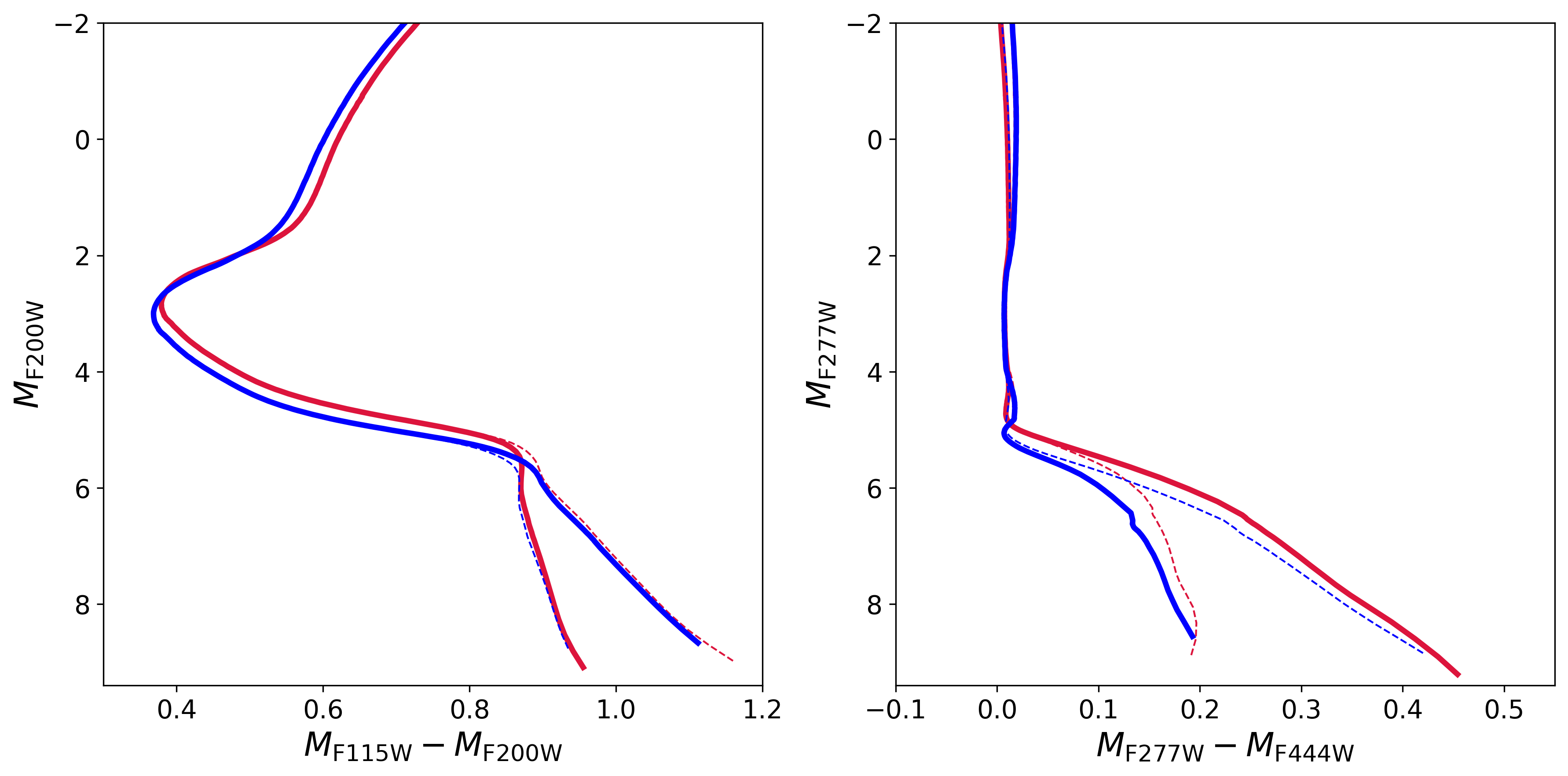}
\caption{Dartmouth isochrones \citep{dotter2008a} for an age of 13\,Gyr, [Fe/H]$=-1.5$, and [$\alpha$/Fe]$=0.4$. The models cover stellar masses $M > 0.1\,M_\odot$. Crimson curves correspond to a primordial helium abundance ($Y=0.246$), while blue curves represent helium-enhanced models ($Y=0.33$). The solid blue and dashed red curves additionally assume carbon and oxygen depletions of 0.5 and 0.9\,dex, respectively, and a nitrogen enhancement of 1.2\,dex relative to the other models. The remaining isochrones adopt [O/Fe]$=0.4$ and solar carbon and nitrogen abundances.}
    \label{fig:iso}
\end{figure}

Figure\,\ref{fig:iso} shows the isochrones {from \citep{dotter2008a}} of 1P and 2P stars in the $M_{\rm F200W}$ vs.\,$(M_{\rm F115W}-M_{\rm F200W})$ and $M_{\rm F277W}$ vs.\,$(M_{\rm F277W}-M_{\rm F444W})$ CMDs (crimson and blue solid lines, respectively). We also include isochrones {from the same database} computed by changing only the helium abundance while keeping the C, N, and O content of 1P stars fixed (blue dashed lines), and by changing only the CNO abundances while maintaining the 1P helium content (crimson dashed lines).

Above the MS knee, the split observed in the left-panel CMD is primarily driven by differences in helium abundance, with helium-rich stars forming a bluer MS locus. Below the MS knee, the MSs instead reflect variations in oxygen abundance, with O-rich stars appearing bluer than O-poor stars at fixed luminosity \citep{milone2012b, marino2024a, marino2024b}.

In contrast, in the right-panel CMD the upper MS of 1P and 2P stars is nearly vertical and largely overlapping. The sequence separation becomes apparent only below the MS knee, where 1P stars appear redder than 2P stars due to their higher oxygen content. The sharp change in the MS slope makes this CMD particularly effective for identifying the location of the MS knee. We also note that the magnitude of the MS knee is sensitive to helium abundance, with helium-rich (2P) stars exhibiting a fainter MS knee than helium-normal (1P) stars.

\subsection{Data reduction}\label{subsec:data}
Stellar photometry and astrometry were performed using the KS2 software package developed by Jay Anderson, which represents an advanced implementation of the reduction techniques originally introduced for ACS/WFC data \citep{anderson2008a}. KS2 simultaneously processes all available exposures and adopts multiple measurement strategies optimized for stars of different brightness levels, an approach that has been successfully applied in numerous previous studies \cite[see e.g.][for details on KS2]{sabbi2016a, bellini2017a, milone2023a}.

For relatively bright stars, we used a PSF-fitting procedure in which stellar fluxes and positions are independently measured in each exposure. These measurements are obtained by fitting a spatially variable effective point-spread function \citep[ePSF,][]{anderson2000a, anderson2022a} that accounts for the dependence of the PSF shape on the detector position. The local sky background is estimated from an annular region surrounding each source, and the final photometric and astrometric quantities are derived by averaging the results from all images.

Fainter stars, for which PSF fitting in individual exposures is not sufficiently robust, are measured using aperture-based techniques after subtracting the contribution of neighboring sources. 
This method, specifically designed for extremely faint sources in crowded fields, relies on weighted-aperture photometry after subtraction of neighboring stars. For NIRCam and WFC3/IR images, we used a small aperture with a radius of 0.75 pixels and estimated the local sky background within an annulus extending from 2 to 4 pixels from the source position measured during the finding stage. For the remaining {\it HST} images, we adopted a $5\times5$ pixel box and estimated the sky background from an annulus between 4 and 8 pixels from the stellar centre. For both bright and faint sources, measurements from the individual images are combined to obtain the final magnitudes and positions.

KS2 provides a set of diagnostic parameters that quantify the quality of the photometric and astrometric measurements. These include the RADXS parameter, a shape diagnostic that measures the excess flux relative to the best-fitting PSF; the quality-of-fit parameter, which quantifies the goodness of the PSF fit; and the root-mean-square scatter of the magnitude measurements \citep[][]{anderson2008a, bedin2008a}.
 These diagnostics were used to select stars that are well isolated and well described by the ePSF model, following the selection criteria established by \citet{milone2023a}. This procedure ensures a clean sample with high-precision measurements suitable for detailed analysis of fine details of the CMD and of stellar proper motions.

The instrumental magnitudes were calibrated to the Vega photometric system by applying encircled-energy corrections and photometric zero points released by STScI for the used NIRCam filters\footnote{\url{https://jwst-docs.stsci.edu /jwst-near-infrared-camera/nircam-performance/nircam-absolute-flux-calibration-and-zeropoints}}. Corrections for pixel-area variations were applied, and stellar positions were adjusted for geometric distortion using the most recent calibration solutions available provided by Jay Anderson. Astrometric catalogs from different epochs were placed onto a common reference frame, allowing the computation of relative proper motions. These were used in this paper to distinguish cluster members from field stars. For the {\it HST} images, we used the distortion solutions provided by \citet{anderson2006a, anderson2022a, bellini2009a} and \cite{bellini2011a}.

To assess photometric uncertainties and completeness, and to construct synthetic photometric diagrams, we performed extensive artificial-star tests following the prescriptions adopted in previous studies \citep[e.g.][]{anderson2008a, milone2023a}. A total of 300,000 artificial stars were injected into the images of each cluster, reproducing the observed radial distribution and luminosity function. The artificial stars were placed along the fiducial sequences, from the lower MS to the red giant branch, and were analyzed using the same reduction procedures applied to real stars. The comparison between the input and recovered properties of the artificial stars was then used to quantify measurement errors and the level of completeness of our sample \citep[see][for details]{milone2012a}.

\section{Photometric diagrams}\label{sec:cmds}

\begin{figure*}
    \centering
    \includegraphics[width=0.9\linewidth]{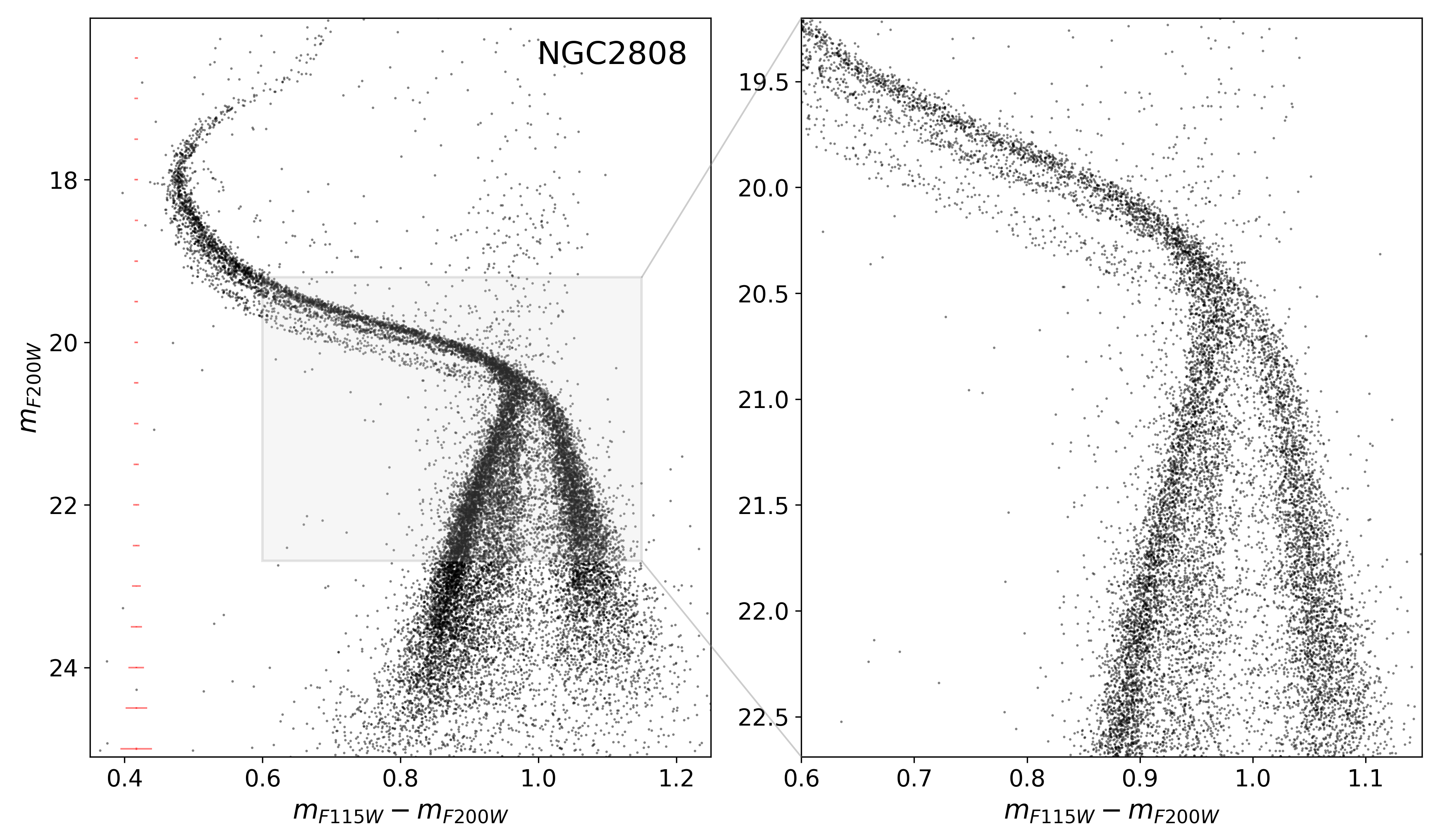}
\caption{CMD of NGC\,2808 from short-wavelength channel photometry (left). 
Red error bars indicate the average photometric uncertainties in colour and magnitude computed in different magnitude bins. The right panel shows a zoom of the region of the left-panel CMD around the MS knee.}
    \label{fig:ngc2808twopanel}
\end{figure*}

As an example, the left panel of Fig.\,\ref{fig:ngc2808twopanel} shows the $m_{\rm F200W}$ versus $m_{\rm F115W}-m_{\rm F200W}$ CMD for stars in the field of view of NGC\,2808, while the right panel provides a zoom-in around the MS knee. This region of the CMD clearly emphasises the distinct behaviour of multiple stellar populations across different evolutionary regimes.

In agreement with previous optical studies, the upper MS is mainly affected by variations in helium abundance. The bluest sequence corresponds to the most chemically extreme population, the reddest sequence to stars with near-primordial helium content, while intermediate sequences trace progressively increasing helium enhancement \citep{dantona2005a, piotto2007a, milone2012b, milone2015a}.

Below the MS knee, the morphology is instead dominated by differences in oxygen abundance, with oxygen-rich stars appearing bluer than oxygen-poor stars at fixed luminosity \citep{milone2012b, marino2024b}. As 2P stars are both helium-enhanced and oxygen-depleted relative to 1P stars, the corresponding sequences intersect near the MS knee, producing an inversion in their relative colours across the transition.

This behaviour, first identified by \cite{milone2012a} using HST/WFC3 infrared observations, is now recovered with much greater clarity thanks to the superior precision of the NIRCam photometry shown in Fig.\,\ref{fig:ngc2808twopanel}. For a detailed discussion of the impact of helium and light-element variations on NIRCam isochrones, we refer the reader to \citet{salaris2019a, milone2023a, ziliotto2023a, marino2024a}.

\begin{figure*}
    \centering
    \includegraphics[width=0.32\linewidth]{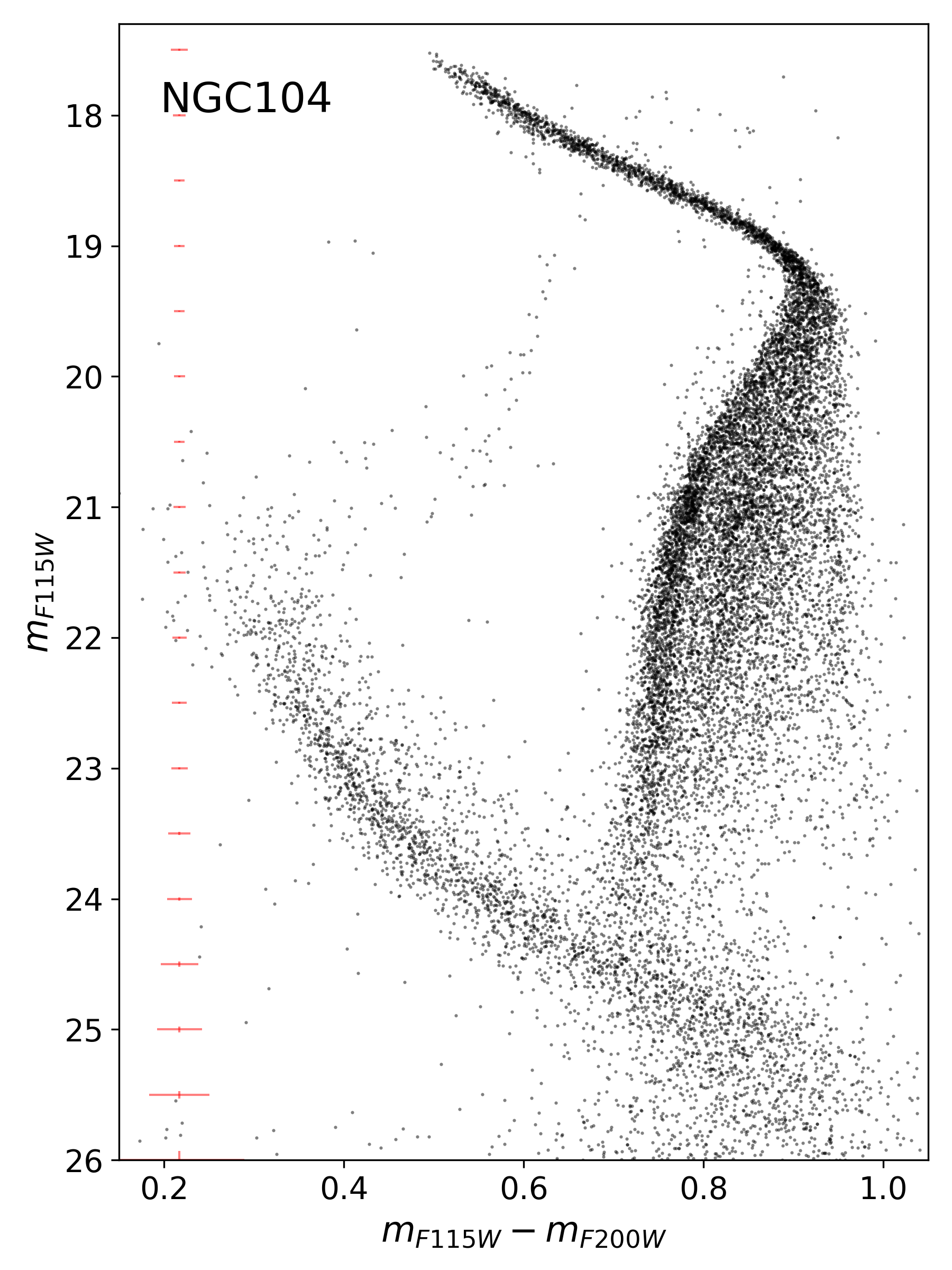}
    \includegraphics[width=0.32\linewidth]{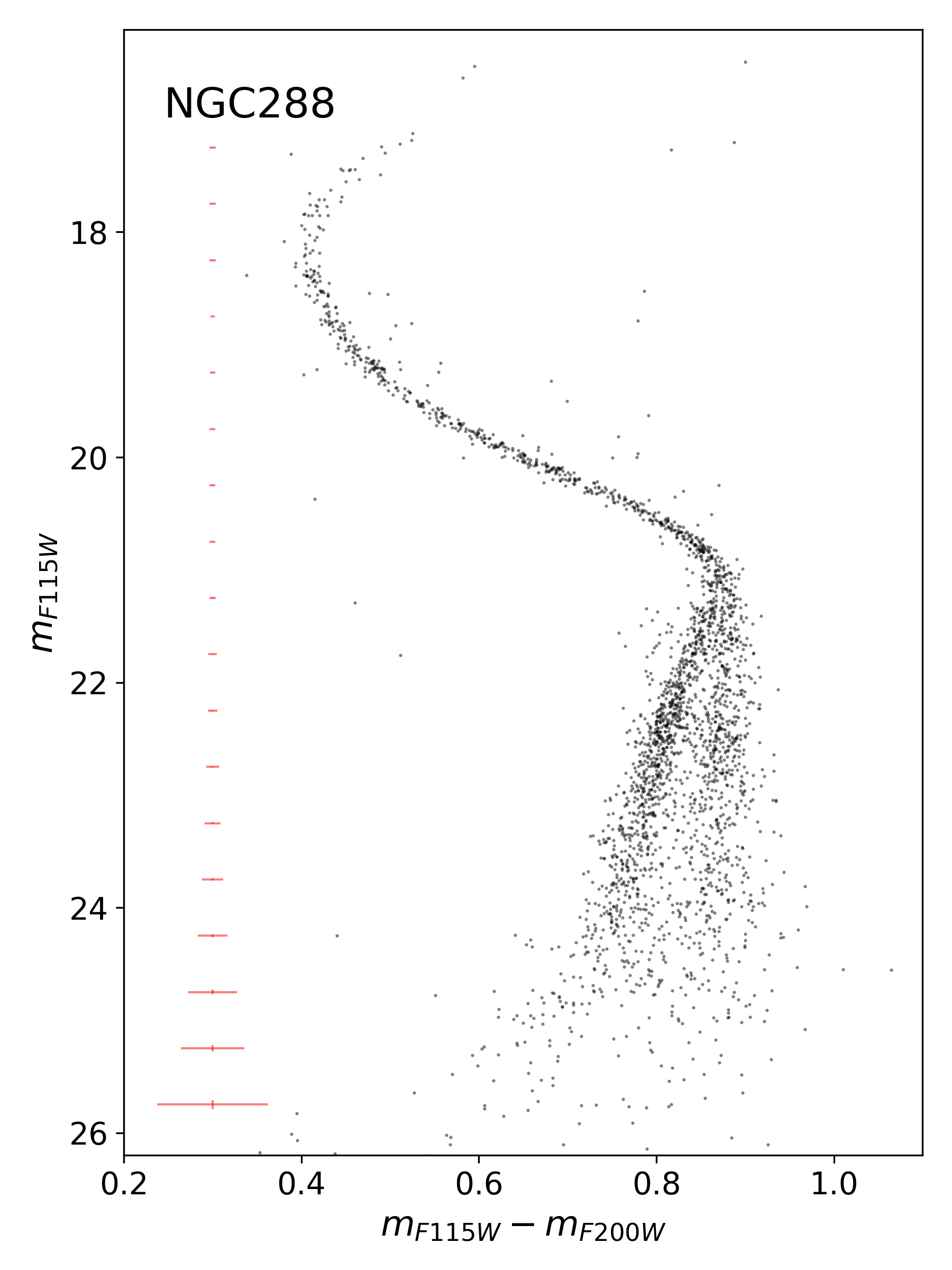}
    \includegraphics[width=0.32\linewidth]{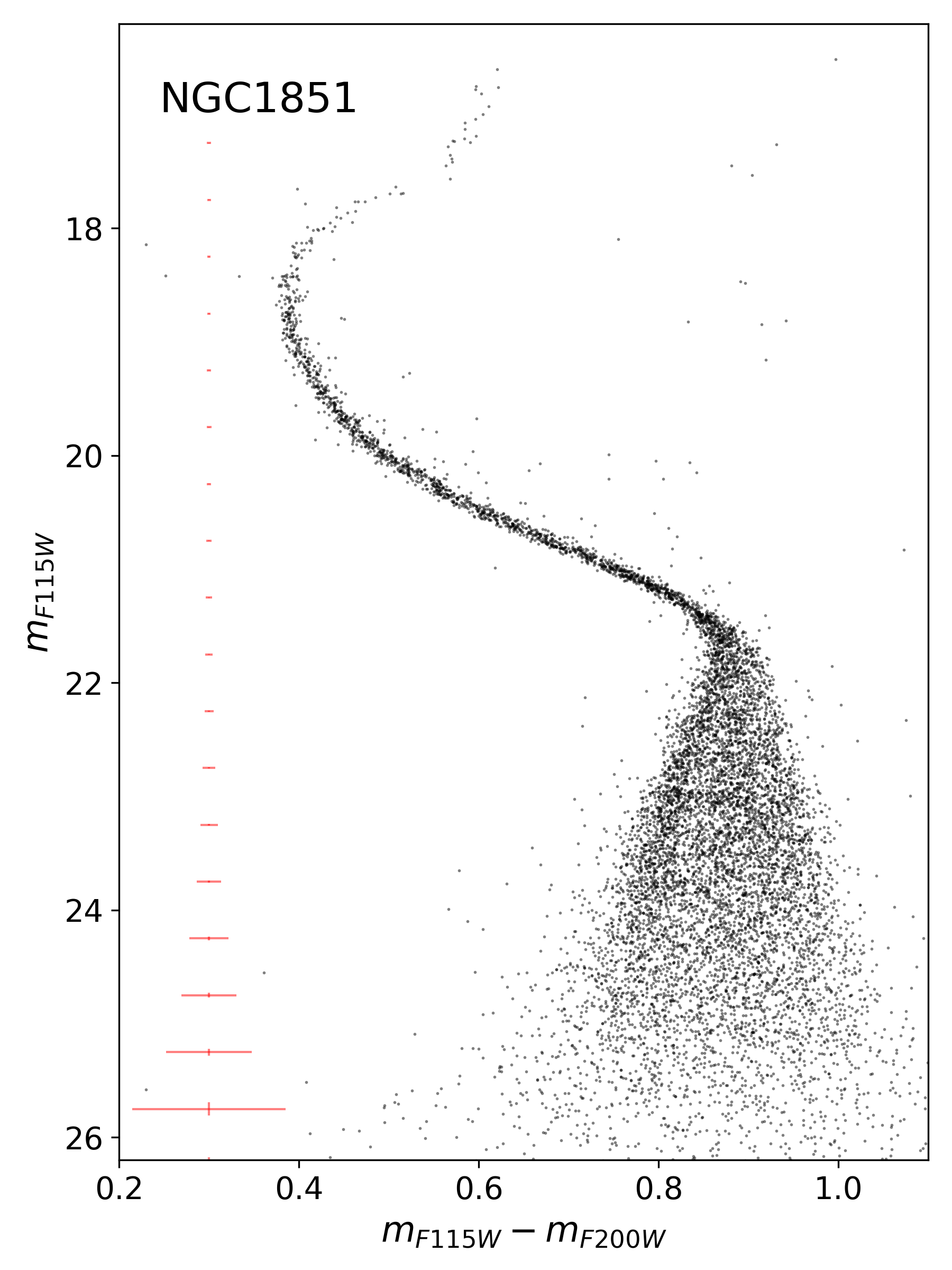}
    \includegraphics[width=0.32\linewidth]{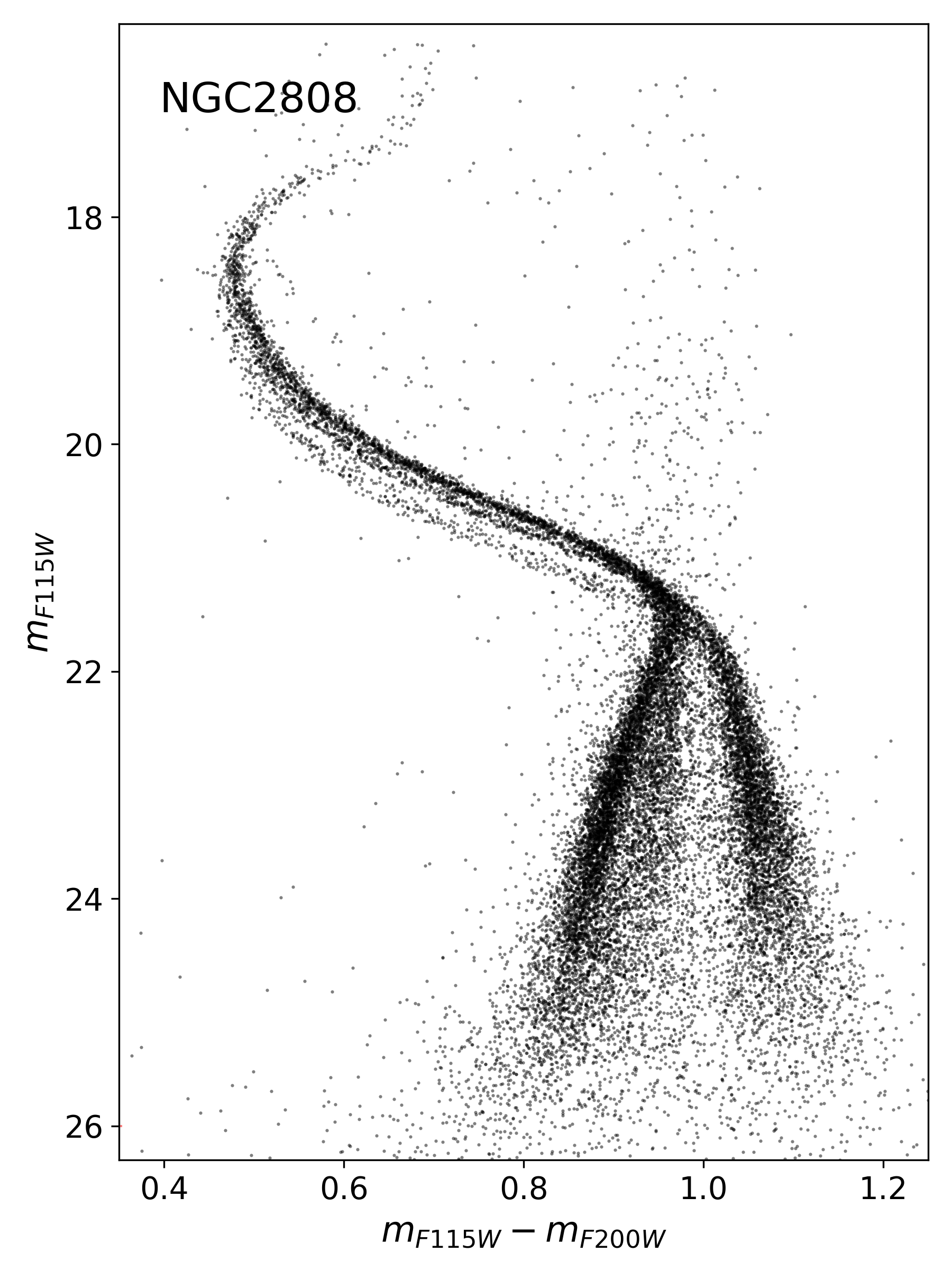}
    \includegraphics[width=0.32\linewidth]{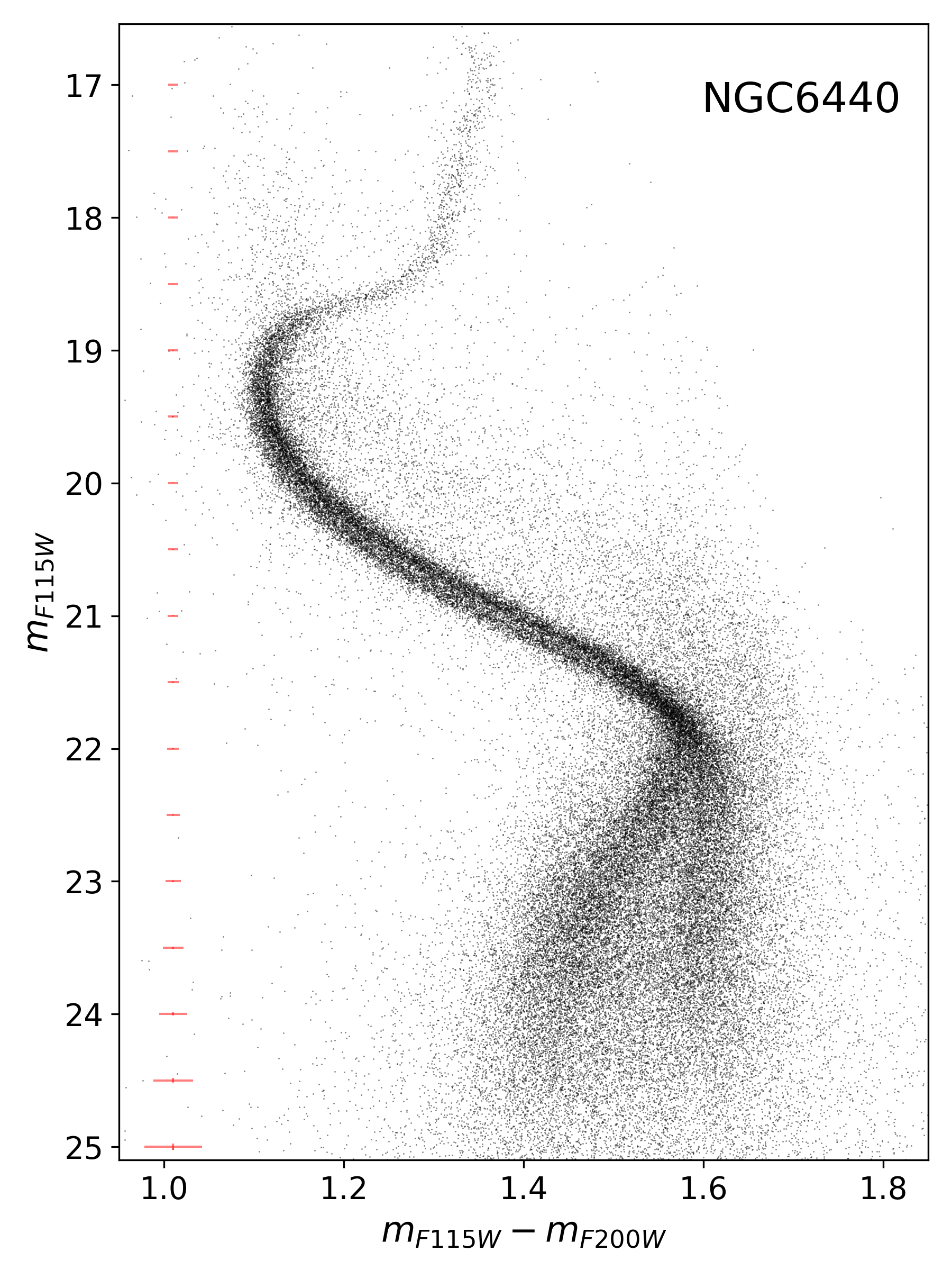}
    \includegraphics[width=0.32\linewidth]{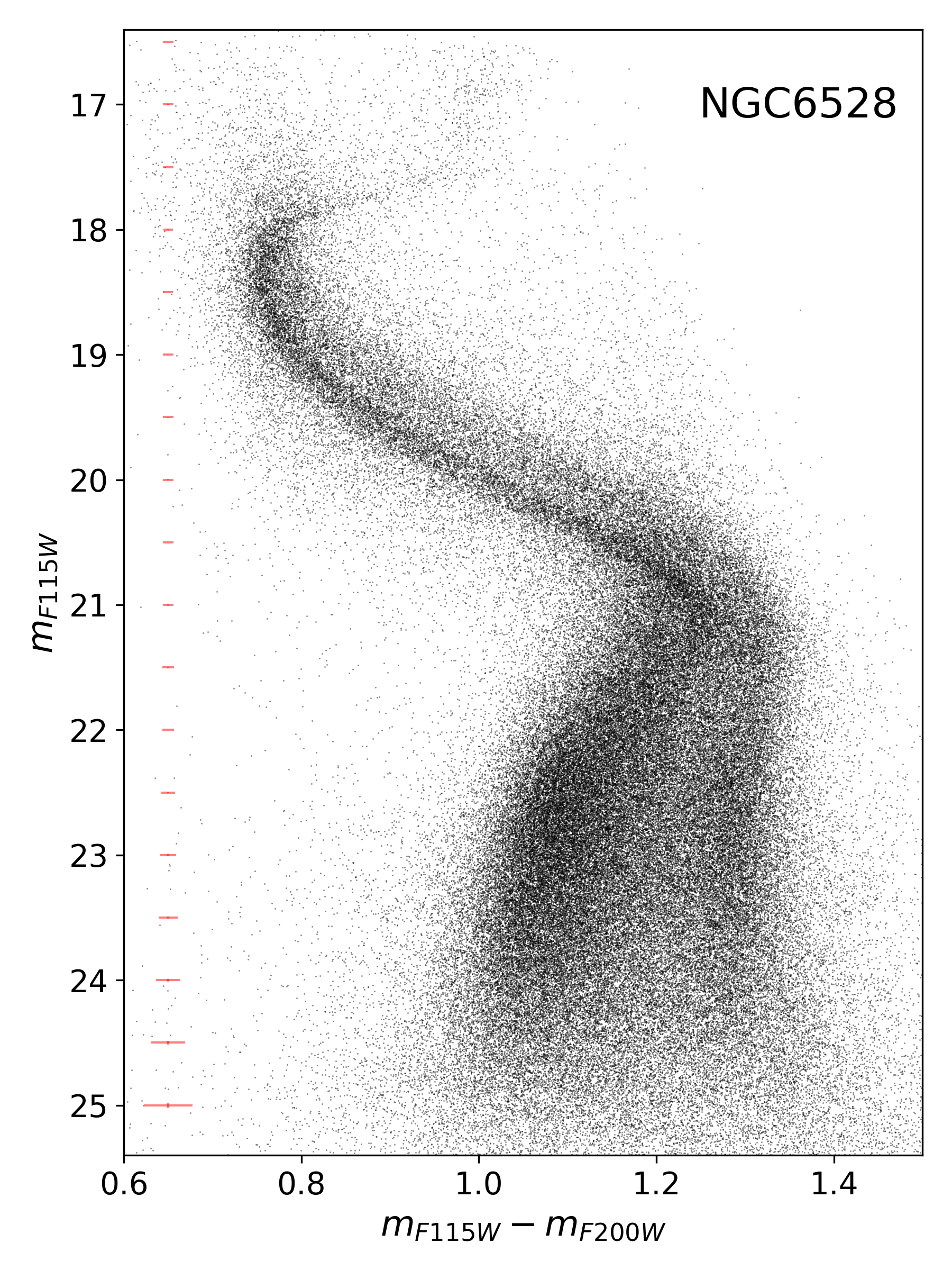}
    \includegraphics[width=0.32\linewidth]{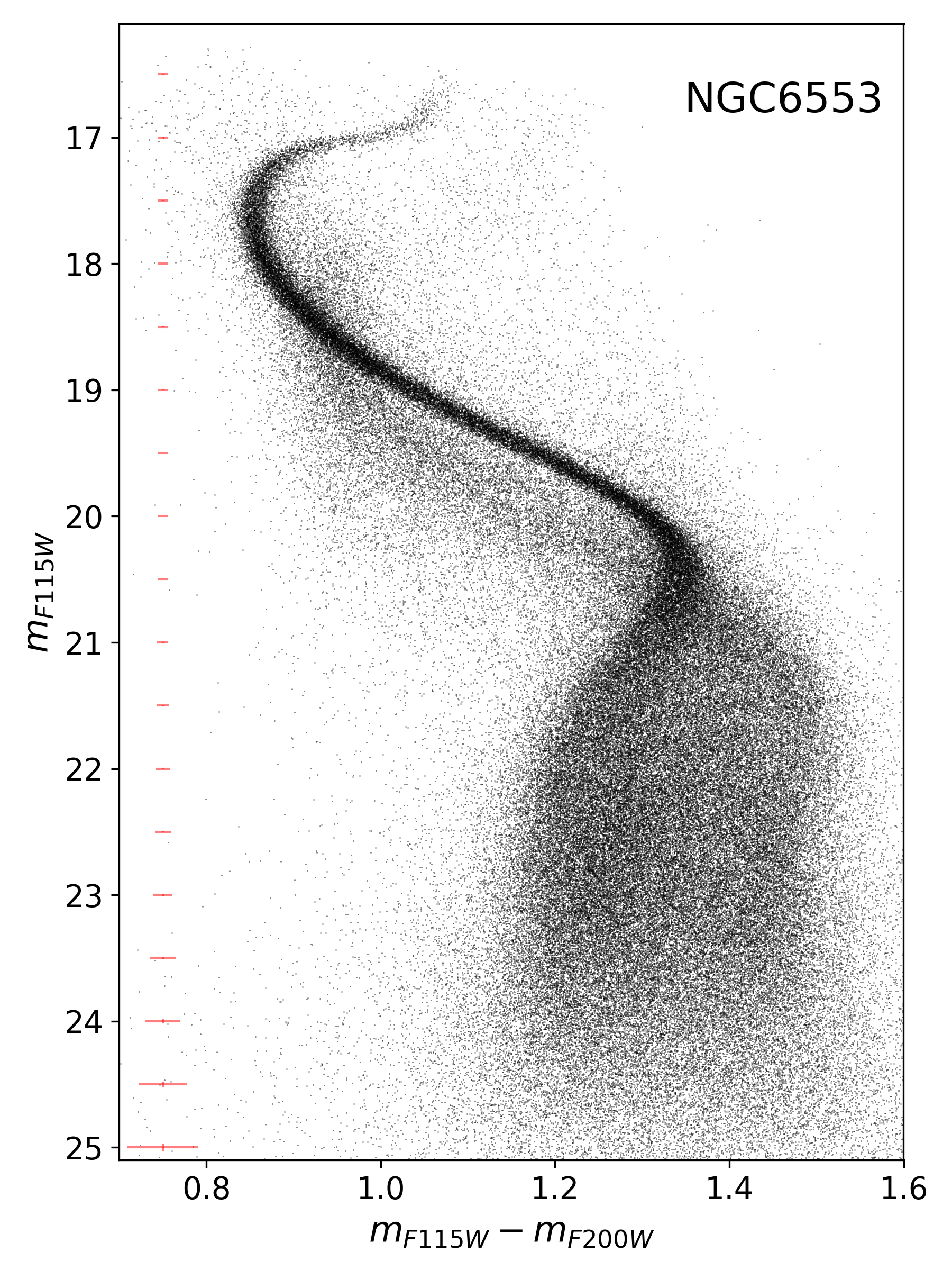}
    \includegraphics[width=0.32\linewidth]{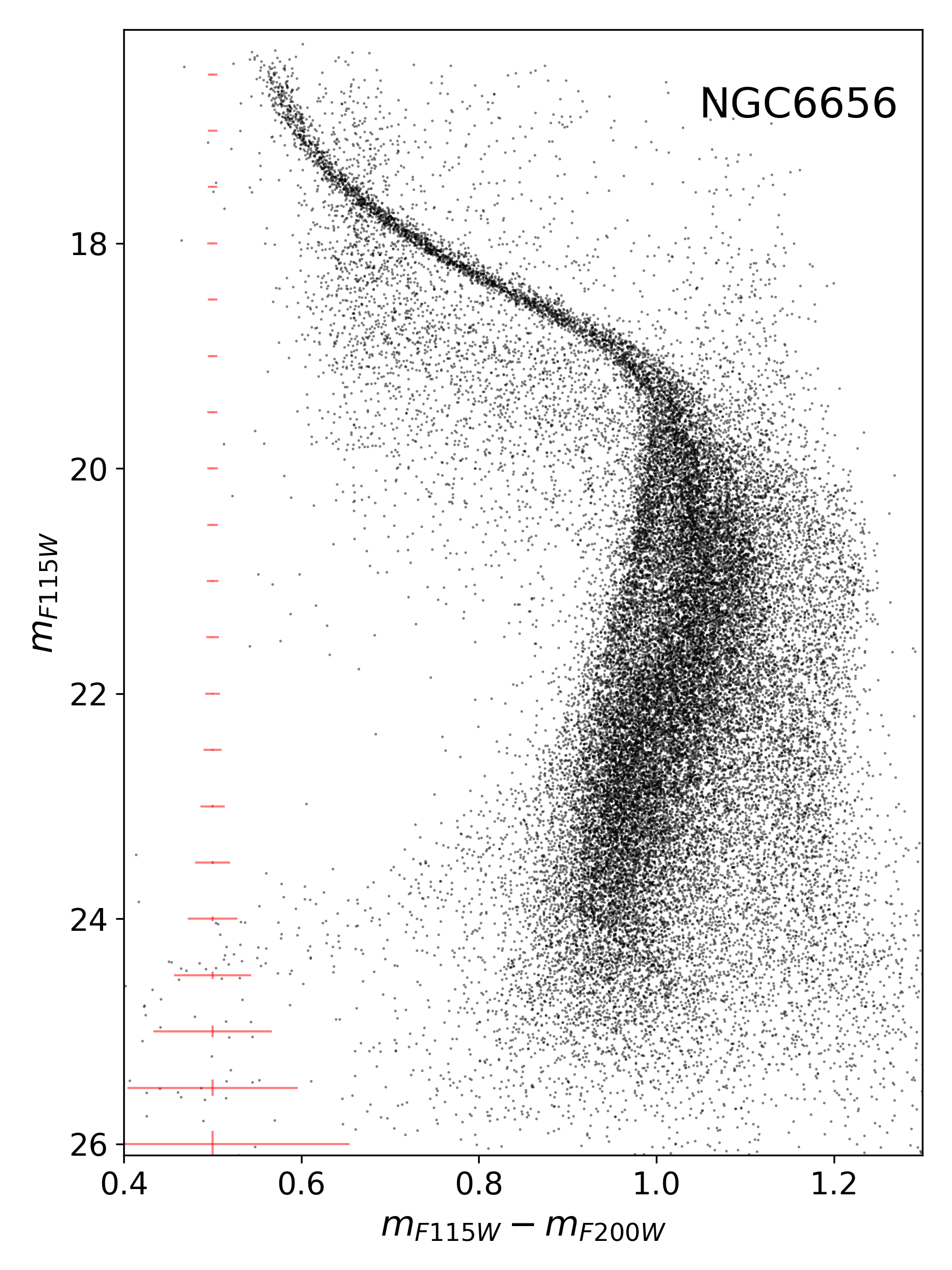}
    \includegraphics[width=0.32\linewidth]{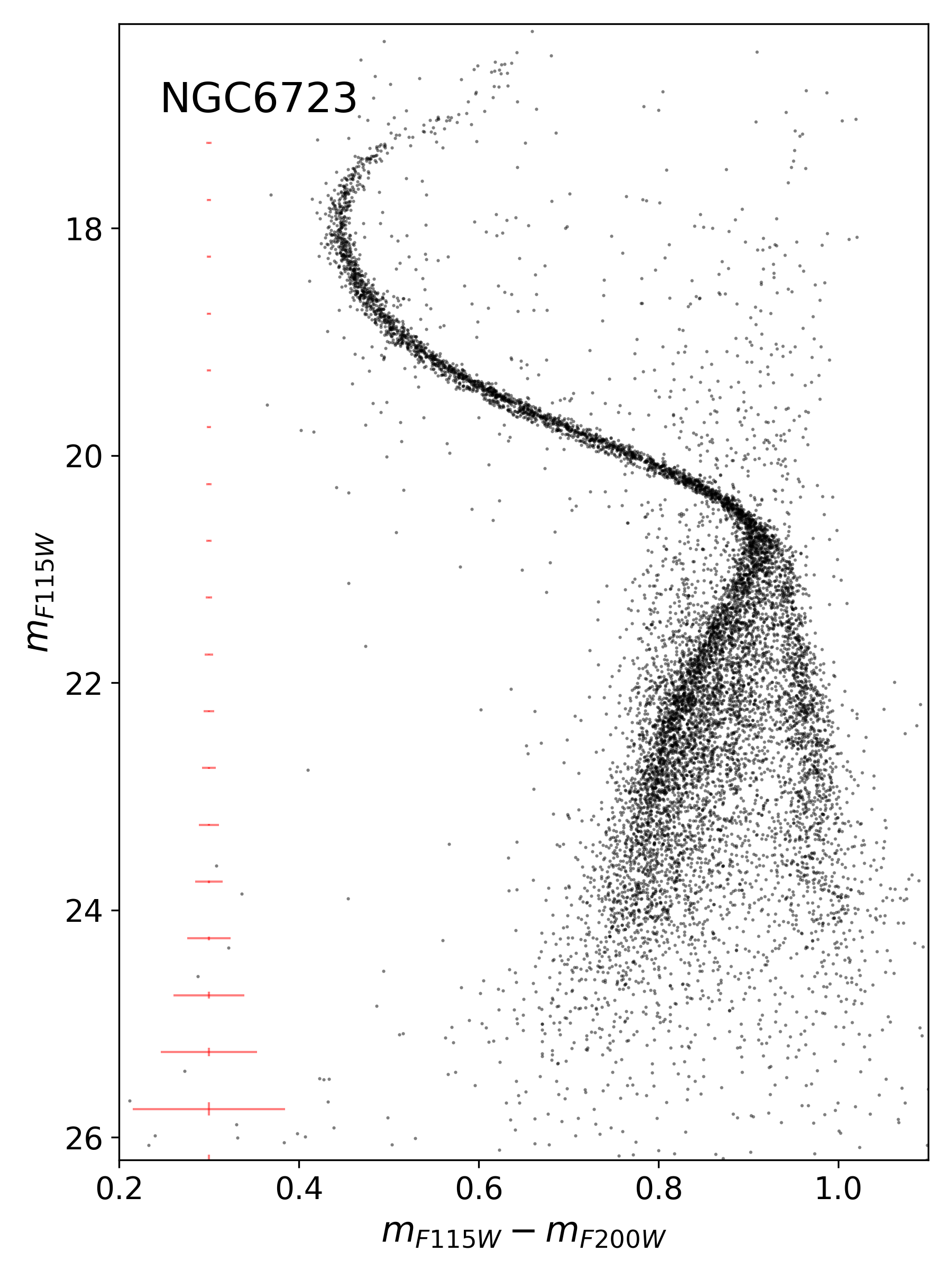}    
\caption{$m_{F115W}$ versus $m_{F115W}-m_{F200W}$ CMDs corrected for differential reddening of stars in the fields of view of the studied GCs, sorted in alphabetical order.  The average color and magnitude uncertainties, calculated for stars in different magnitude bins, as a function of magnitude are indicated by the red error bars plotted on the left side of each diagram.}
    \label{fig:CMDsSW}
\end{figure*}

\begin{figure*}
    \centering   
    \includegraphics[height=0.45\linewidth]{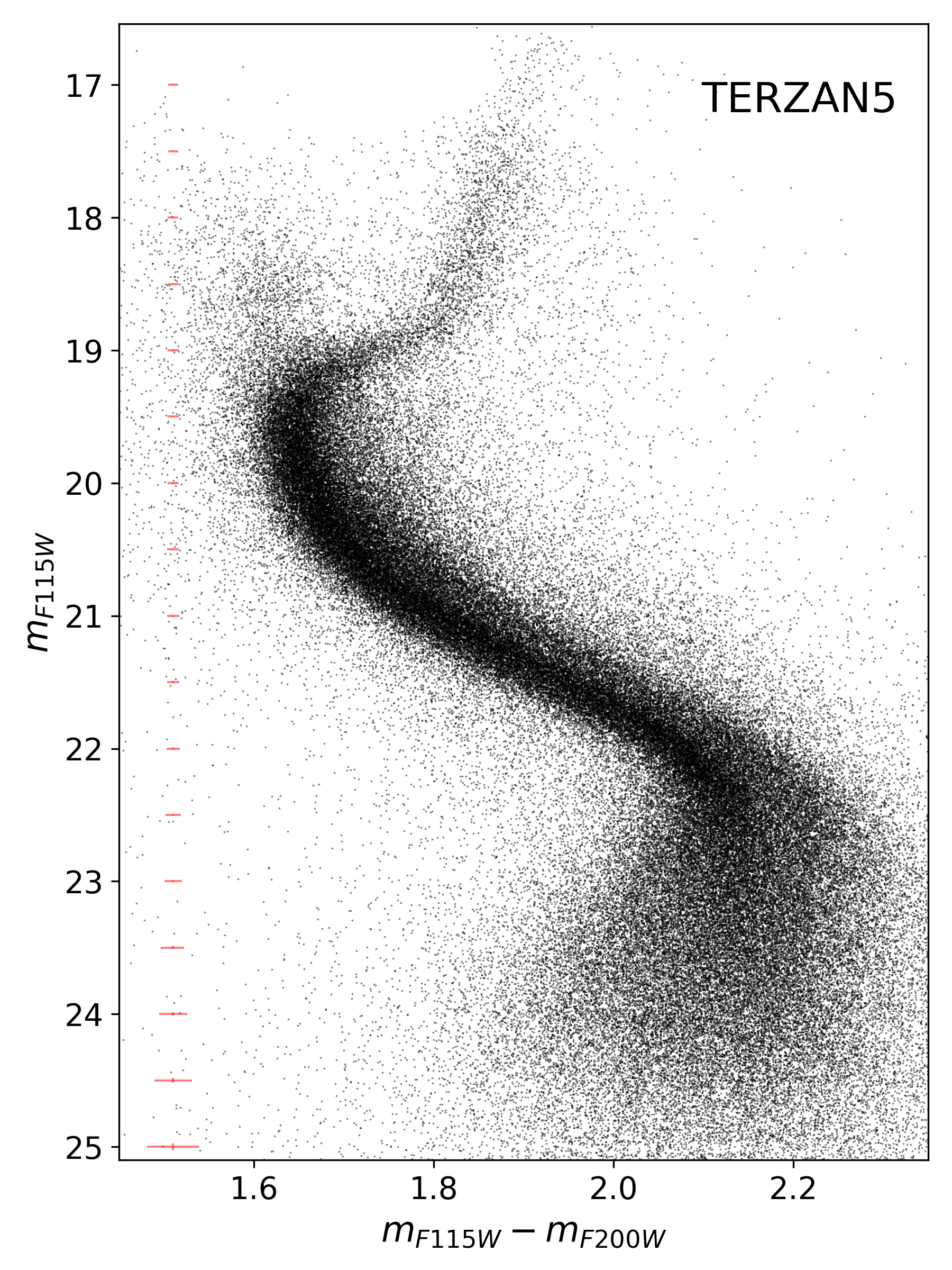}    
    \includegraphics[height=0.45\linewidth]{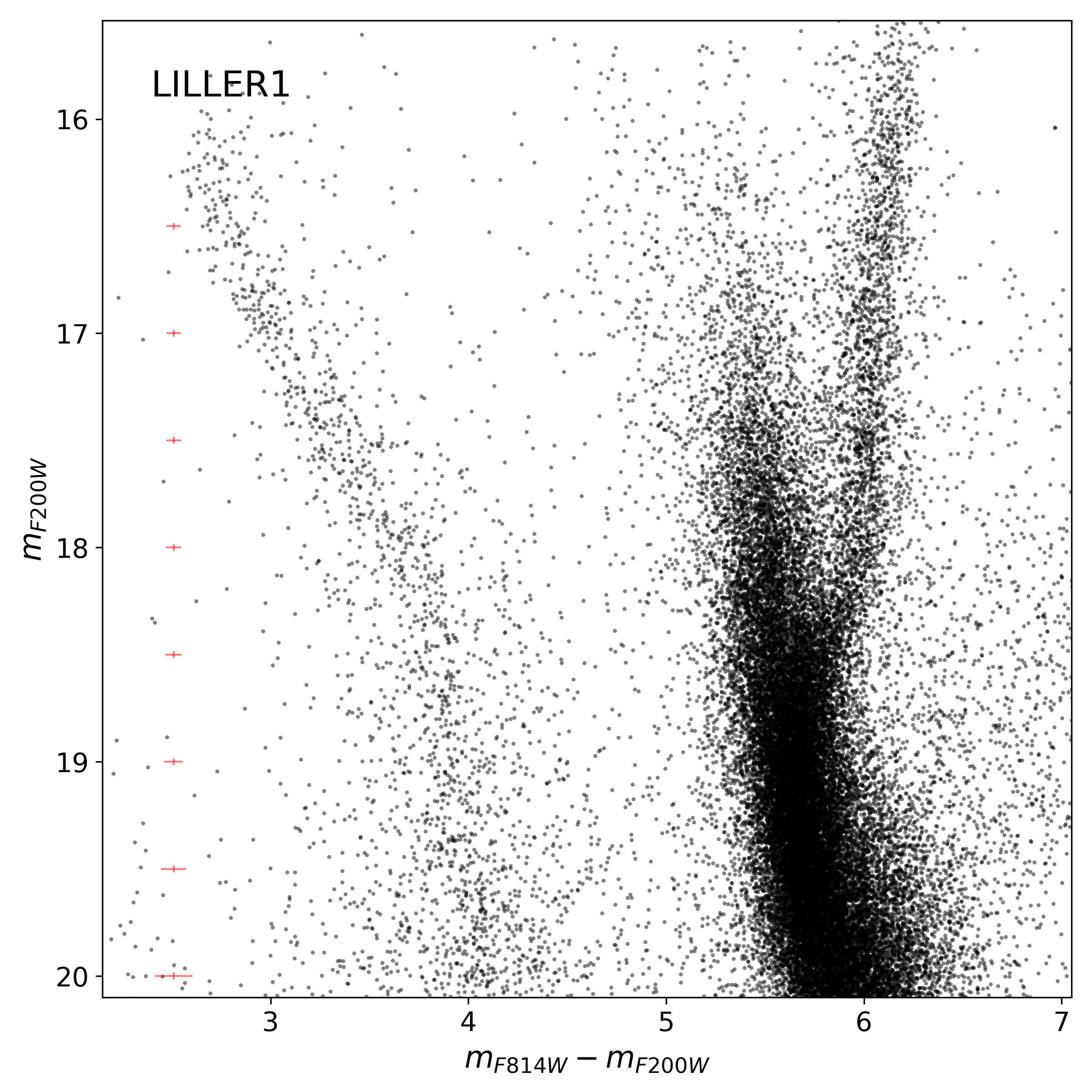}    
\caption{Differential-reddening-corrected CMDs of stars in the fields of view of Terzan\,5 (left) and Liller\,1 (right). Left: $m_{\rm F115W}$ versus $m_{\rm F115W}-m_{\rm F200W}$. Right: $m_{\rm F200W}$ versus $m_{\rm F814W}-m_{\rm F200W}$.
}
    \label{fig:CMDsSWb}
\end{figure*}

\begin{figure*}
    \centering
    \includegraphics[width=0.32\linewidth]{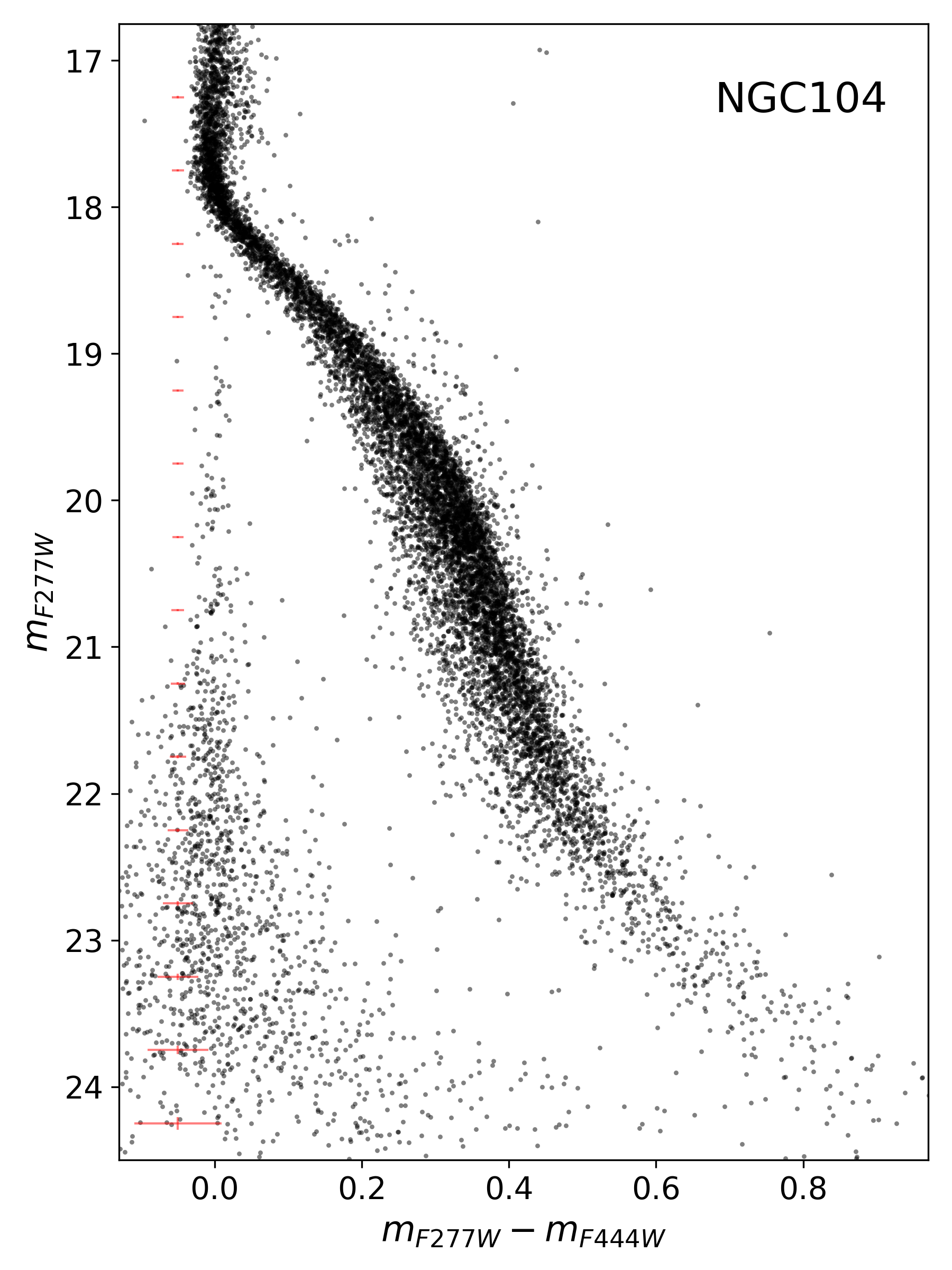}
    \includegraphics[width=0.32\linewidth]{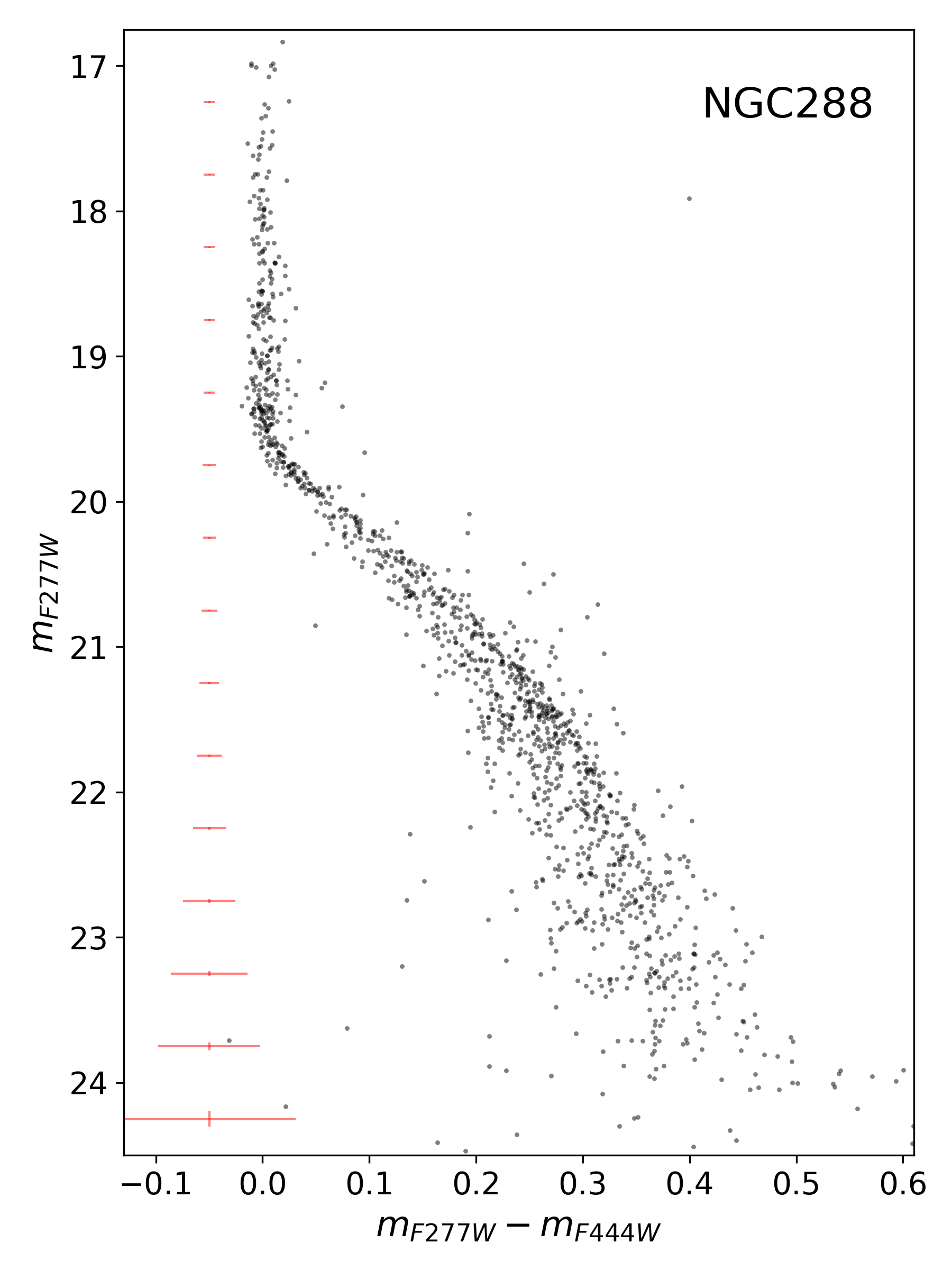}
    \includegraphics[width=0.32\linewidth]{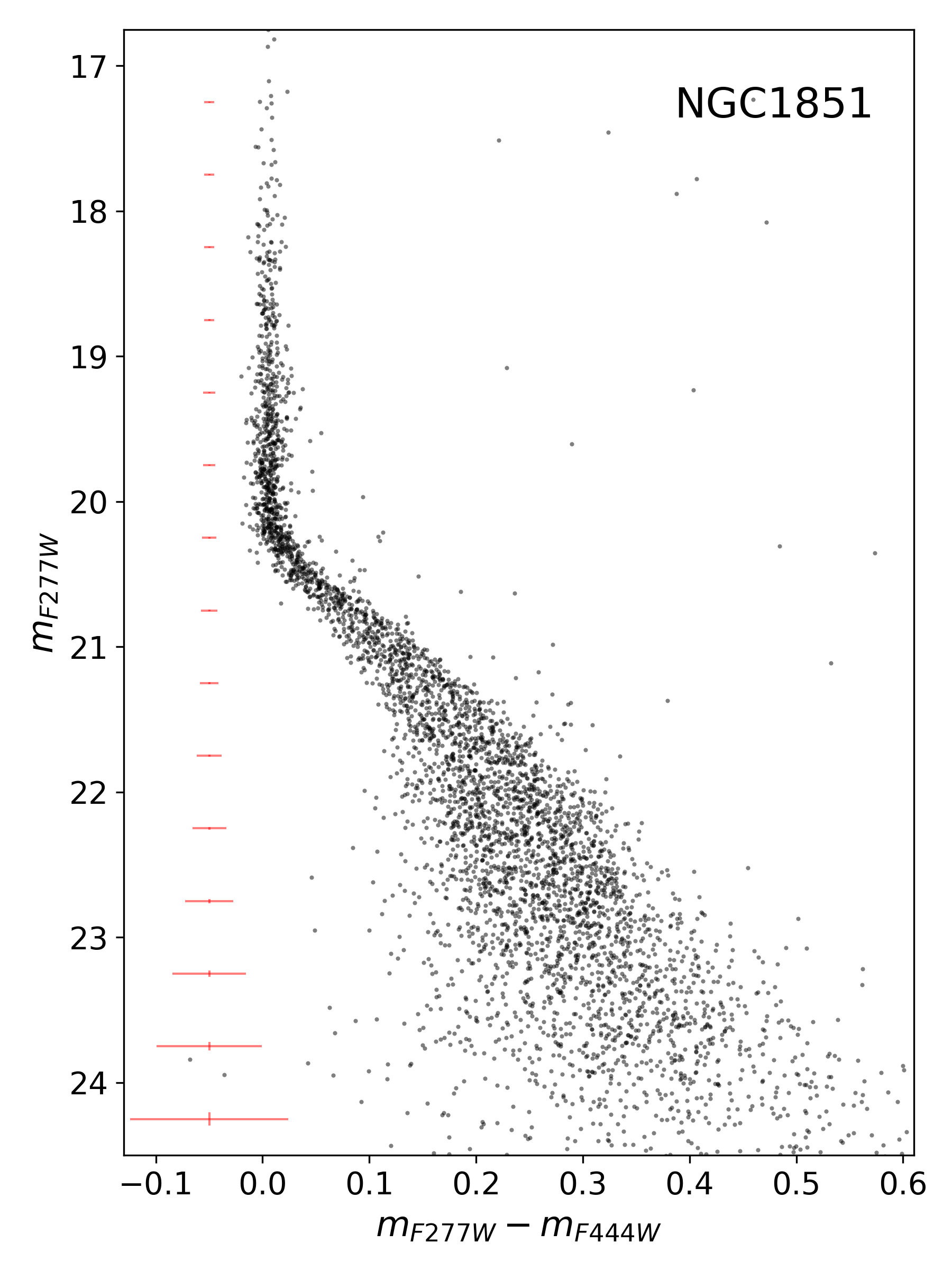}
    \includegraphics[width=0.32\linewidth]{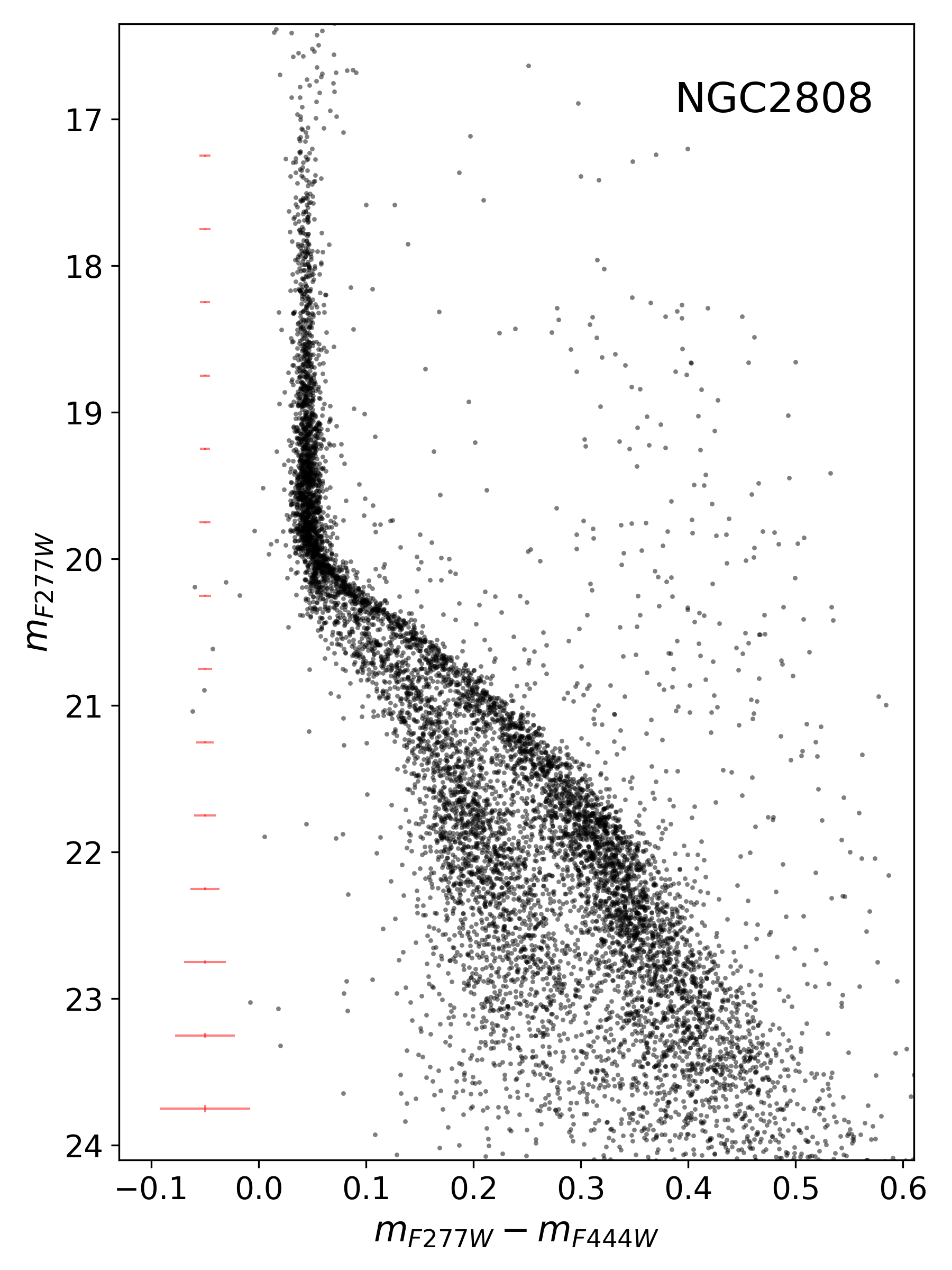}
    \includegraphics[width=0.32\linewidth]{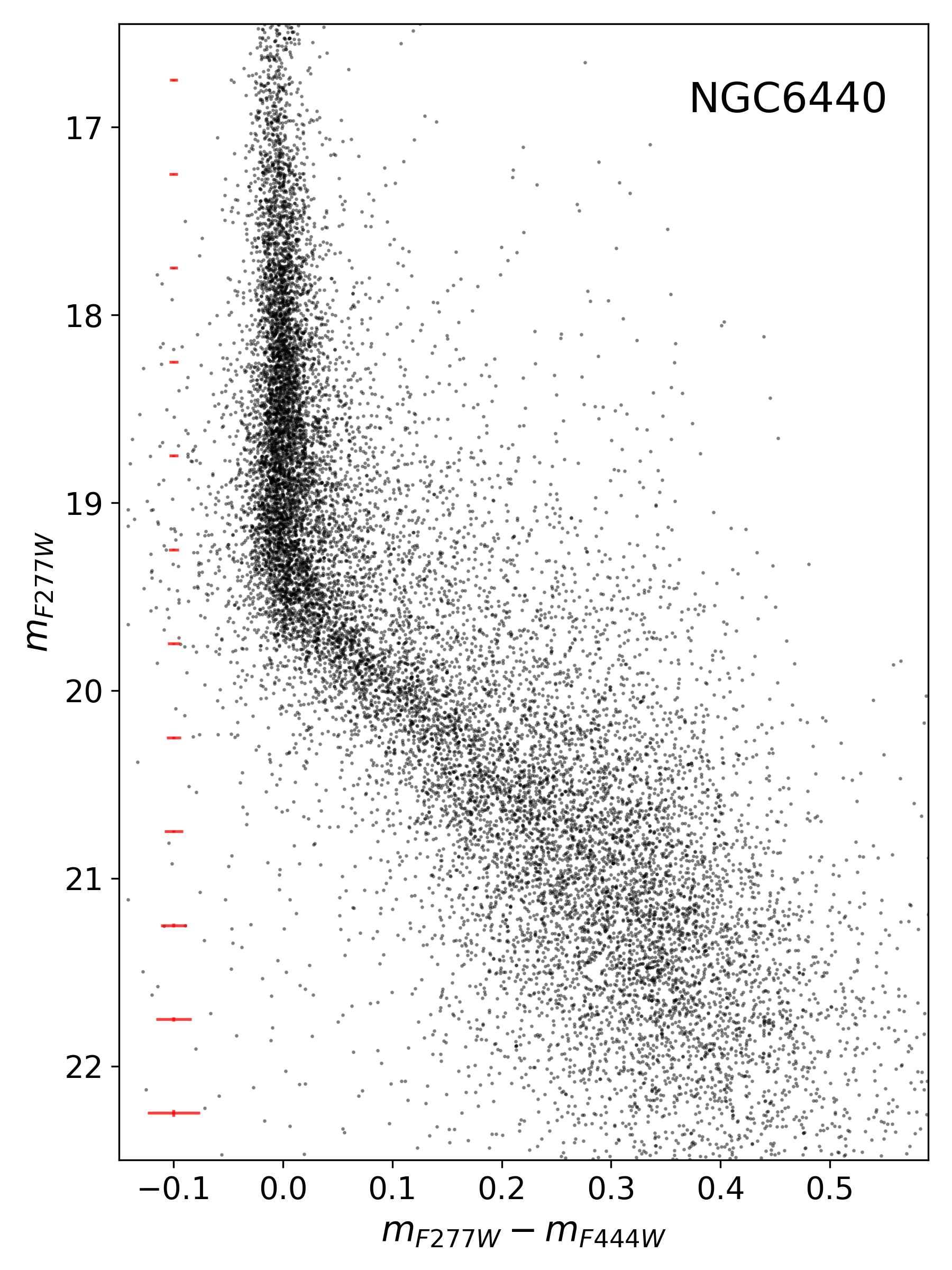}    
    \includegraphics[width=0.32\linewidth]{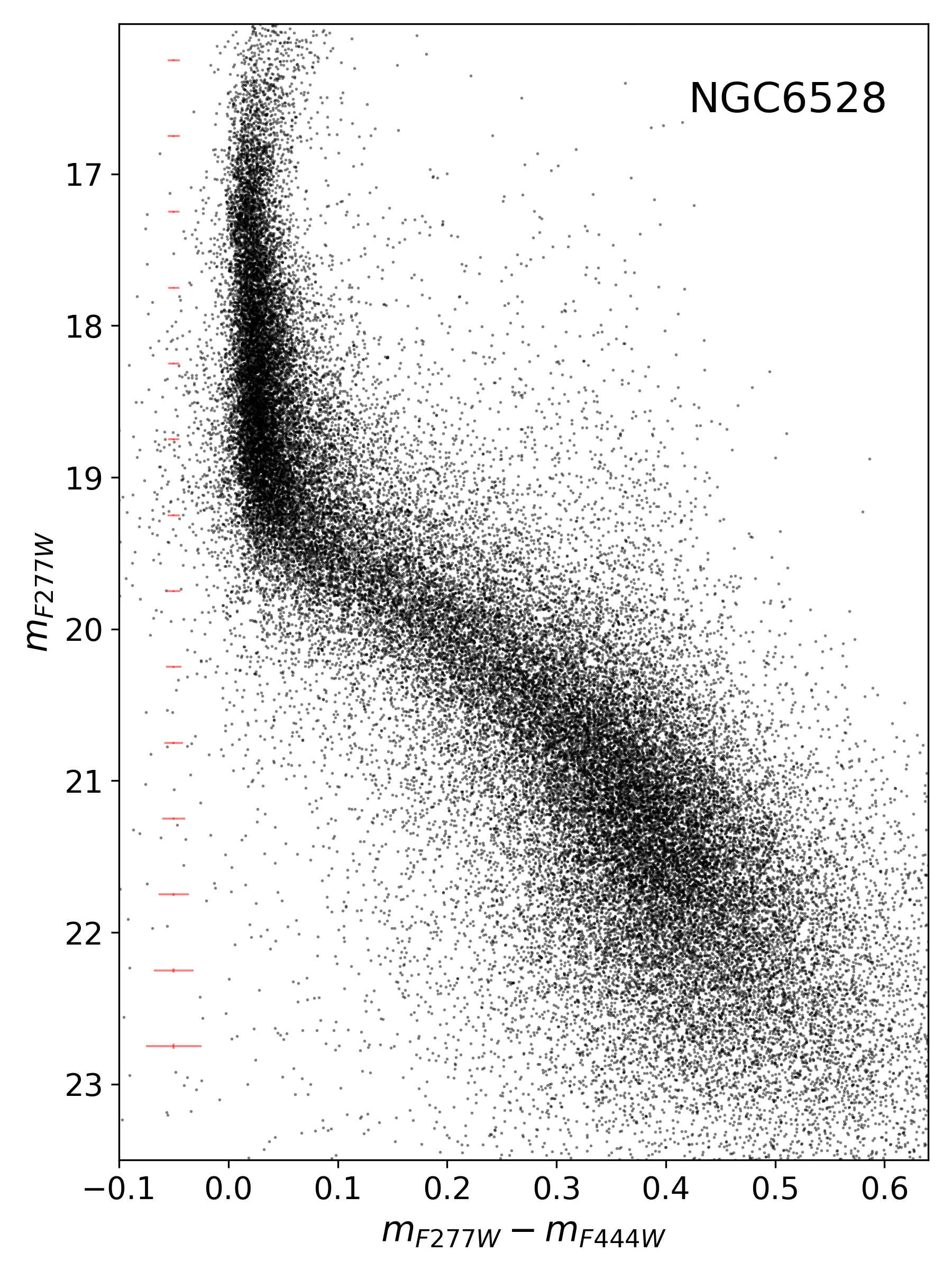}    
    \includegraphics[width=0.32\linewidth]{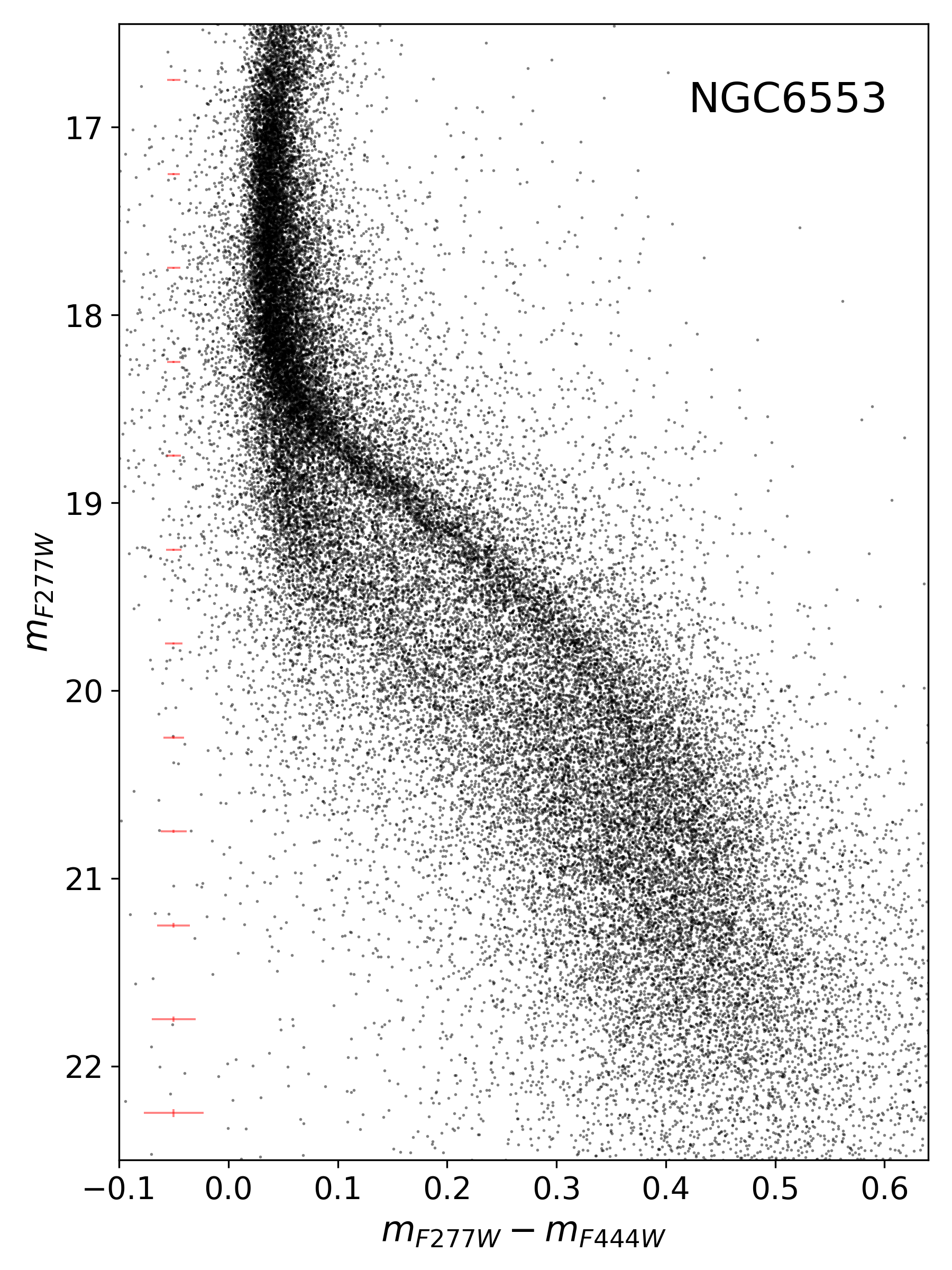}
    \includegraphics[width=0.32\linewidth]{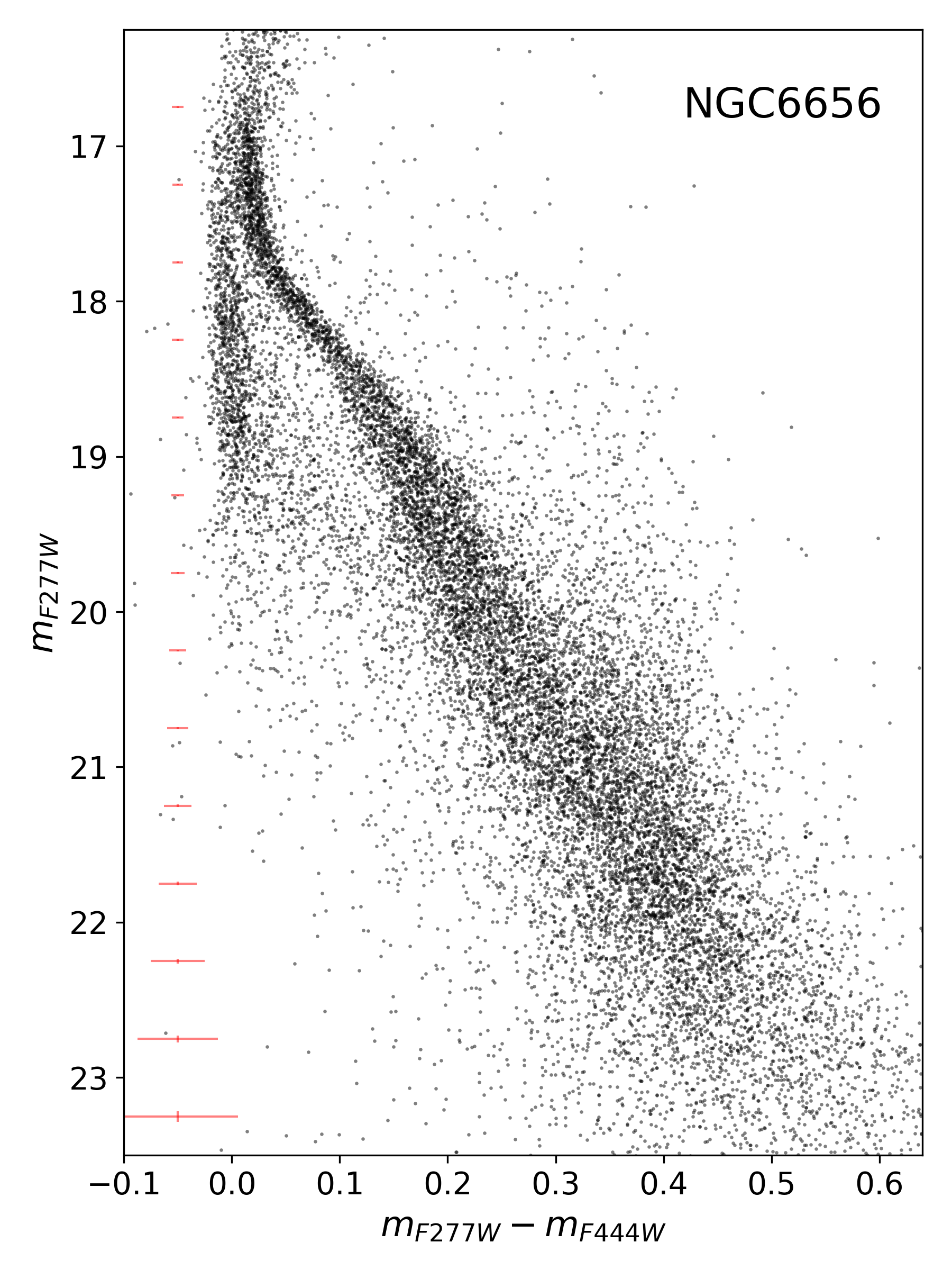}
    \includegraphics[width=0.32\linewidth]{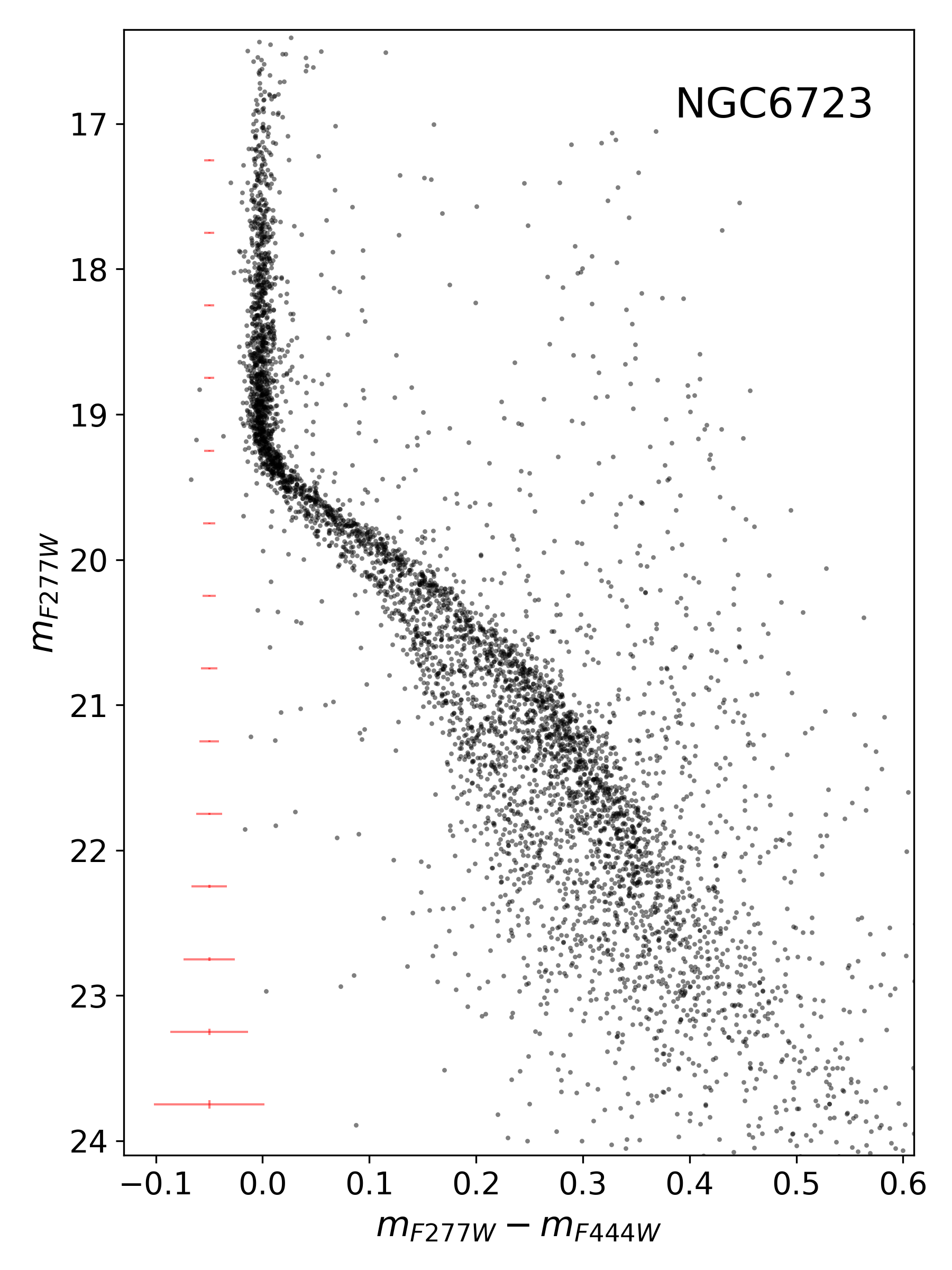}
\caption{ $m_{F277W}$ versus $m_{F277W}-m_{F444W}$ CMDs of stars in the field of view of the studied clusters.}
    \label{fig:CMDsLW}
\end{figure*}

The collection of CMDs obtained from photometry in the short-wavelength channel of NIRCam is shown in Fig.~\ref{fig:CMDsSW} and \ref{fig:CMDsSWb}, whereas the the CMDs based on long-wavelength images are provided in Fig.\,\ref{fig:CMDsLW}. A visual inspection of these diagrams reveals significant field-star contamination in the GCs projected toward the Galactic bulge, namely NGC\,6440, NGC\,6528, NGC\,6553, NGC\,6656, Terzan\,5 and Liller\,1.
 For all the other clusters, the CMDs already show a clear color broadening, or even a split, among stars fainter than the MS knee, as expected for multiple stellar populations with different oxygen abundances. This feature contrasts with the much narrower upper MS.

\begin{figure*}
    \centering
    \includegraphics[width=0.32\linewidth]{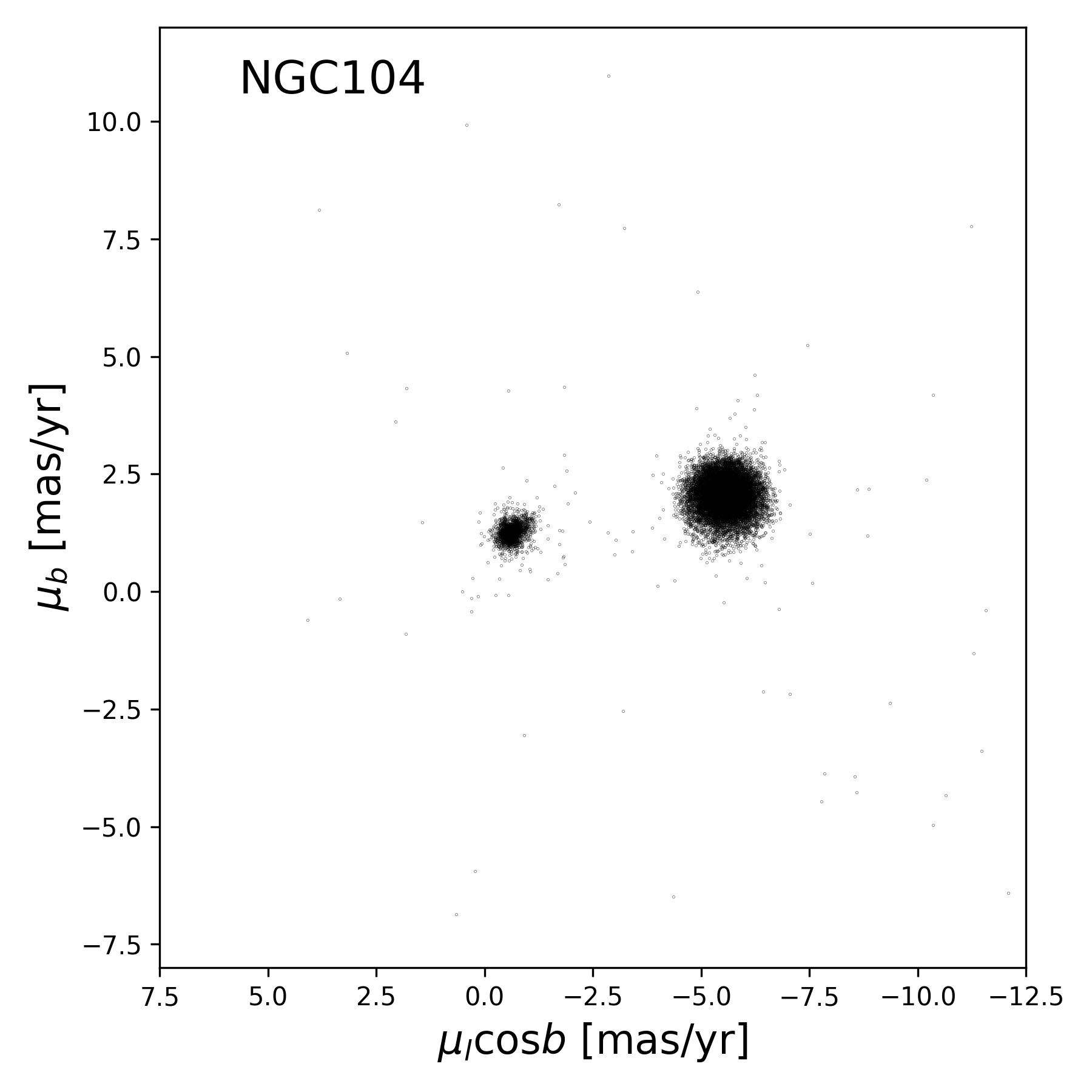}
    \includegraphics[width=0.32\linewidth]{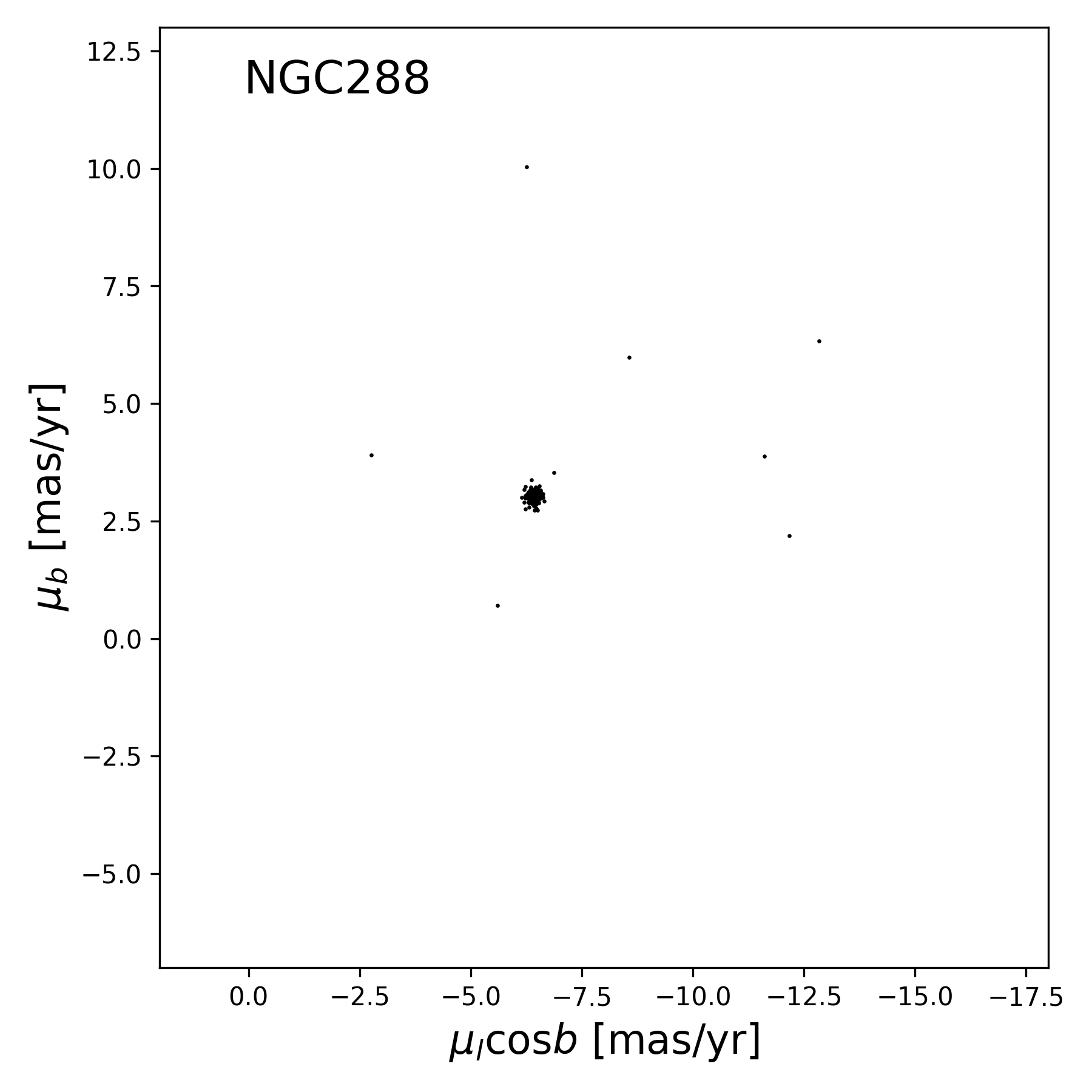}
    \includegraphics[width=0.32\linewidth]{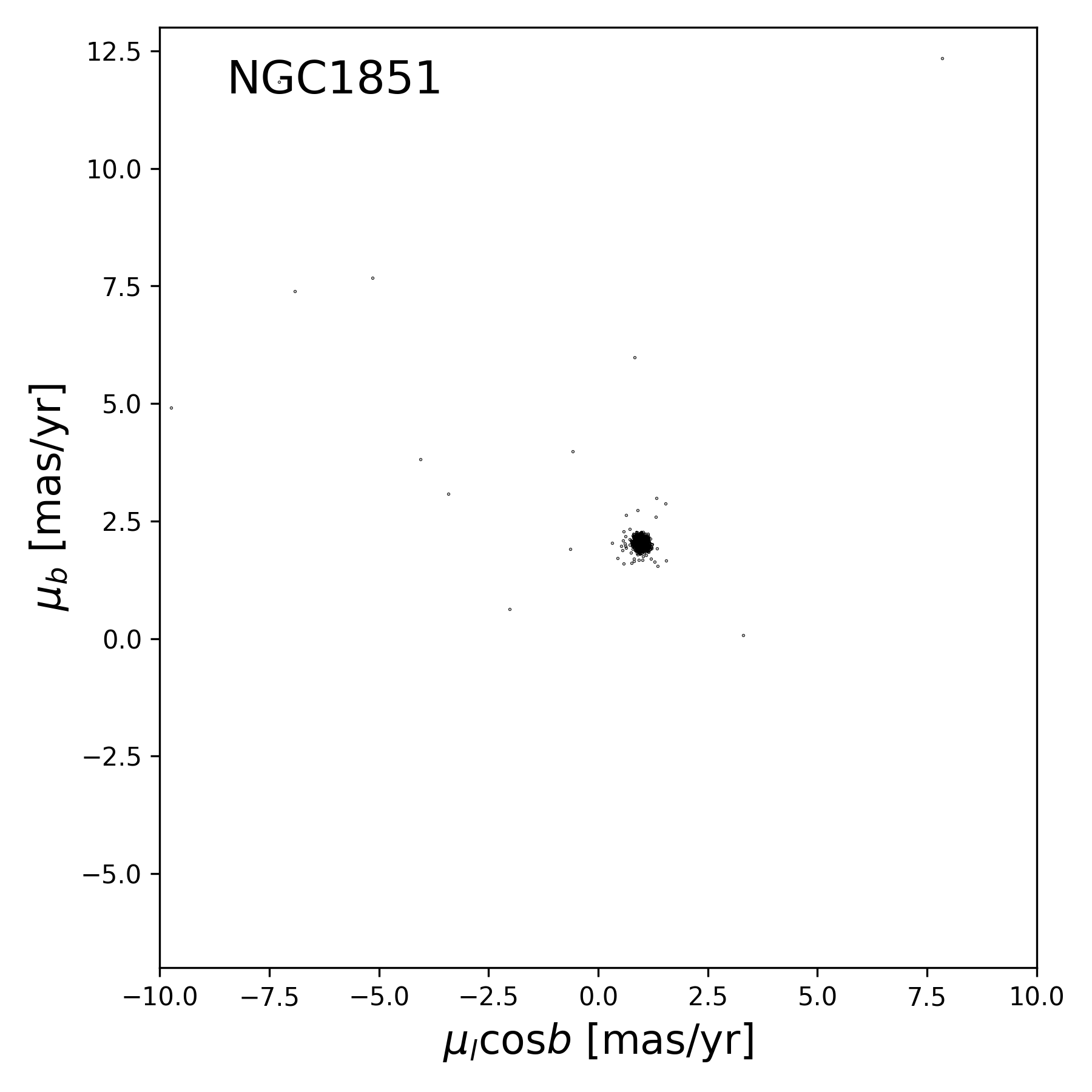}
    \includegraphics[width=0.32\linewidth]{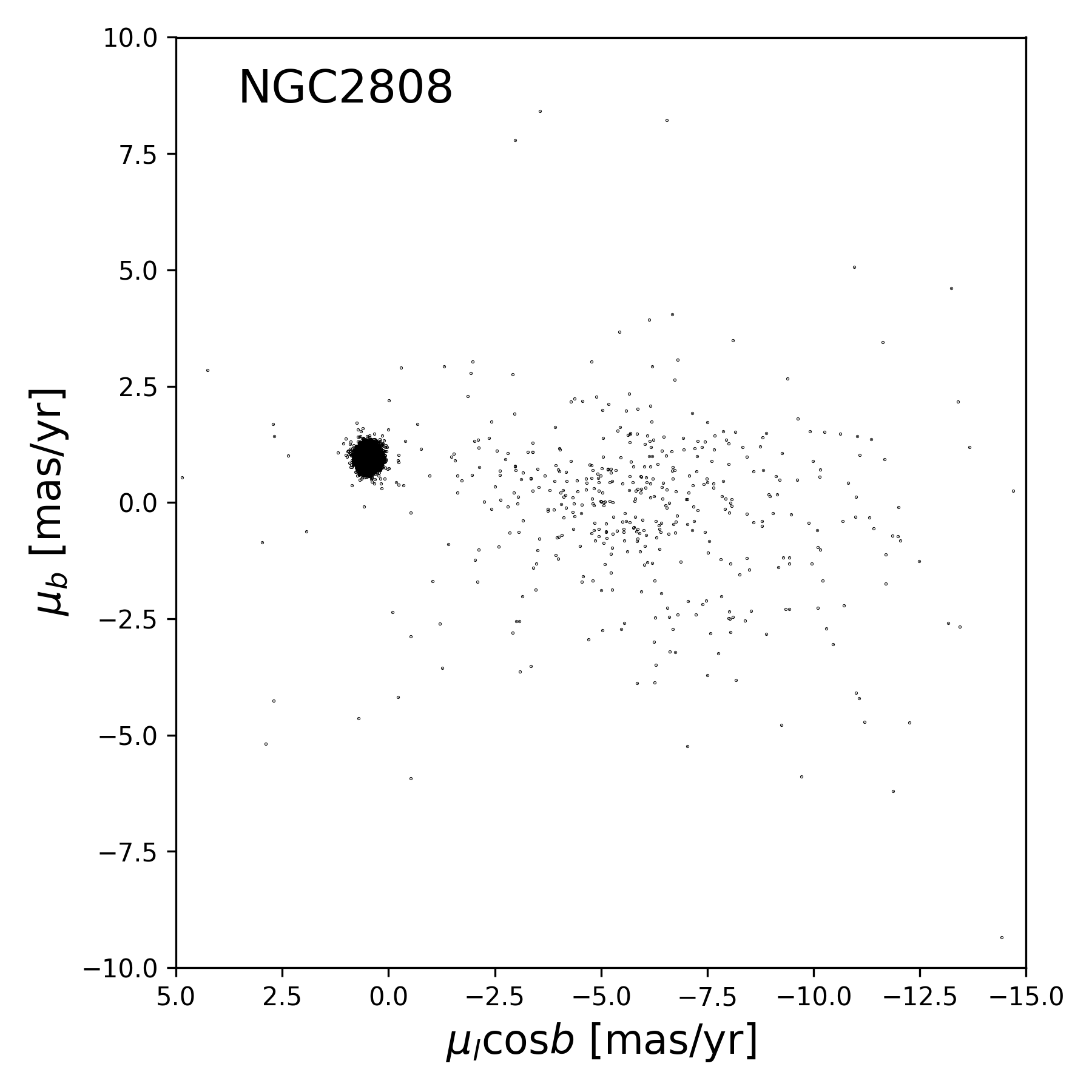}    
    \includegraphics[width=0.32\linewidth]{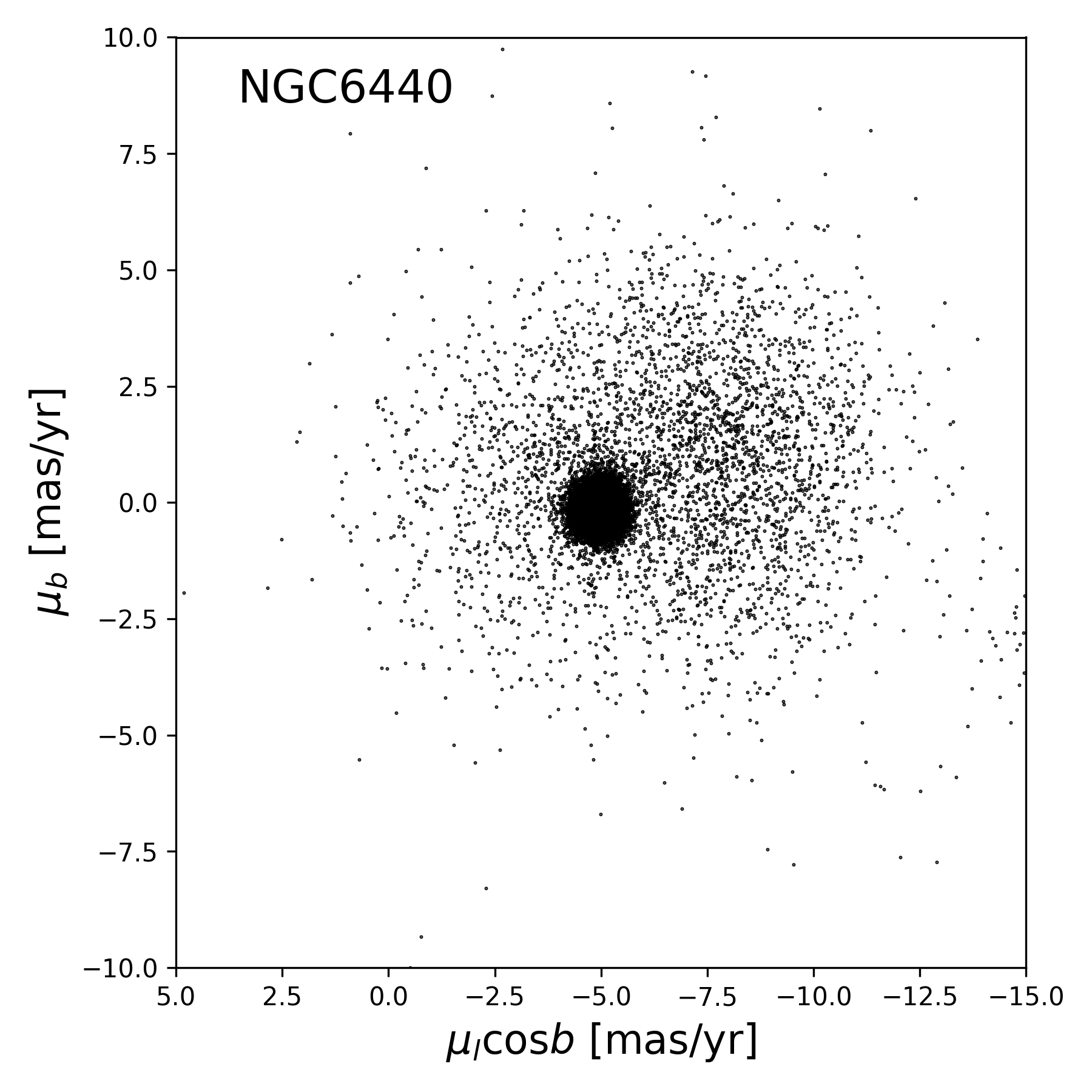}
    \includegraphics[width=0.32\linewidth]{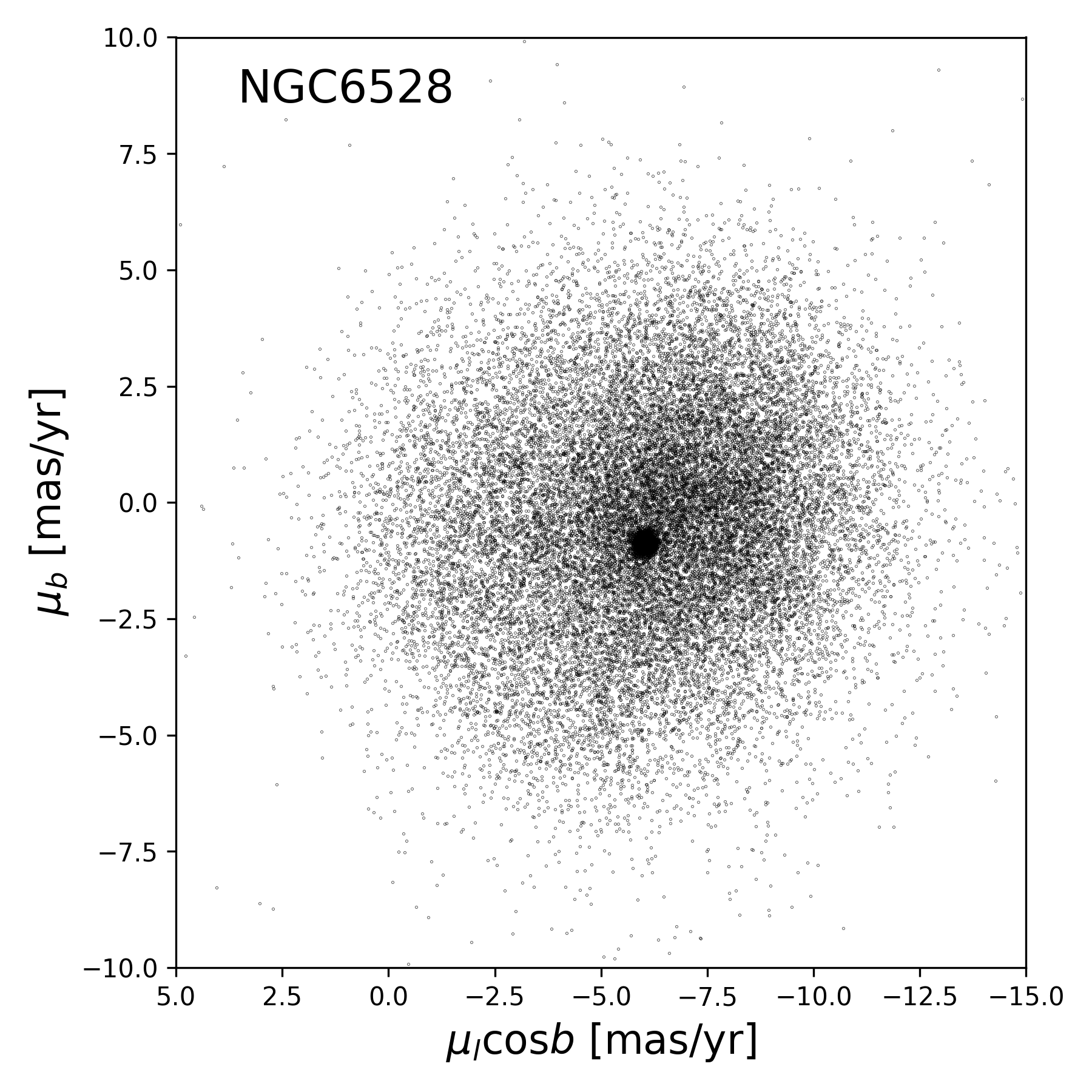}    
    \includegraphics[width=0.32\linewidth]{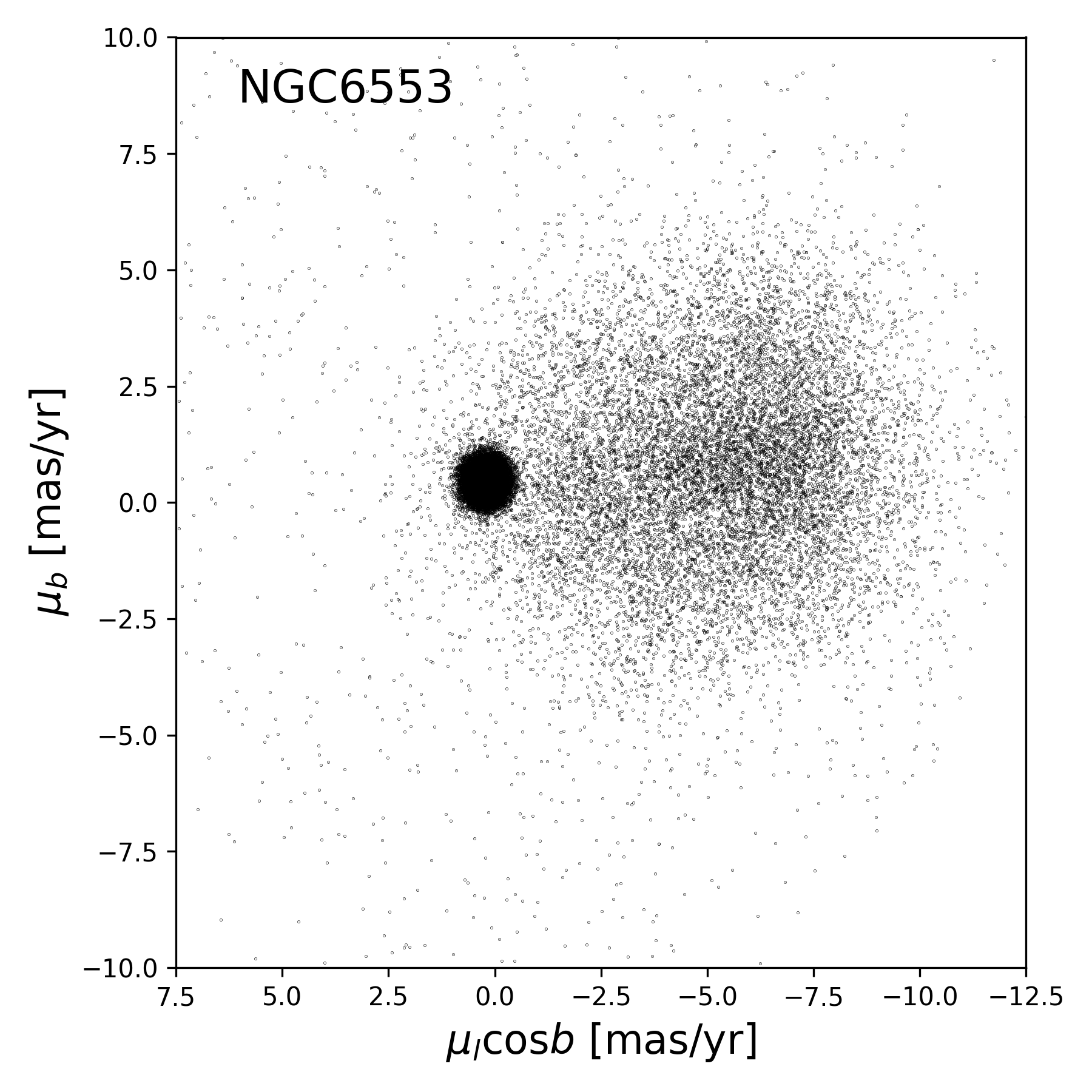}    
    \includegraphics[width=0.32\linewidth]{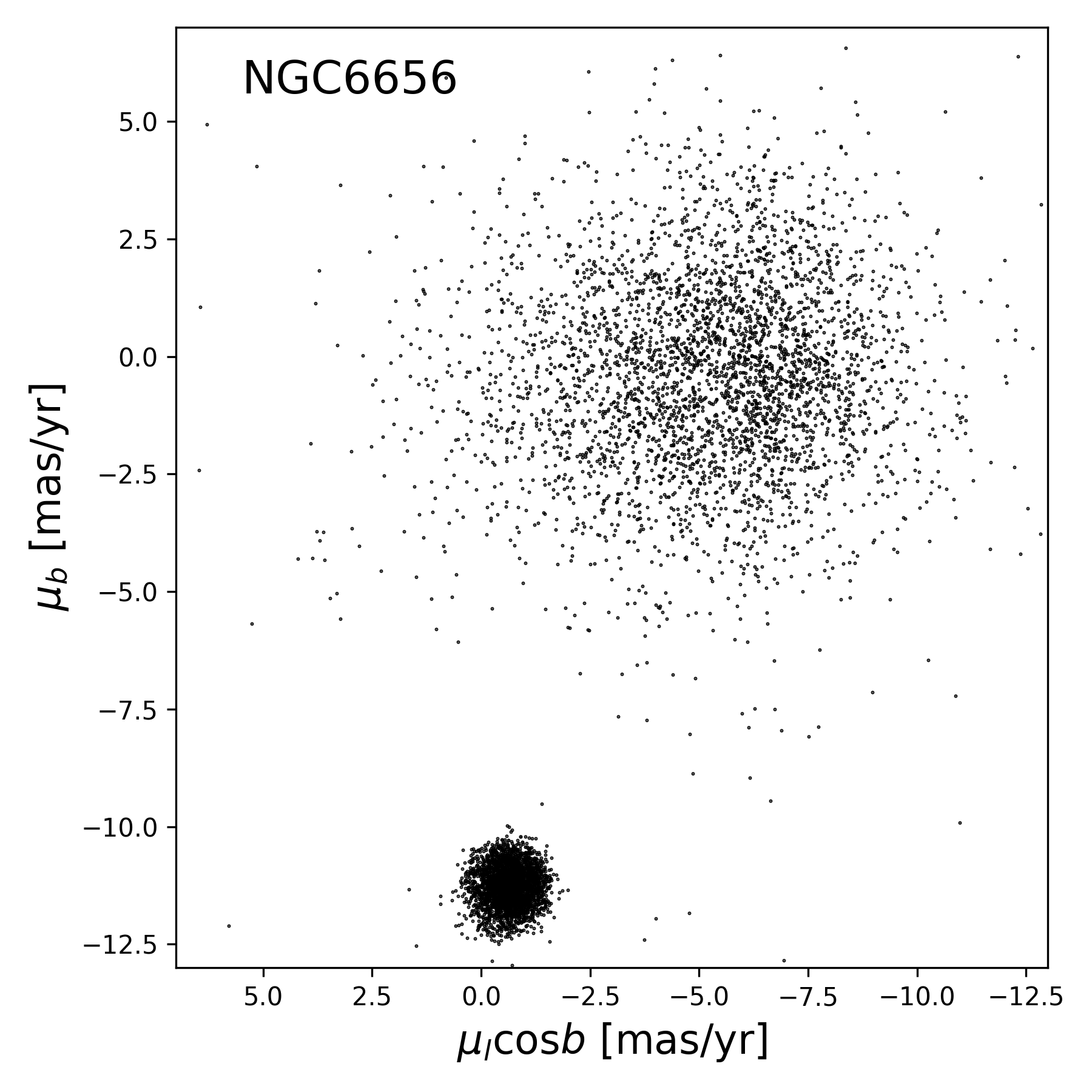}
    \includegraphics[width=0.32\linewidth]{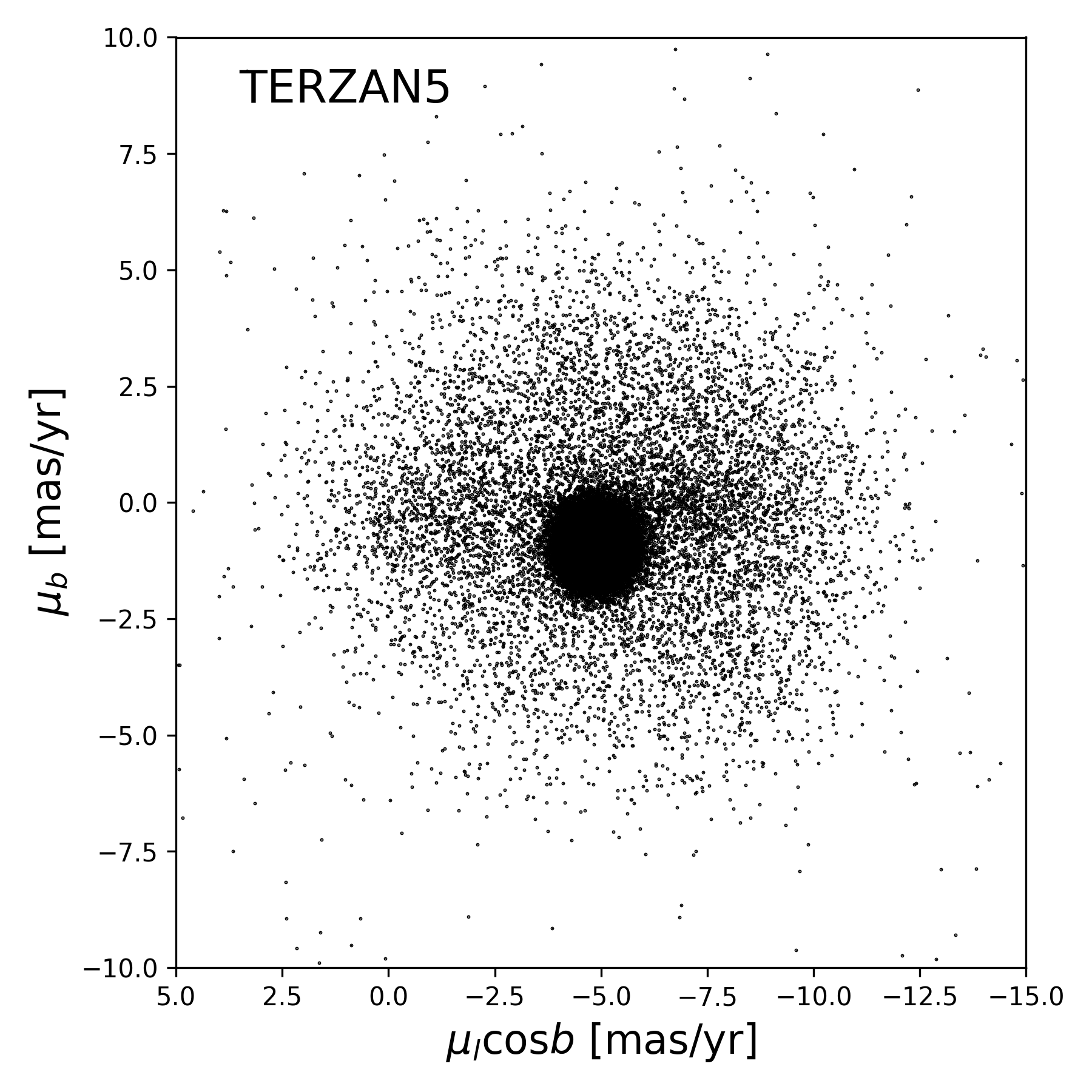}
\caption{Proper motion diagrams for stars in the field of view of the studied clusters.}
    \label{fig:PMs}
\end{figure*}

\subsection{Proper motions}

Relative proper motions were derived following standard
high-precision astrometric procedures \citep[e.g.][]{anderson2003a, piotto2012a, bellini2014a, libralato2022a, milone2023a, ziliotto2025a, ziliotto2026a}, based on comparing stellar positions measured at
multiple epochs by JWST, HST, and Gaia.

For each cluster, we generated independent photometric and
astrometric catalogs for all filters and epochs. A master frame
was defined using the F200W NIRCam images,
oriented with the $X$ axis toward the west and the $Y$ axis
toward the north. Stellar positions from each catalog were
transformed into this reference system via six-parameter linear
transformations, using bright, unsaturated cluster members as
reference stars.

The selection of reference stars was performed iteratively.
We first identified likely members from their location in the
 CMD and derived preliminary proper motions. Membership was then refined using both the CMD and the proper-motion diagram, and the improved member sample was adopted to recompute the transformations
and proper motions. Stellar positions were corrected for the
measured displacements and the transformations recalculated,
reducing residual alignment errors.

To account for small-scale residual distortions, we applied
local transformations. For each
target, the transformation was computed from the nearest $N$
reference stars, excluding the target itself. The choice of $N$
and the magnitude range was optimized to balance sensitivity
to local systematics and statistical robustness. For stars at the
bright and faint ends, we adopted an adaptive magnitude
selection to ensure a sufficient number of neighbors.

Proper motions were finally obtained by fitting weighted
least-squares linear relations to the transformed $x$ and $y$
positions as a function of epoch. The slopes of these fits
provide the proper-motion components, and their formal
uncertainties define the associated errors.
Relative proper motions were placed on an absolute reference
frame by comparison with absolute stellar proper motions from
Gaia DR3 \citep{gaia2021a}, following the procedure described in
\citet{milone2023a}. A compilation of proper-motion diagrams is presented in Fig.~\ref{fig:PMs} for all clusters with available multi-epoch HST and {\it JWST} imaging, with the exception of Liller~1, whose proper-motion diagram is discussed separately in Sect.~\ref{sec:MPs}.

\begin{figure*}
    \centering
    \includegraphics[width=0.9\linewidth]{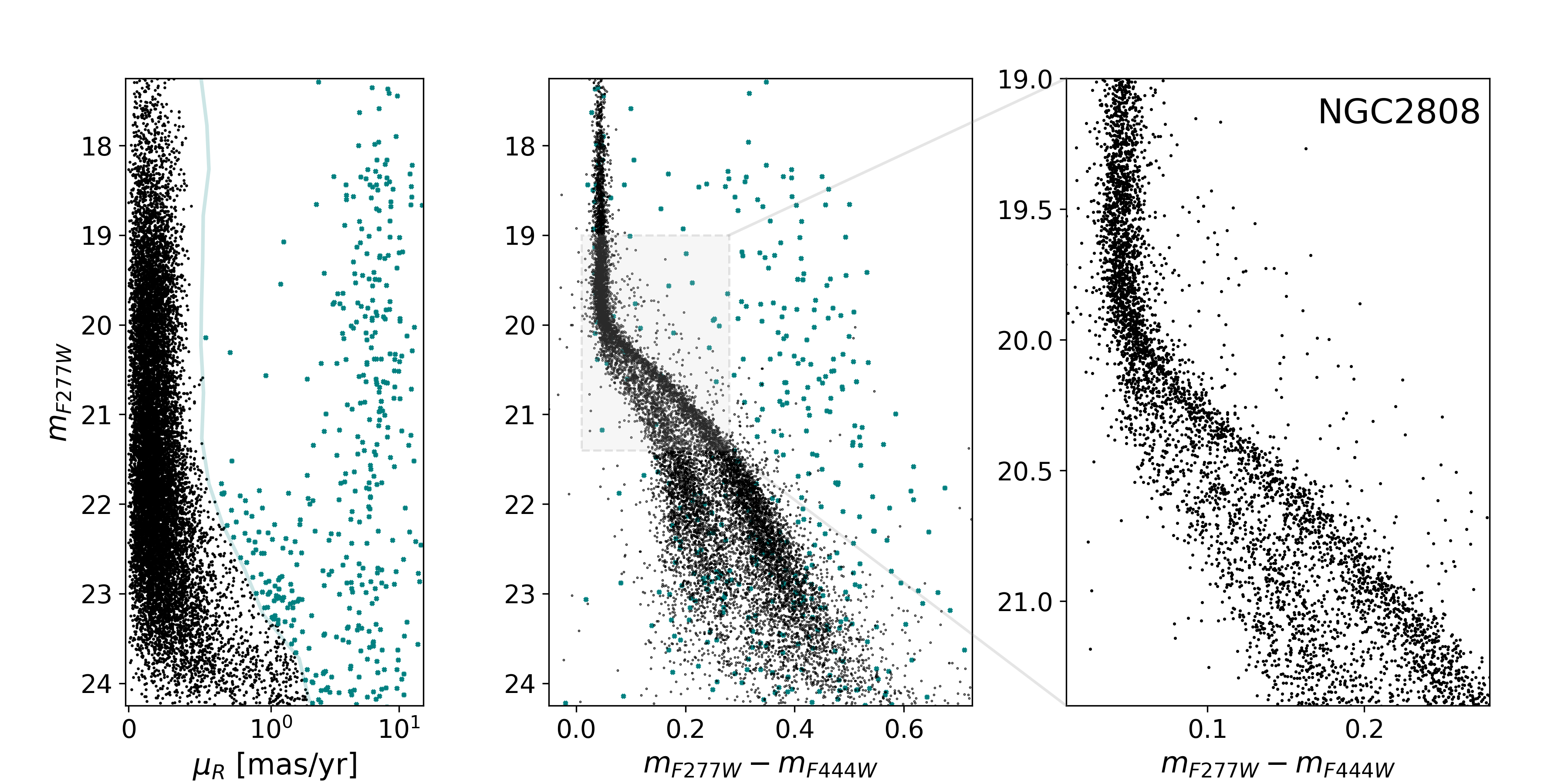}
\caption{Illustration of the procedure adopted to select probable cluster members in NGC\,2808. \textit{Left panel:} $m_{\rm F277W}$ magnitude as a function of the total proper motion, $\mu_{\rm R}$, relative to the cluster mean motion. The teal line marks the boundary between probable cluster members (black points) and field stars (teal symbols). 
\textit{Middle panel:} $m_{\rm F277W}$ vs.\ $m_{\rm F277W}-m_{\rm F444W}$ CMD from long-wavelength photometry, with the same colour coding as in the left panel. \textit{Right panel:} Zoom of the CMD around the MS knee.} 
    \label{fig:ngc2808pm}
\end{figure*}

The main steps of the procedure used to select probable cluster members are illustrated in Fig.~\ref{fig:ngc2808pm}. The left panel shows the $m_{\rm F277W}$ magnitude as a function of the total proper motion, $\mu_{\rm R}$, relative to the mean motion of NGC\,2808. The teal line separates the most probable cluster members (black points) from field stars (teal symbols); it is defined by shifting the mean cluster motion by four times the proper-motion dispersion, computed in 0.5-mag bins.

The middle panel displays the $m_{\rm F277W}$ vs.\ $m_{\rm F277W}-m_{\rm F444W}$ CMD obtained from long-wavelength channel photometry. Field stars and cluster members are indicated with the same symbols as in the left panel. The right panel shows a zoom of the CMD around the MS knee. A prominent feature is the presence of three distinct MSs of M-dwarfs, which converge at the MS knee, together with a nearly vertical MS for brighter stars.

\begin{figure*}
    \centering
    \includegraphics[width=0.32\linewidth]{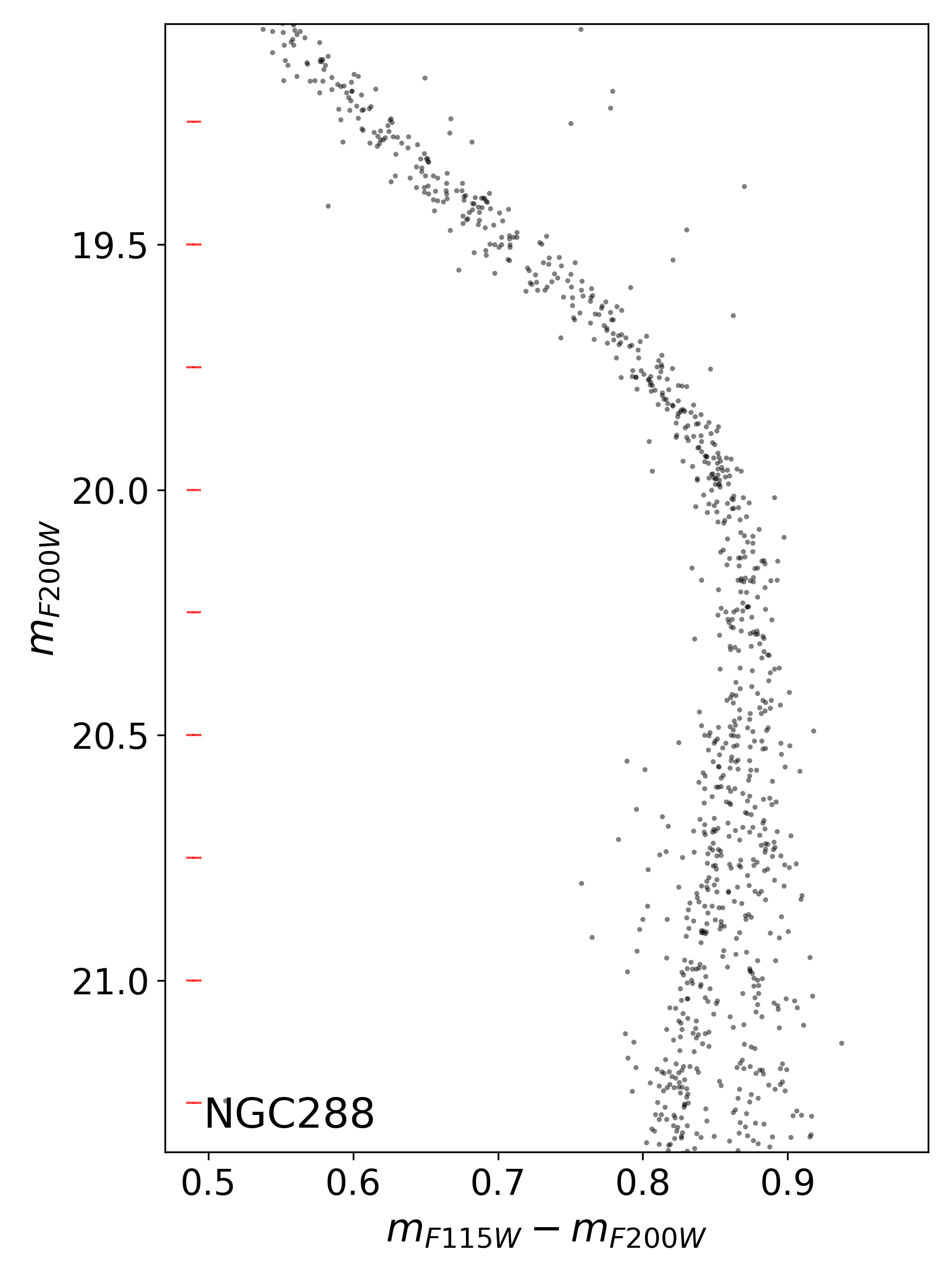}    
    \includegraphics[width=0.32\linewidth]{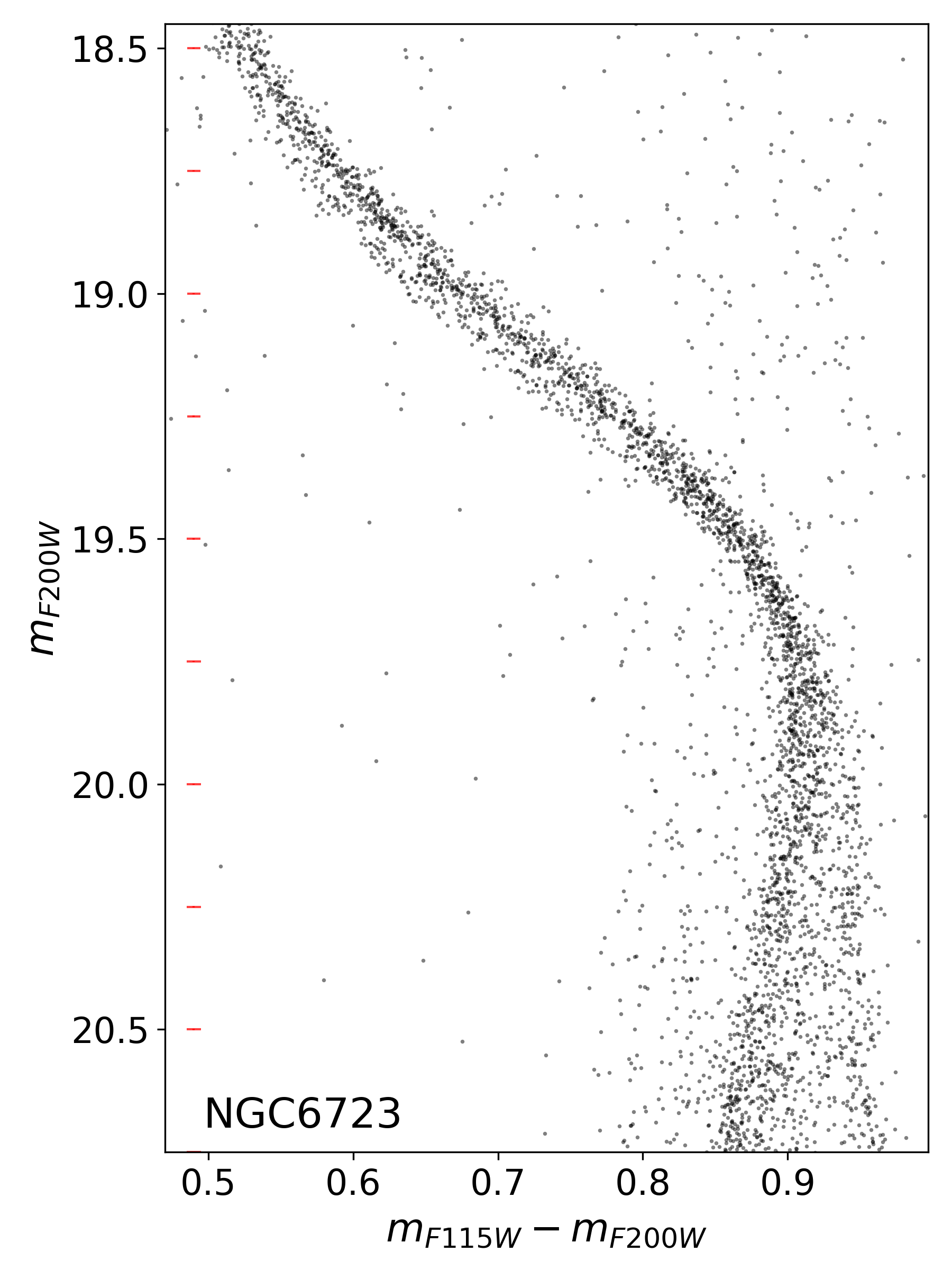}    
    \includegraphics[width=0.32\linewidth]{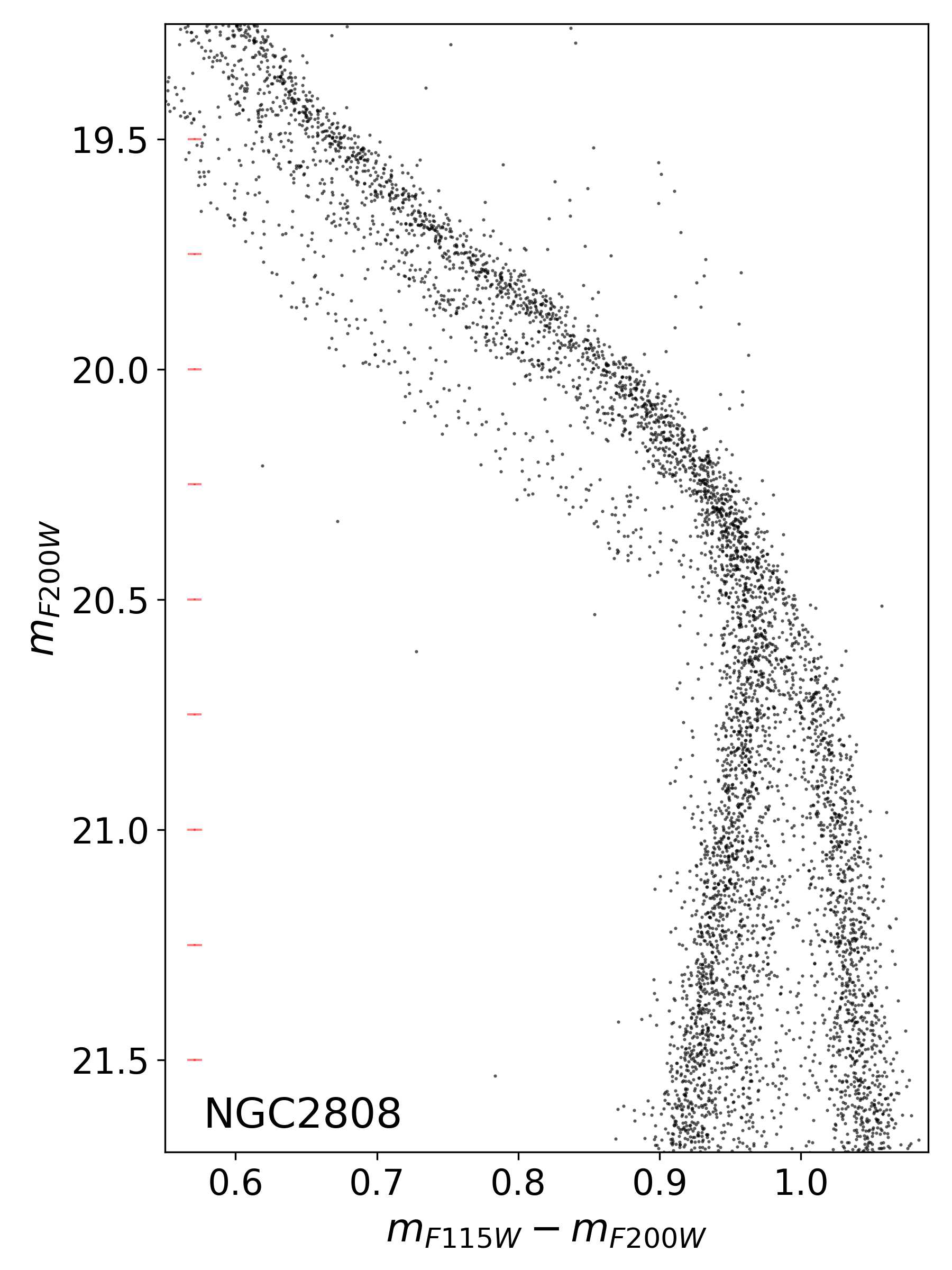}    
        \includegraphics[width=0.32\linewidth]{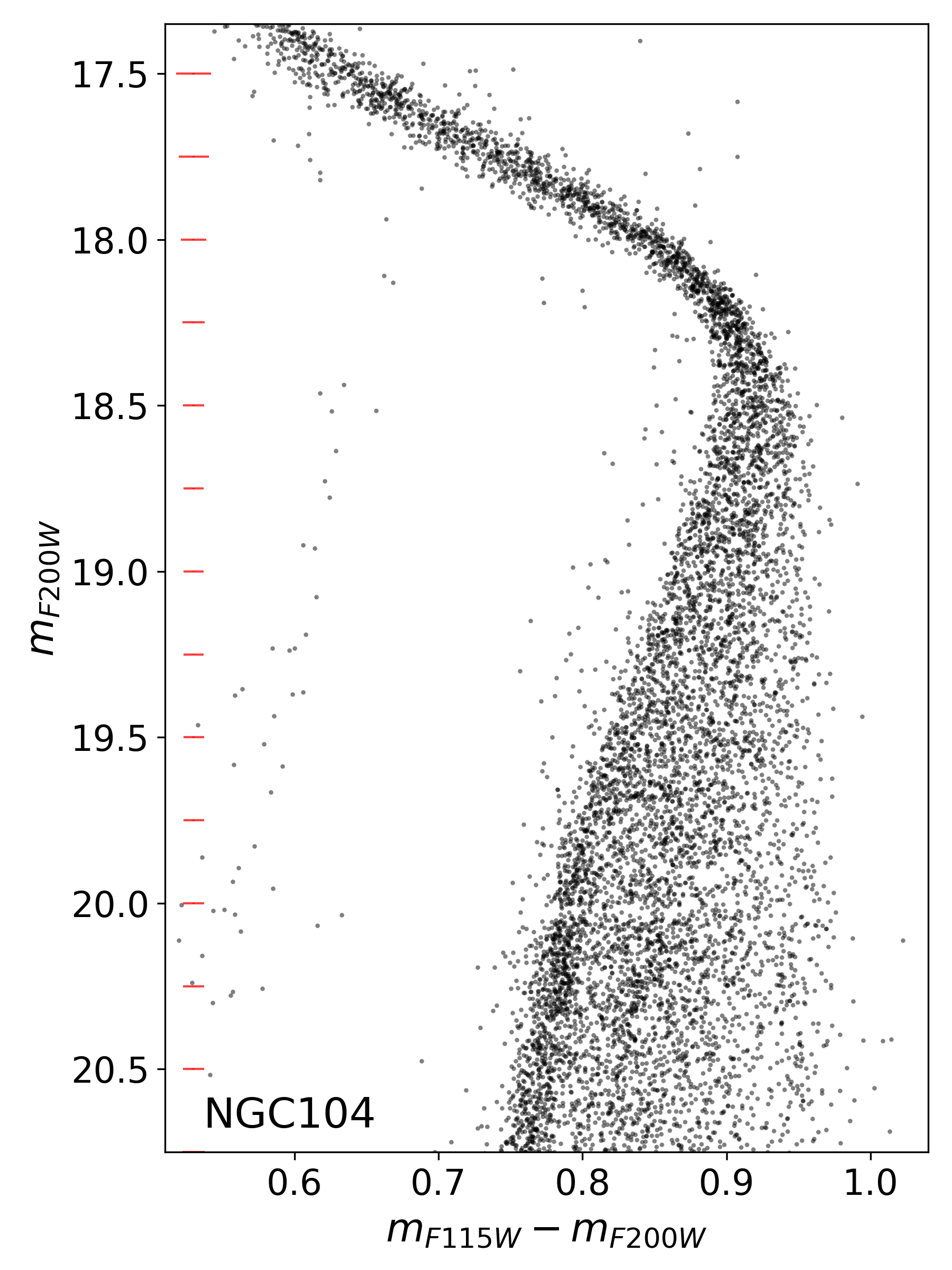} 
        \includegraphics[width=0.32\linewidth]{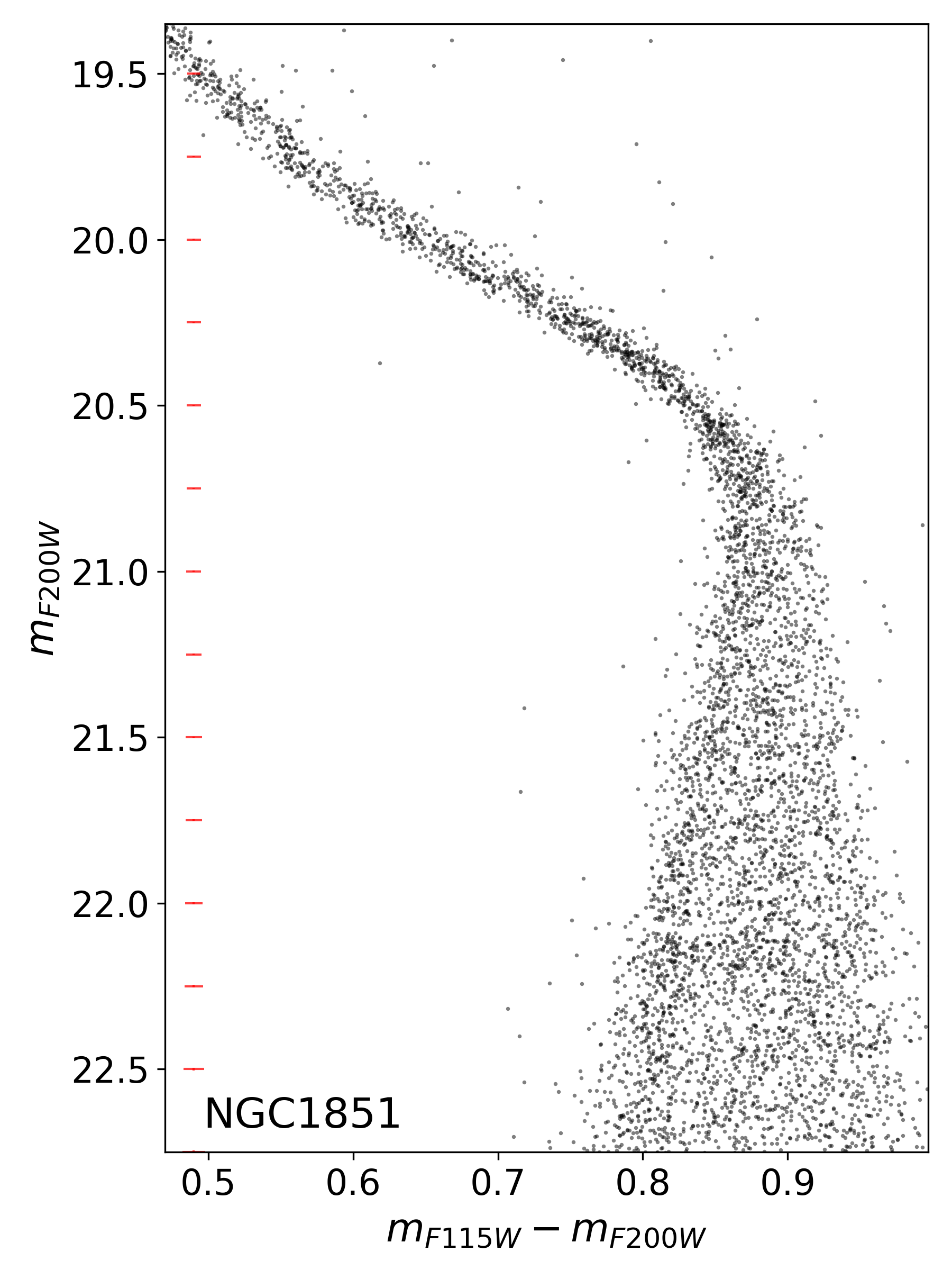} 
        \includegraphics[width=0.32\linewidth]{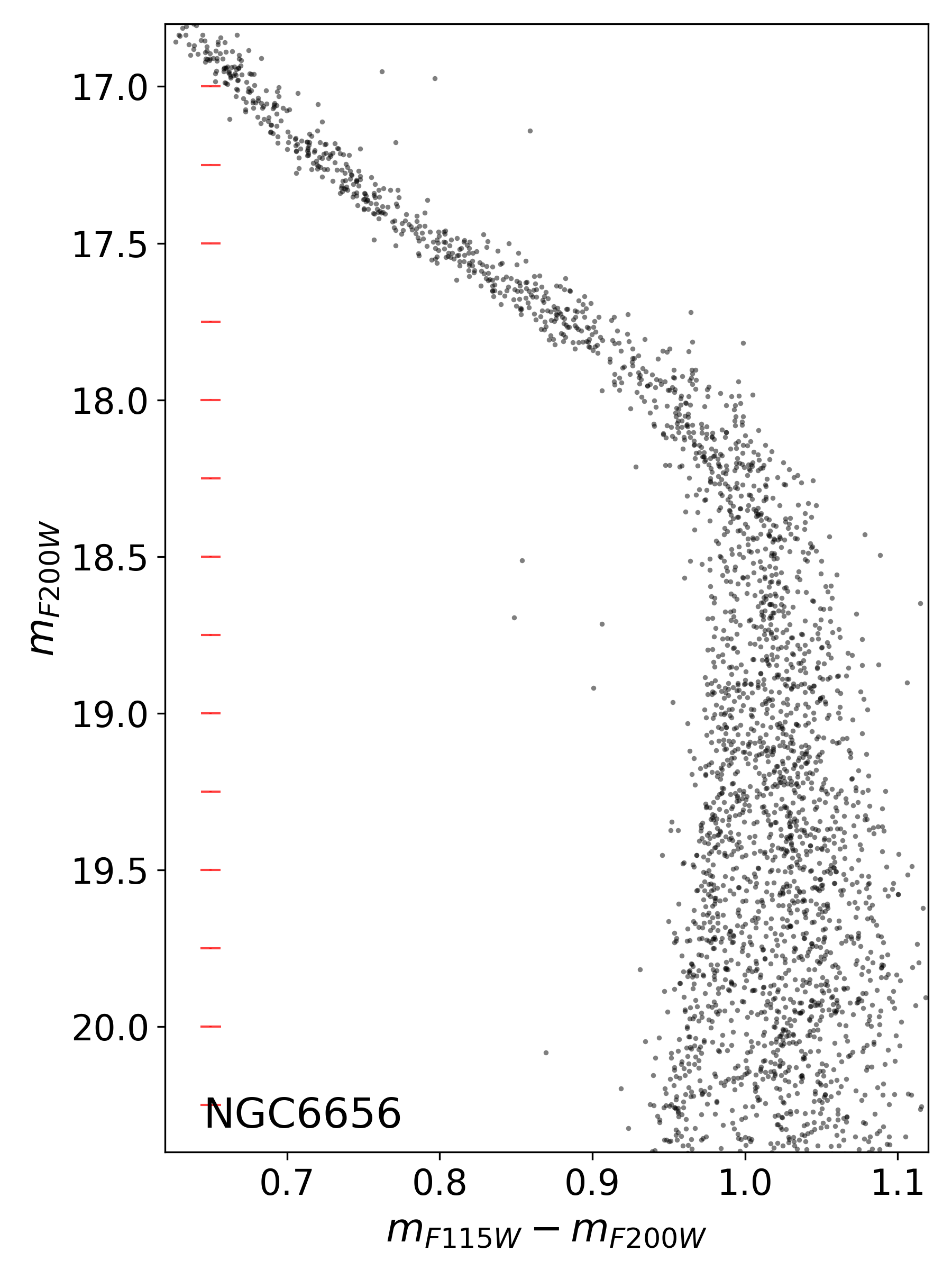} 
    \includegraphics[width=0.32\linewidth]{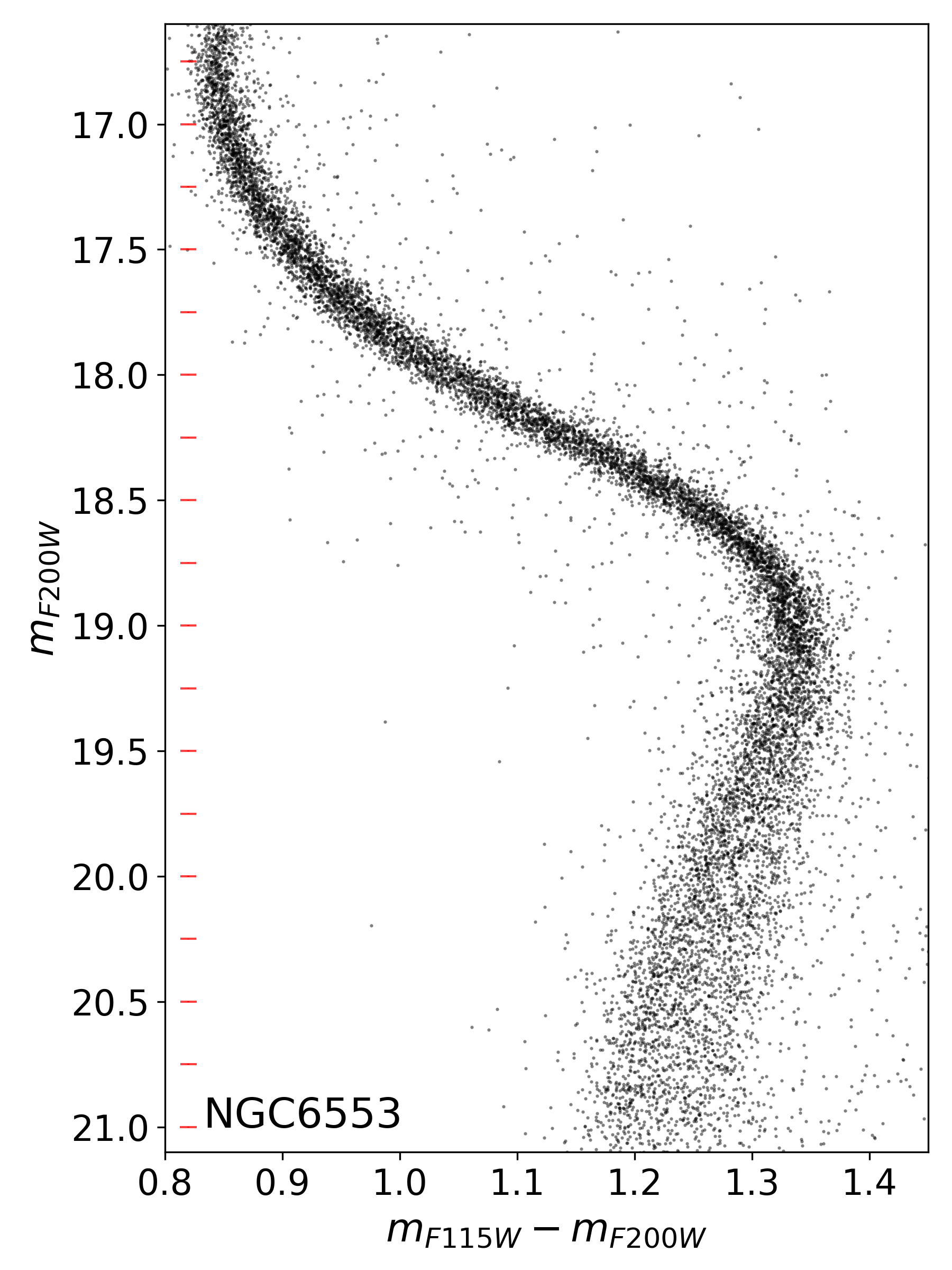} 
    \includegraphics[width=0.32\linewidth]  {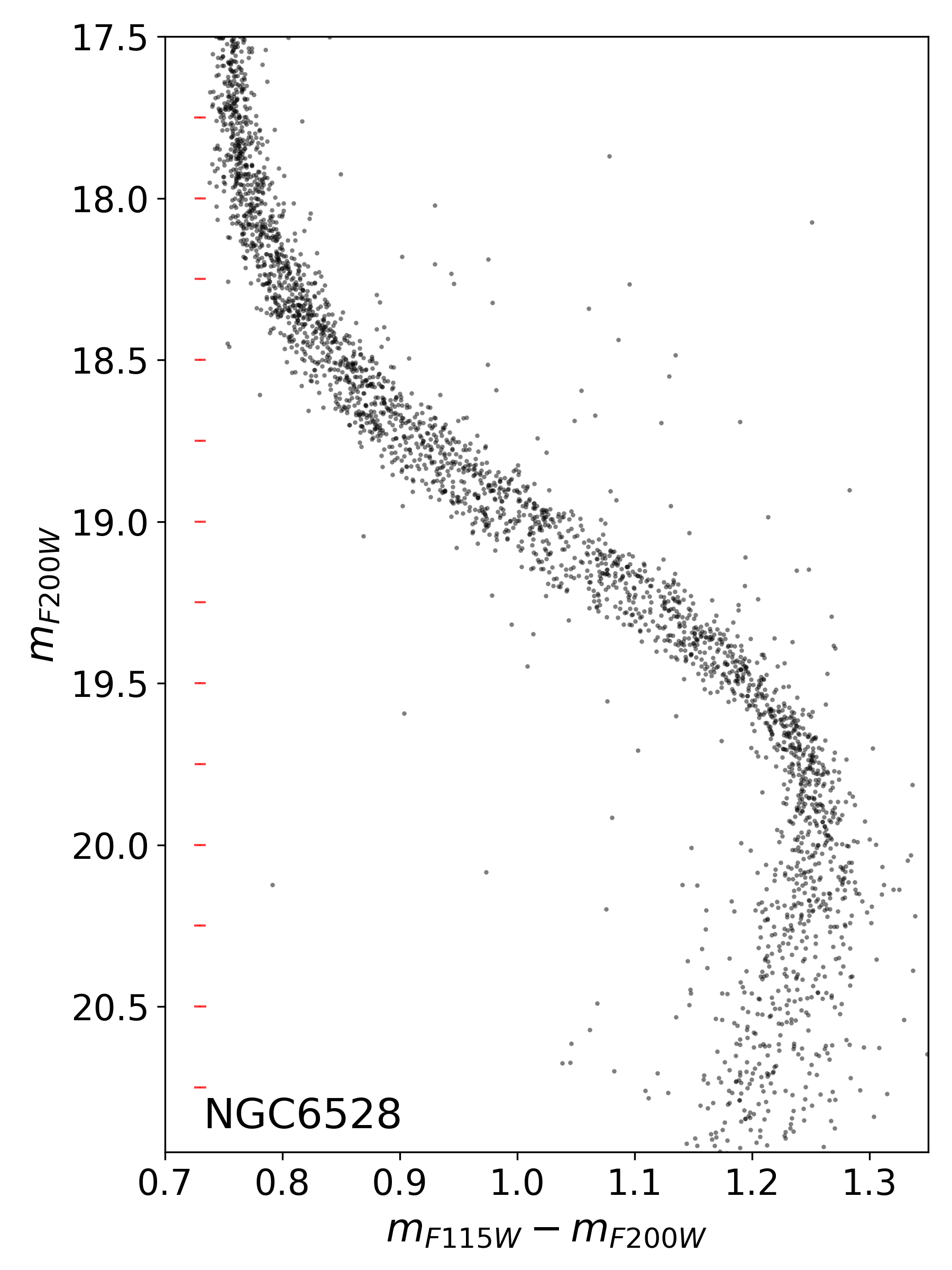}    
    \includegraphics[width=0.32\linewidth]  {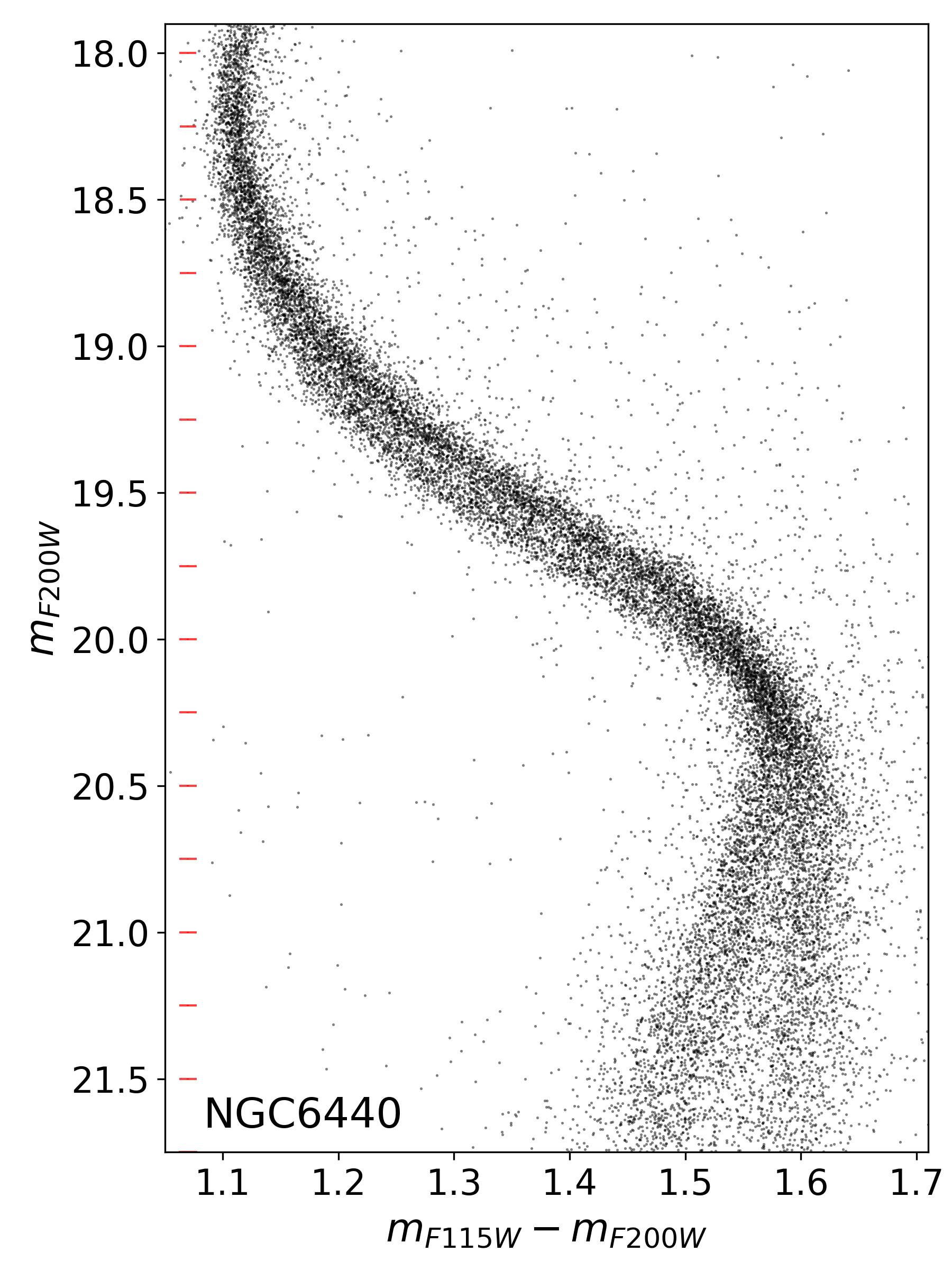}    

\caption{$m_{\rm F200W}$ vs.\,$m_{\rm F115W} - m_{\rm F200W}$ CMDs of the studies GCs, zoomed in on the MS. For NGC\,2808, NGC\,6553, NGC\,6528, NGC\,6440, and NGC\,6656, which are significantly affected by field-star contamination, only proper-motion-selected probable cluster members are shown.}
    \label{fig:CMDsSWzoom}
\end{figure*}

 \begin{figure*}
    \centering
    \includegraphics[width=0.32\linewidth]{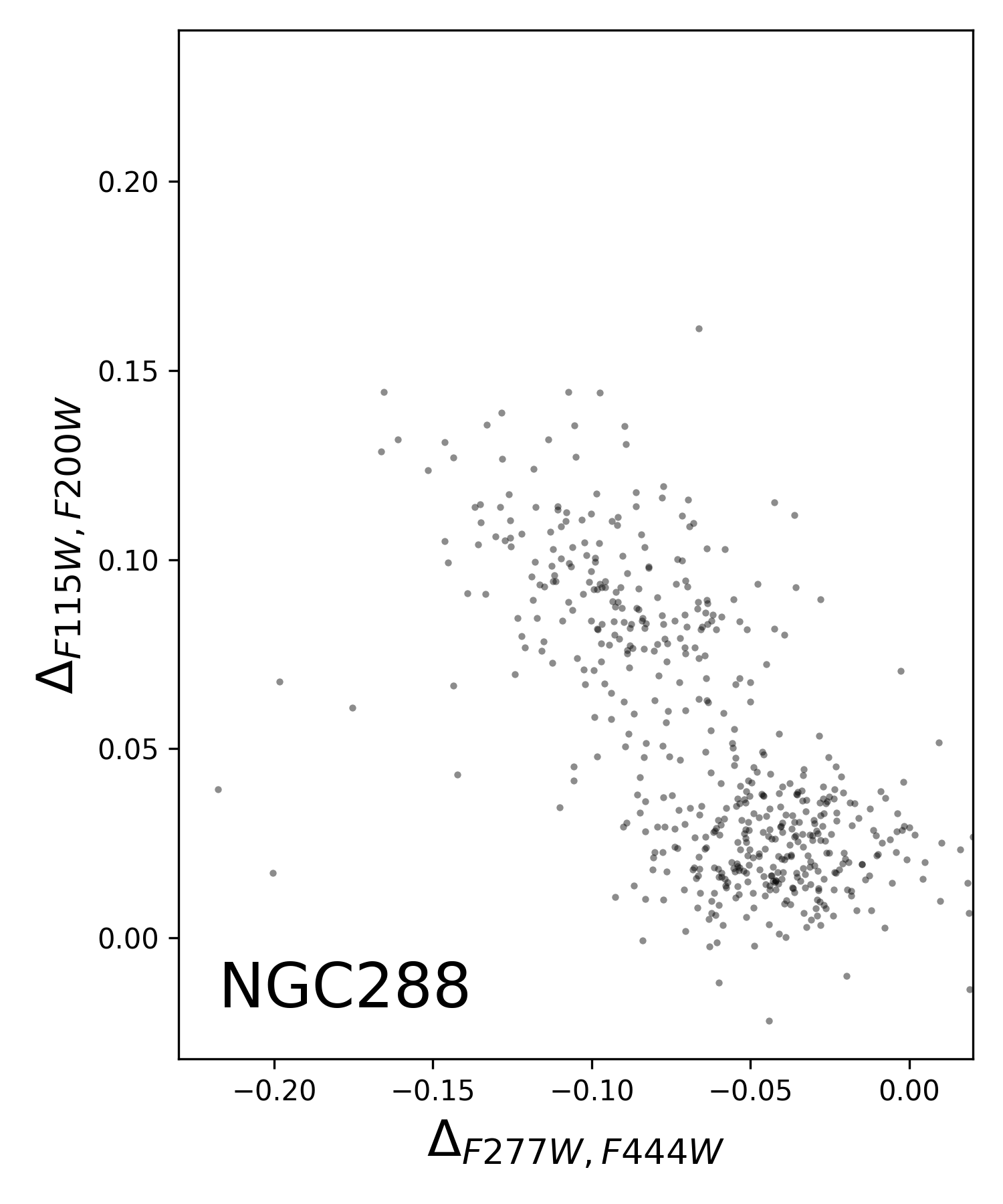}
    \includegraphics[width=0.32\linewidth]{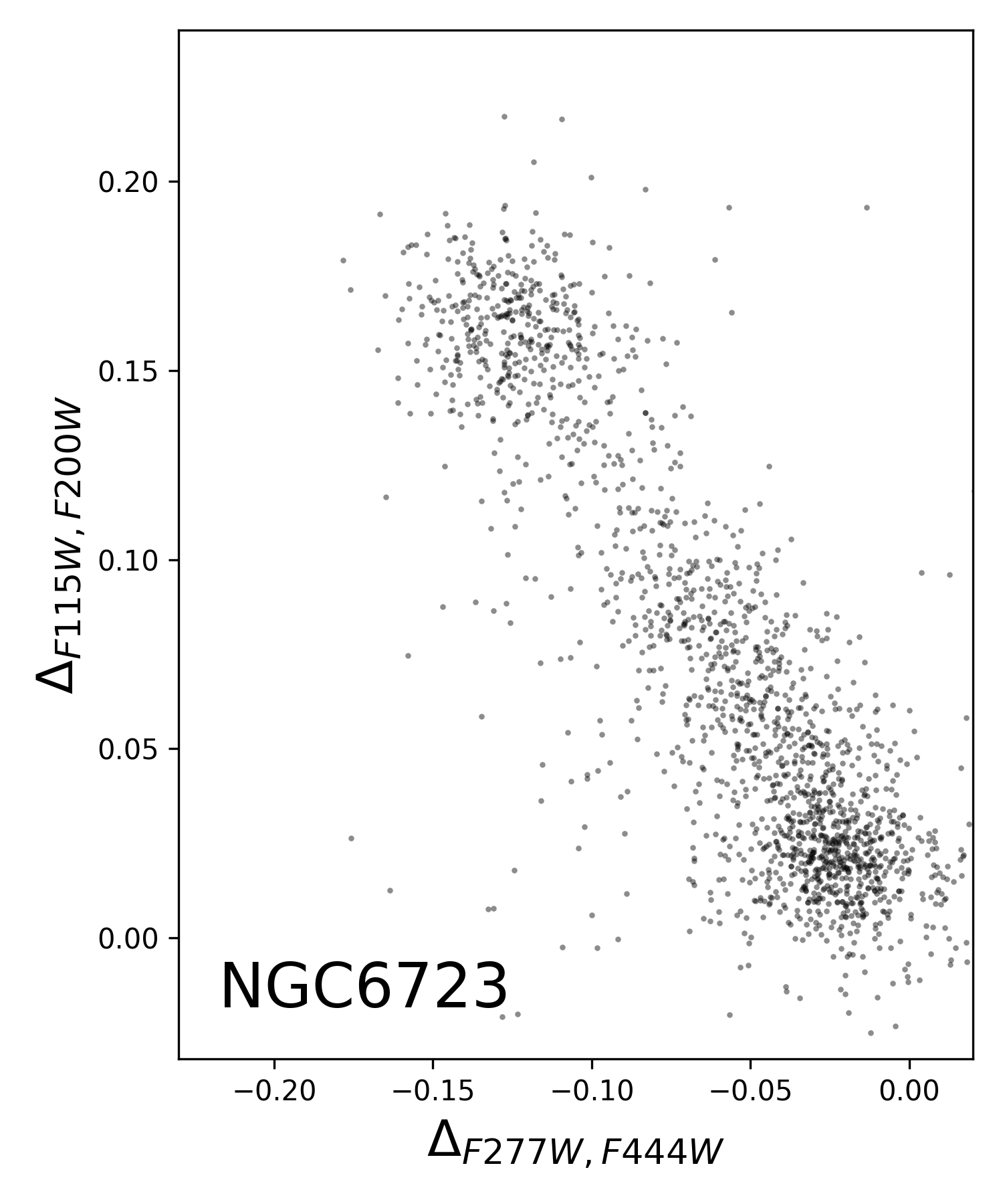}
    \includegraphics[width=0.32\linewidth]{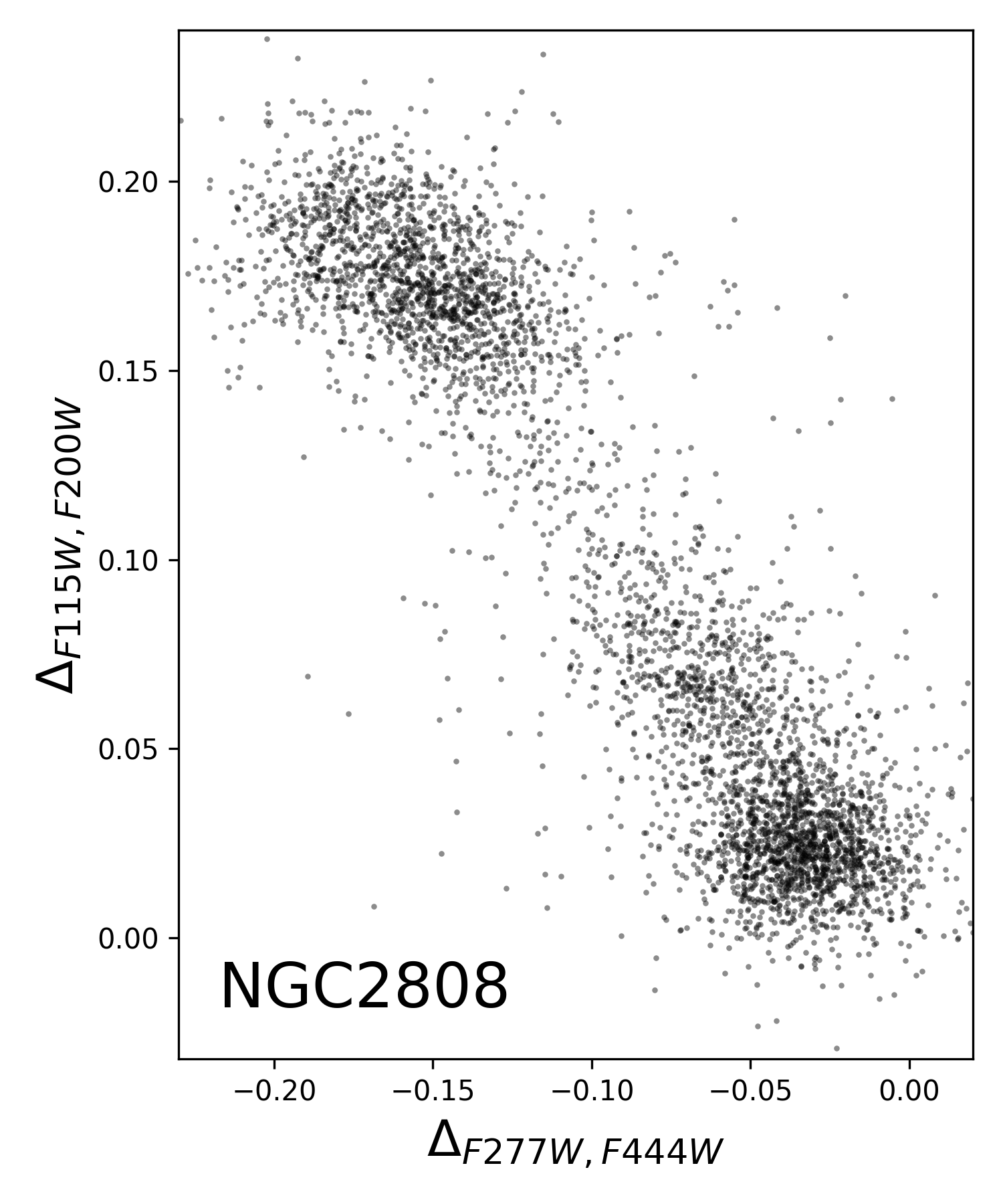}
    \includegraphics[width=0.32\linewidth]{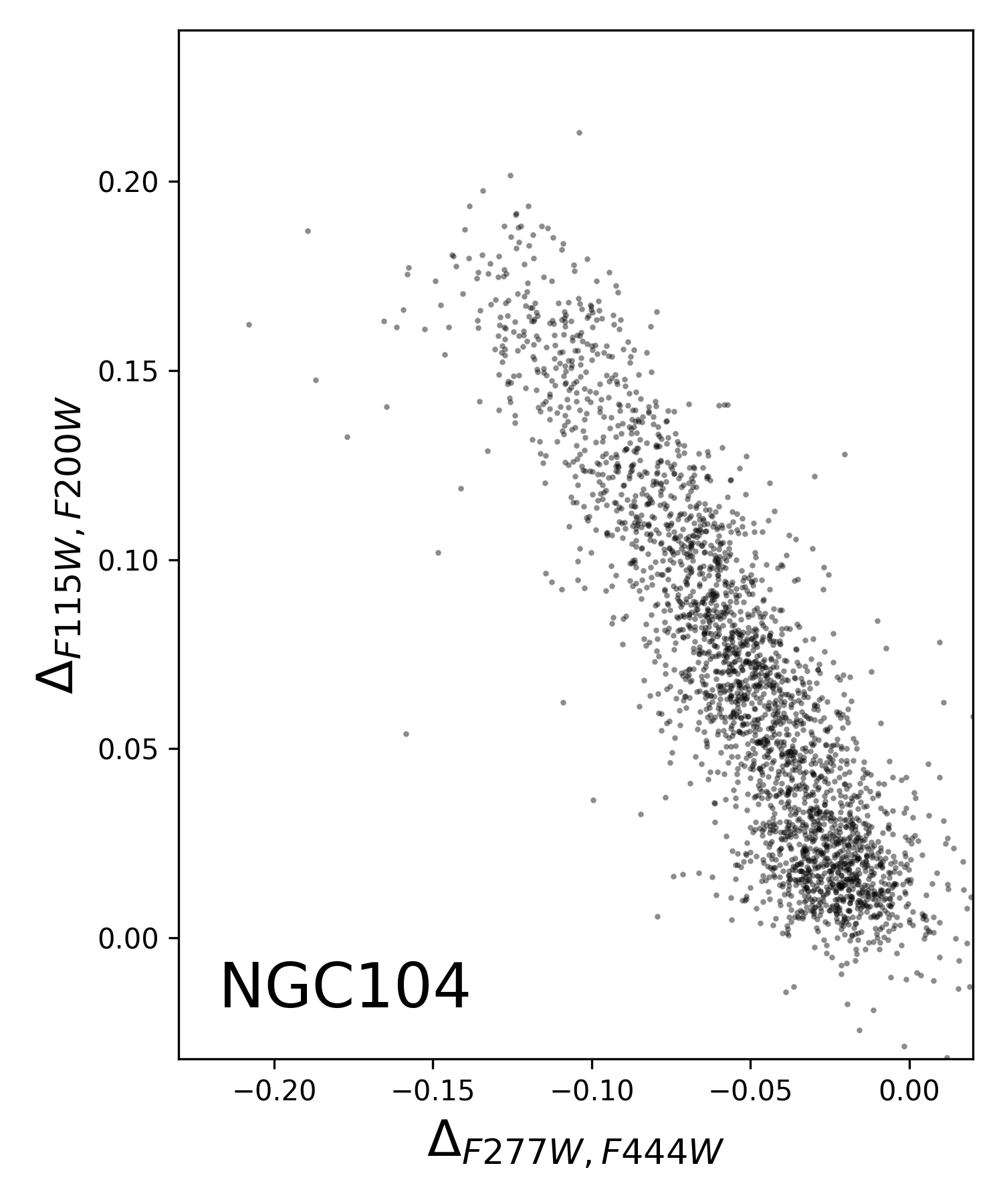}
    \includegraphics[width=0.32\linewidth]{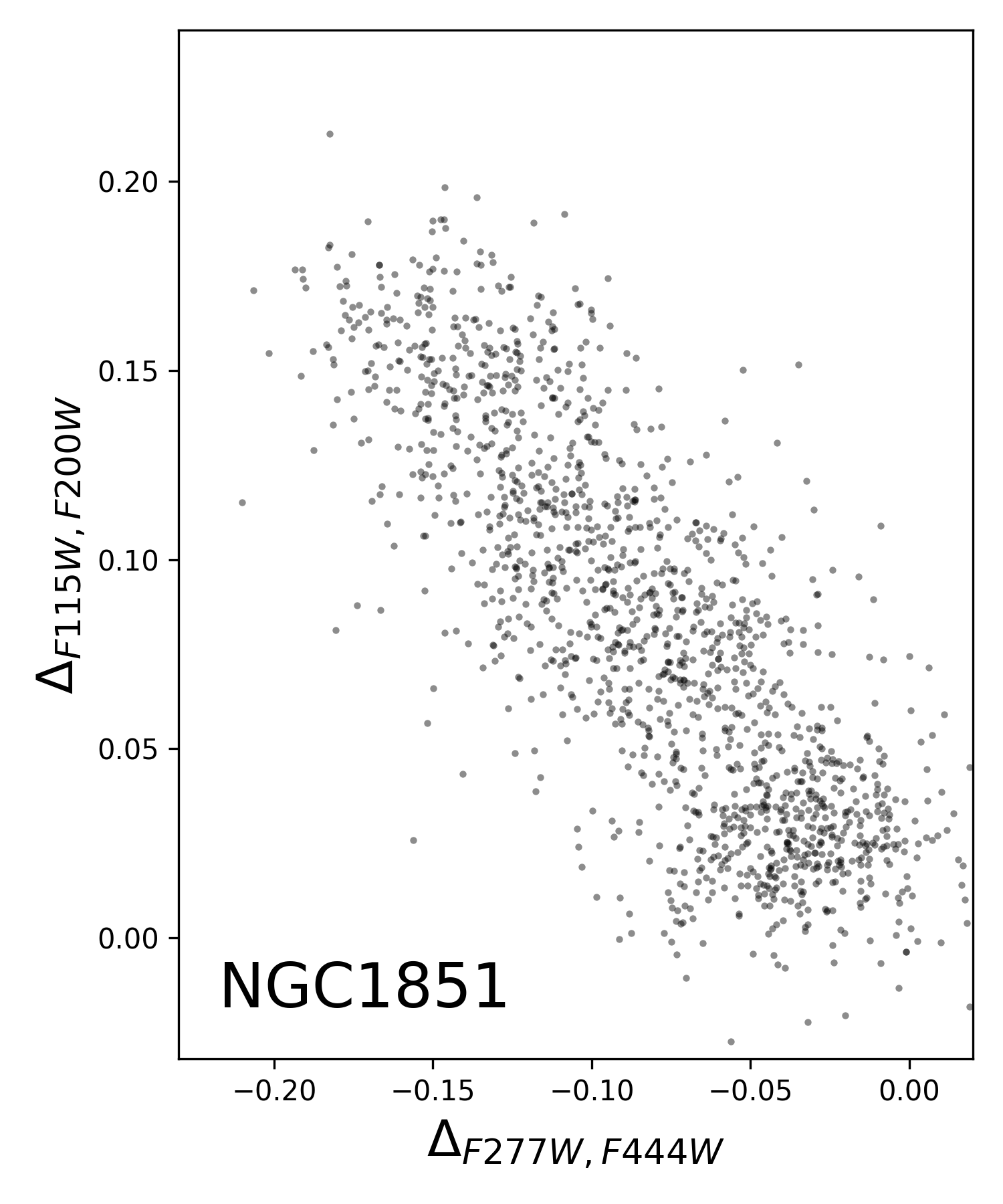}
    \includegraphics[width=0.32\linewidth]{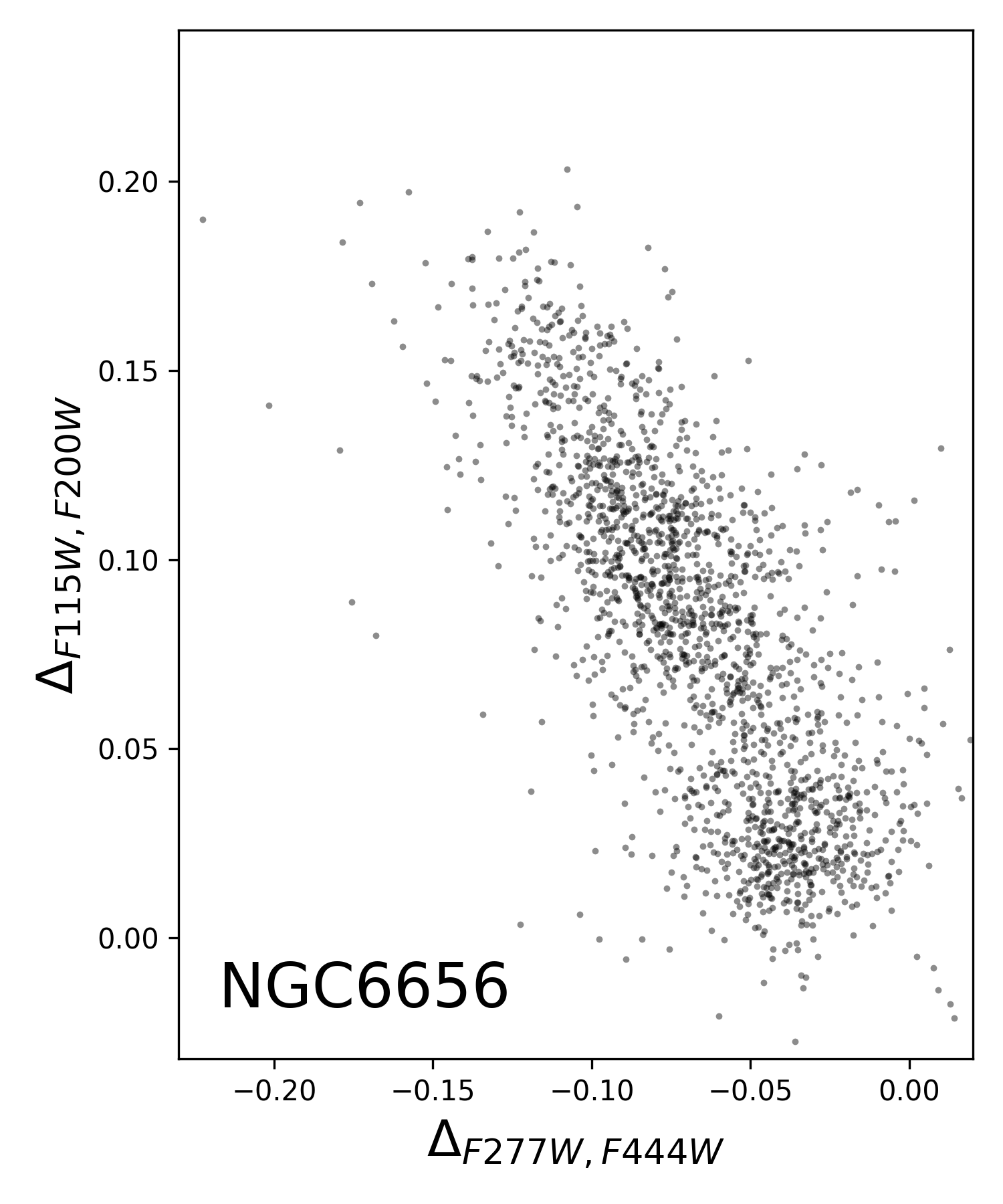}
\caption{ ChMs of M-dwarfs for the Type\,I GCs, NGC\,288, NGC\,6723, NGC\,2808 and NGC\,104 and for the Type\,II GCs NGC\,1851 and NGC\,6656.}
    \label{fig:ChMs}
\end{figure*}

\begin{figure*}
    \centering
    \includegraphics[width=1.0\linewidth]{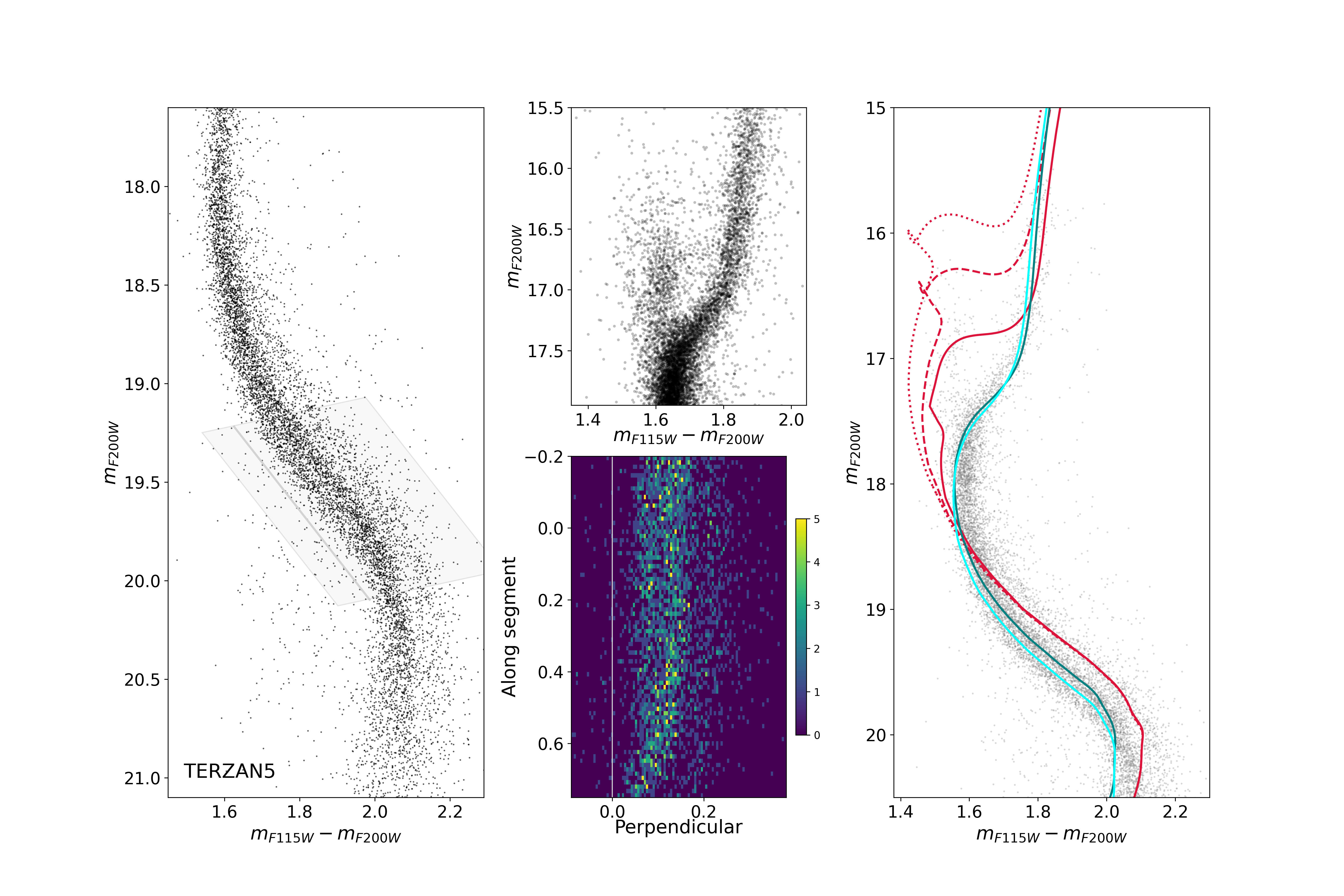}    
\caption{CMDs of proper-motion--selected stars in Terzan~5, corrected for differential reddening. The right panel shows stars at radial distances $>50\arcsec$ from the cluster center, while the bottom-middle panel presents the Hess diagram for stars above the knee. In this case, the reference frame is rotated so that the vertical axis aligns with the gray segment shown in the left panel. The top-middle panel zooms in on the SGB region. The right panel compares the observed CMD with isochrones from the BaSTI database: teal and cyan curves are $\alpha$-enhanced, 13\,Gyr isochrones with [Fe/H] $=-0.4$ and helium abundances $Y=0.25$ and $0.30$, respectively, while crimson curves are solar-scaled isochrones with {[Fe/H]=0.5 and} ages of 5.0, 3.5, and 2.0\,Gyr.}
    \label{fig:Terzan5}
\end{figure*}

\begin{figure*}
    \centering
    \includegraphics[width=1.0\linewidth]{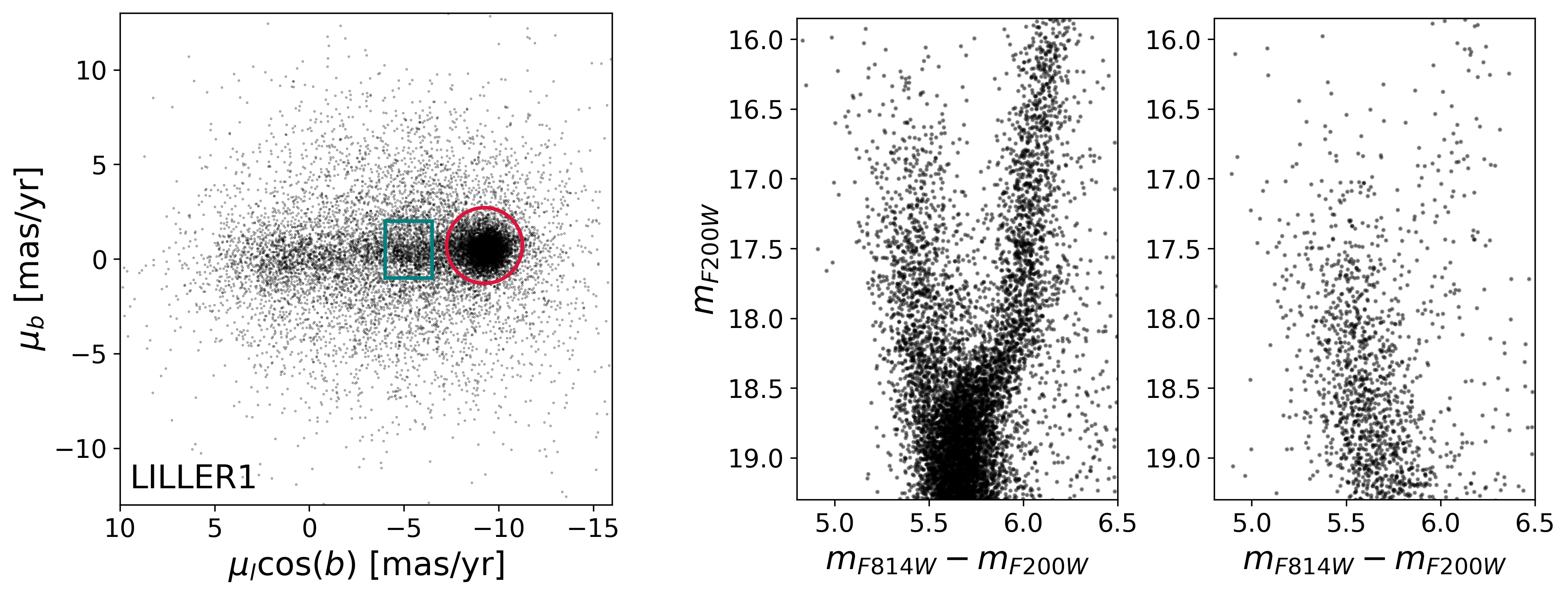}    
    \caption{Proper-motion diagram of stars in the field of view of Liller 1 (left panel). The middle and right panels show the CMDs corresponding to stars selected in the left panel: probable cluster members within the crimson circle, and field stars within the teal rectangle, respectively.}
    \label{fig:LILLER1PM}
\end{figure*}

\section{Multiple Stellar Populations}\label{sec:MPs}
To better disentangle the main behaviour of the multiple populations, we present in Fig.~\ref{fig:CMDsSWzoom} zoomed $m_{\rm F200W}$ vs.\ $m_{\rm F115W} - m_{\rm F200W}$ CMDs. These diagrams focus on the MS region around the knee, thereby emphasizing the multiple sequences both above and below it. 

\subsection{The Type\,I GCs NGC\,288, NGC\,6723, and NGC\,2808}
The top panels display the CMDs of the Type\,I GCs NGC\,288, NGC\,6723, and NGC\,2808. These diagrams confirm the trend of increasing complexity in the multiple-population patterns previously identified along the RGB \citep{milone2017a}. The MS stars brighter than the knee show a moderate color broadening that exceeds what is expected from observational uncertainties, and can therefore be ascribed to star-to-star helium variations. In NGC\,6723, we detect a split upper MS, with the blue component comprising approximately one third of the total MS population. NGC\,2808 exhibits a well-defined triple MS. 
Notably, at fixed magnitude, the blue MS displays a larger color spread than the middle and red sequences, suggesting the presence of an intrinsic helium dispersion among the blue-MS stars. Furthermore, at a given magnitude level relative to the turn off, the color broadening of bright MS stars increases progressively from NGC\,288 to NGC\,6723 and NGC\,2808.

All three GCs clearly exhibit split or multiple MSs among M-dwarfs. To better highlight the presence of multiple populations in the low-mass regime, we show in Fig.~\ref{fig:ChMs} the $\Delta_{F115W,F200W}$ vs.\ $\Delta_{F277W,F444W}$ ChMs for M-dwarfs \footnote{{The ChMs were constructed following the procedure described by \citet{milone2017a} and adopted in several subsequent papers by our team. Briefly, we derived the blue and red fiducial lines along the MS in the $m_{\rm F200W}$ vs.,$m_{\rm F115W}-m_{\rm F200W}$ and $m_{\rm F200W}$ vs.,$m_{\rm F277W}-m_{\rm F444W}$ CMDs. We then verticalized both diagrams with respect to these fiducial lines, which define the blue and red boundaries of the MS, and combined the resulting normalized color offsets to construct the ChM. The ChM width was normalized to the color separation between the blue and red fiducial lines, measured 1.5 magnitudes below the MS knee in the F200W band.}}. 
 The ChMs of NGC\,288, NGC\,6723, and NGC\,2808 display increasing extensions, from the least to the most massive cluster \footnote{NGC\,288, NGC\,6723, and NGC\,2808 have masses of 1.16$\pm$0.03$\times$10$^{5} M_{\odot}$, 1.57$\pm$0.13$\times 10^{5} M_{\odot}$, and 7.42$\pm$0.05$\times 10^{5} M_{\odot}$, respectively \citep{baumbardt2018a}.}. A particularly striking feature is the presence of nearly discrete stellar clumps in the ChMs of all three clusters. Nevertheless, the stellar density never drops to zero across the entire ChM; instead, prominent minima are observed around $\Delta_{F115W,F200W} \sim 0.05$ for NGC\,288 and $\Delta_{F115W,F200W} \sim 0.12$ for both NGC\,6723 and NGC\,2808. These minima give the visual impression of split MSs in the CMDs shown in Figs.~\ref{fig:CMDsSW} and \ref{fig:CMDsSWzoom}.

\subsection{NGC\,104 and the Type\,II GCs NGC\,1851 and NGC\,6656}
The CMDs of the Type~I GC NGC\,104 and of the Type~II GCs NGC\,1851 and NGC\,6656 exhibit several similarities. In particular, they show a moderate color broadening among MS stars brighter than the knee, and a more pronounced color spread among M-dwarfs. Both the ChMs and the CMDs display a continuous color distribution, with no evidence of significant gaps, in contrast to what is observed in NGC\,288, NGC\,6723, and NGC\,2808.

In contrast to the behavior observed among RGB stars, where the ChMs of NGC\,1851 and NGC\,6656 split into two distinct sequences populated by stars with different total C$+$N$+$O content and varying abundances of some heavy elements \citep{milone2017a, marino2019a, dondoglio2025a}, the ChMs of these Type~II GCs {shown in Figure\,\ref{fig:ChMs}} exhibit a single sequence. This is a consequence of the limited sensitivity of the $\Delta_{F115W,F200W}$ and $\Delta_{F277W,F444W}$ pseudo-colors to variations in C$+$N$+$O and metallicity. For example, a difference in [Fe/H] of 0.15~dex, corresponding to the iron-abundance difference between the two stellar groups in NGC\,6656 \citep{marino2009a, marino2011a}, translates into a variation of only $\sim$0.03~mag in $\Delta_{F115W,F200W}$ and leaves $\Delta_{F277W,F444W}$ nearly unchanged.

A common feature of most CMDs in Fig. \ref{fig:CMDsSWzoom} is the narrow color range spanned by stars in the vicinity of the MS knee. This behavior primarily arises because, in monometallic GCs, the sequences of 1P and 2P stars intersect and reverse their relative $m_{\rm F115W}-m_{\rm F200W}$ colors. In this context, NGC\,6656 represents a notable exception, as it exhibits a significantly broader color spread around the knee. This characteristic of the CMD is qualitatively consistent with internal metallicity variations, as indicated by spectroscopic studies of RGB stars \citep{marino2009a, marino2011a, mckenzie2022a}.

\subsection{The Bulge GCs NGC\,6528, NGC\,6553 and NGC\,6440}
The CMDs of the three bulge clusters exhibit distinct multiple-population patterns. Both NGC\,6528 and NGC\,6553 show only moderate color broadening among stars fainter than the MS knee, consistent with relatively small oxygen variations between 1P and 2P stars of [O/Fe]$\sim$0.1 dex. This contrasts with the much wider $m_{\rm F115W}-m_{\rm F200W}$ color range observed in NGC\,6440. 

{It is worth noting that spectroscopic studies of RGB stars by \citet{munoz2017a,munoz2018a,munoz2020a} reported no significant internal oxygen abundance spread in NGC~6528, NGC~6553, and NGC~6440. The modest oxygen variations inferred from our lower-MS photometry for NGC~6528 and NGC~6553 are therefore qualitatively consistent with these spectroscopic results. In contrast, the broader lower-MS color spread observed in NGC~6440 may suggest a larger oxygen variation than inferred from the available spectroscopy. A quantitative comparison between the oxygen abundances derived from our lower-MS photometry and those measured spectroscopically is beyond the scope of the present paper and will be the subject of a forthcoming study from this series.}

Given the masses of $8.96\pm1.85 \times 10^{4}$, $2.35\pm0.19 \times 10^{5}$, and $4.42\pm0.64 \times 10^{5}\,M_{\odot}$ for NGC\,6528, NGC\,6553, and NGC\,6440, respectively \citep{baumbardt2018a}, this result may suggest the presence of a mass dependence of the internal oxygen variation in these metal-rich GCs.

The multiple-population picture is further complicated by the morphology of the upper MS. NGC\,6553 displays a narrow upper MS, whereas both NGC\,6528 and NGC\,6440 exhibit two distinct MSs. If the observed MS splitting is driven by helium variations, these differences may indicate a smaller internal helium spread in NGC\,6553 compared to NGC\,6528 and NGC\,6440.
 
These findings appear to challenge previous evidence suggesting that the extent of internal helium variation correlates with GC mass, and instead point toward a lack of correlation between helium and oxygen variations.

The elemental abundances inferred from our dataset may help constrain the nature of the polluters responsible for the chemical composition of 2P stars. In particular, theoretical models predict that the oxygen depletion produced by metal-rich massive AGB stars is substantially smaller than that expected from metal-poor AGB stars \citep{ventura2018a, dellagli2018a}. The limited internal oxygen variations observed in the metal-rich GCs NGC\,6528 and NGC\,6553 are therefore in good agreement with these predictions.

\subsection{The multi-age GCs Terzan\,5 and Liller\,1}

The CMD of probable cluster members of Terzan\,5 is shown in Fig.\,\ref{fig:Terzan5}. In the left-panel CMD, we excluded the highly crowded central regions by considering only stars located at radial distances greater than 50 arcsec from the cluster center. A striking  feature {identified here for the first time} is the presence of a sparsely populated red MS that lies on the red side of the bulk of the MS population.

Our analysis further reveals that dominant MS itself is split into two distinct sequences, a feature that is most clearly visible in the CMD region highlighted by the light-gray shaded area. This region is further emphasized in the Hess diagram shown in the bottom-middle panel, where the CMD has been rotated so that the gray segment is aligned vertically \citep{marchuk2026a}. Notably, the two MS components converge at the level of the MS knee, similarly to what is observed in metal-rich Bulge GCs such as NGC\,6528 and NGC\,6440.

The upper-middle panel presents the proper-motion-selected CMD of the entire field of view, zoomed in on the SGB region. The most prominent feature is the presence of multiple SGBs and MS stars near the turn offs (MSTOs), as also reported by \cite{zullo2026a}. In addition to the most populous faint SGB and MSTO, two significant stellar overdensities are visible at approximately $m_{\rm F200W}\sim16.7$ and $m_{\rm F200W}\sim16.3$, together with a population of brighter SGB and MSTO stars extending up to $m_{\rm F200W}\sim15.7$ and beyond.

{Our analysis demonstrates that} these brighter SGB and MSTO stars are associated with the reddest MS component, as illustrated in the right panel of Fig.\,\ref{fig:Terzan5}, where we compare the observations with solar-scaled BaSTI isochrones of [Fe/H]$=+0.5$ and ages of 5.0, 3.5, and 2.0 Gyr. In contrast, the dominant MS population is well reproduced by 13 Gyr isochrones with [Fe/H]$=-0.4$, [$\alpha$/Fe]$=0.4$, and helium abundances of $Y=0.25$ and $Y=0.30$.

\citet{ferraro2009a,ferraro2021a} proposed that Terzan\,5 and Liller\,1 are not genuine GCs, but rather surviving relics of the early Galactic bulge formation process. This interpretation is motivated by observations of high-redshift star-forming galaxies, which frequently host massive UV-bright clumps associated with intense star formation \citep[e.g.,][]{cowie1995a,elmegreen2004a,elmegreen2005a}. These clumps are thought to migrate toward the galaxy centres, contributing to bulge assembly and potentially evolving into metal-rich GC analogues. Owing to their large masses, they may survive for several Gyr, retain stellar ejecta and supernova products, and experience multiple episodes of star formation and chemical enrichment, thereby tracing the assembly history of the bulge \citep{immeli2004a,shapiro2010a}.

Within this framework, {based on our identification of multiple helium populations within the old stellar component of Terzan\,5, we suggest} suggest that multiple helium-enriched populations may also form in dense primordial environments other than classical GCs, such as bulge progenitor fragments. In this context, a fraction of the 2P-like stars observed in the Galactic bulge \citep{schiavon2017a} could originate from such systems.

An alternative scenario is proposed by \cite{mckenzie2018a} \citep[see also][]{bastian2022a}, who suggested that massive clusters in the inner Galaxy may occasionally interact kinematically and spatially with giant molecular clouds, enabling rapid gas accretion and subsequent star formation. Such events are expected to be rare over a Hubble time but more likely in the dense inner few kiloparsecs. In this picture, the old populations in systems such as Terzan\,5 and Liller\,1 correspond to original GC stars, while younger populations form from accreted gas. In this context, the presence of multiple MSs with different helium content in Terzan\,5 is consistent with the interpretation that its old component represents a genuine GC that has subsequently undergone self-enrichment, a phenomenon commonly observed in massive clusters.

The left panel of Fig.\,\ref{fig:LILLER1PM} shows the proper-motion diagram of stars in the field of view of Liller\,1. The distribution reveals three main components: a hot, nearly spheroidal bulge population, the bulk of cluster members enclosed within the crimson circle, and a kinematically cold component consistent with Galactic disk stars. This latter population shows a tight, flattened distribution in proper-motion space, with $\mu_{\rm b}$ narrowly distributed around zero. Notably, the proper motions of Liller\,1 lie along the same locus and may partially overlap with those of the disk population.

The CMD of probable cluster members is shown in the middle panel of Fig.\,\ref{fig:LILLER1PM}. It confirms the coexistence of an old stellar population, defining the faint SGB and RGB, together with a younger MS population.

The right panel presents the CMD of stars enclosed in the teal rectangle in the proper-motion diagram. {We find that} these kinematically cold stars, likely associated with the Galactic inner disk, overlap  in colour-magnitude space with the younger stellar sequence of Liller\,1, suggests similar photometric properties between the two populations.
Noticeably, { we find that a disk-like component is also evident} in the proper-motion diagram of Terzan\,5 (Fig.\,\ref{fig:PMs}), where stars with $-1.5<\mu_{\rm b}<1.5$ mas/yr and $-2.0<\mu_{l}\cos{b}<1.0$ mas/yr occupy a region of the CMD consistent with the young stellar populations of Terzan\,5. 

Our results are qualitatively consistent with the scenario proposed by \cite{mckenzie2018a}. The coexistence of old and young stellar populations in Liller\,1 supports multiple star-formation episodes. In addition, the partial overlap in proper-motion space and photometric properties between the kinematically cold disk component and the young stellar sequence suggests that the cluster resides in a dynamical regime where interactions with Galactic disk material are plausible.

\begin{figure}
    \centering
    \includegraphics[width=1.0\linewidth]{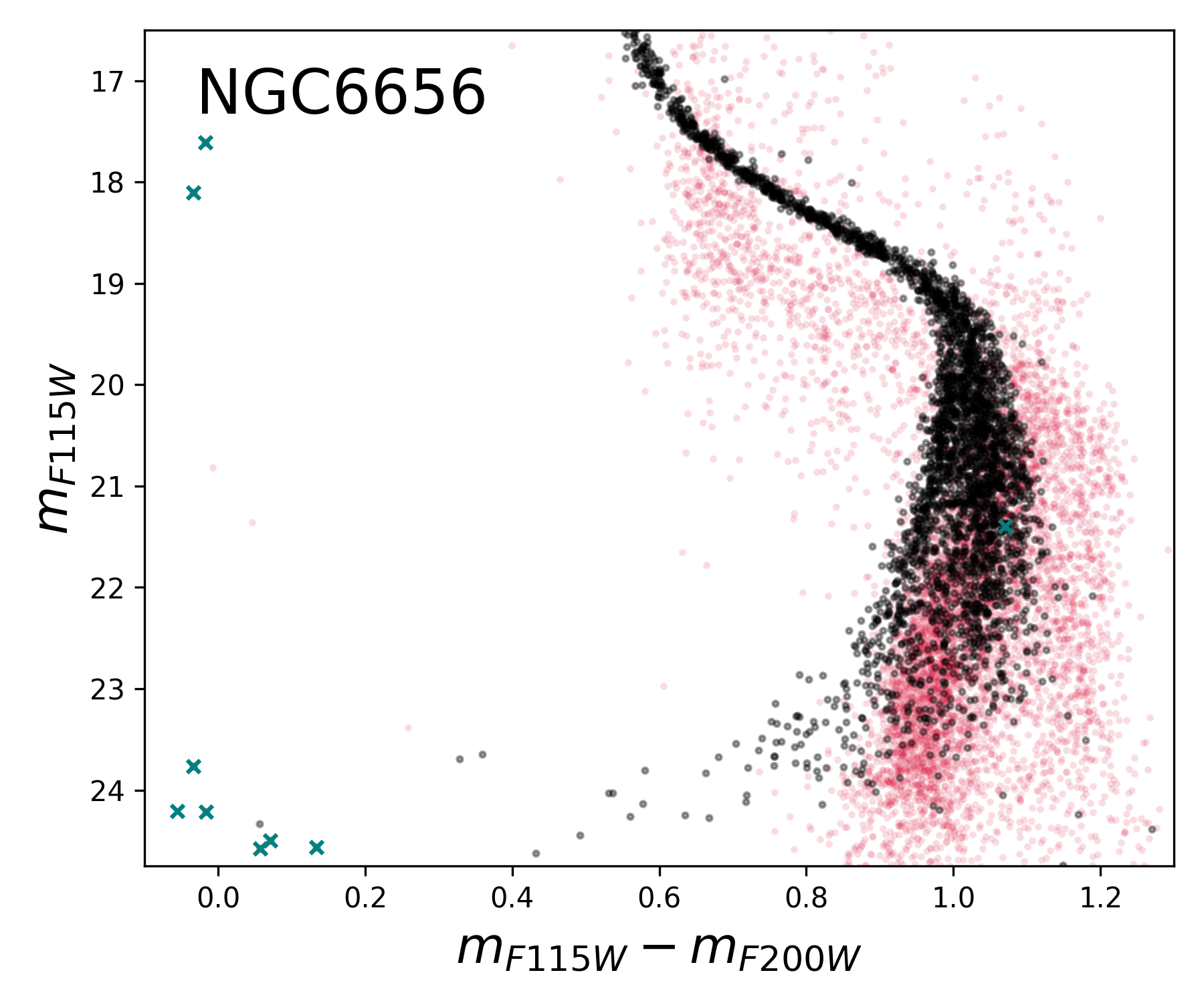}
\caption{CMD of stars with measured proper motions in the field of view of NGC\,6656. Black and red symbols denote cluster members and field stars, respectively. Probable white dwarfs are marked by teal crosses.}
    \label{fig:coda}
\end{figure}

 \begin{figure}
    \centering
    \includegraphics[width=1\linewidth]{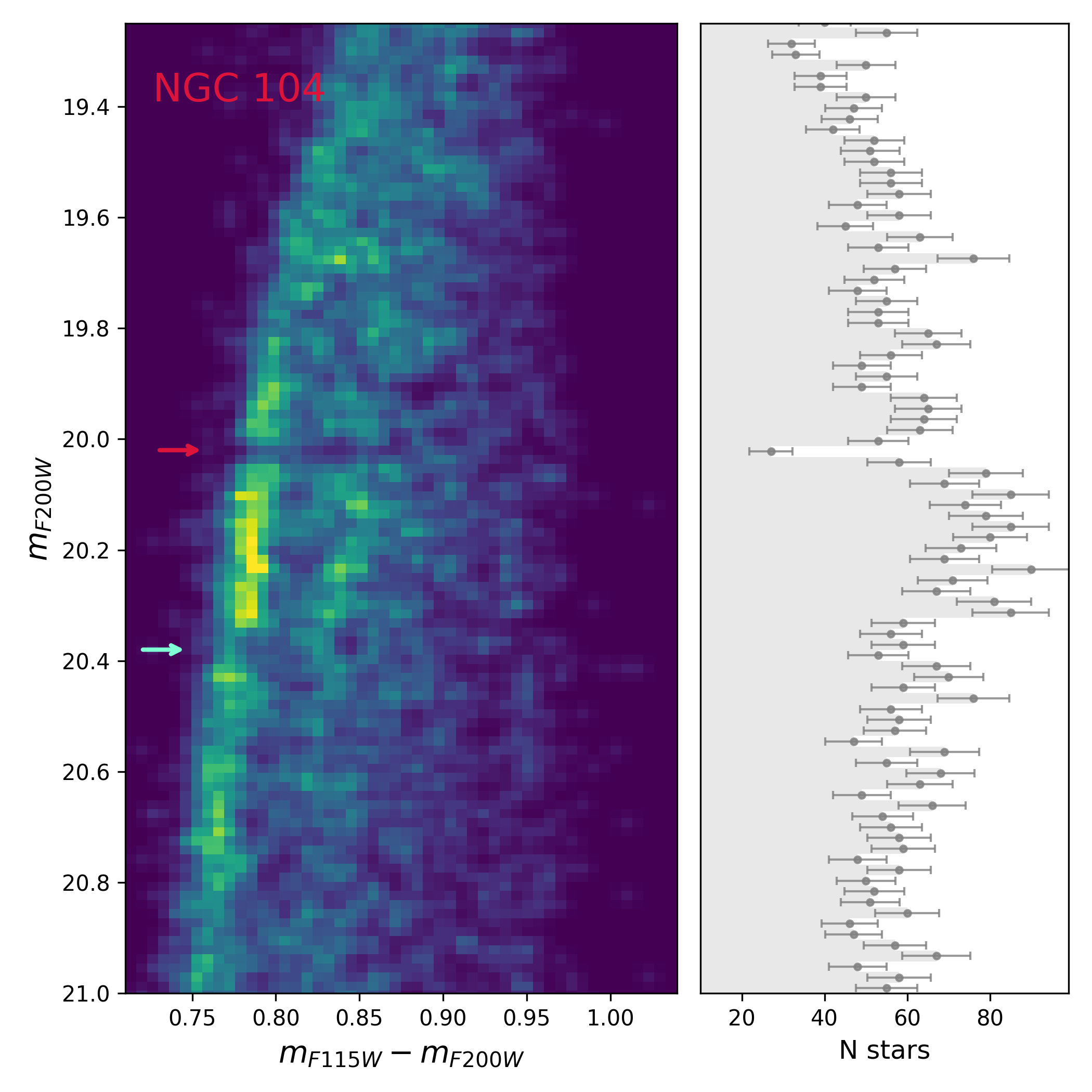}
    \includegraphics[width=1\linewidth]{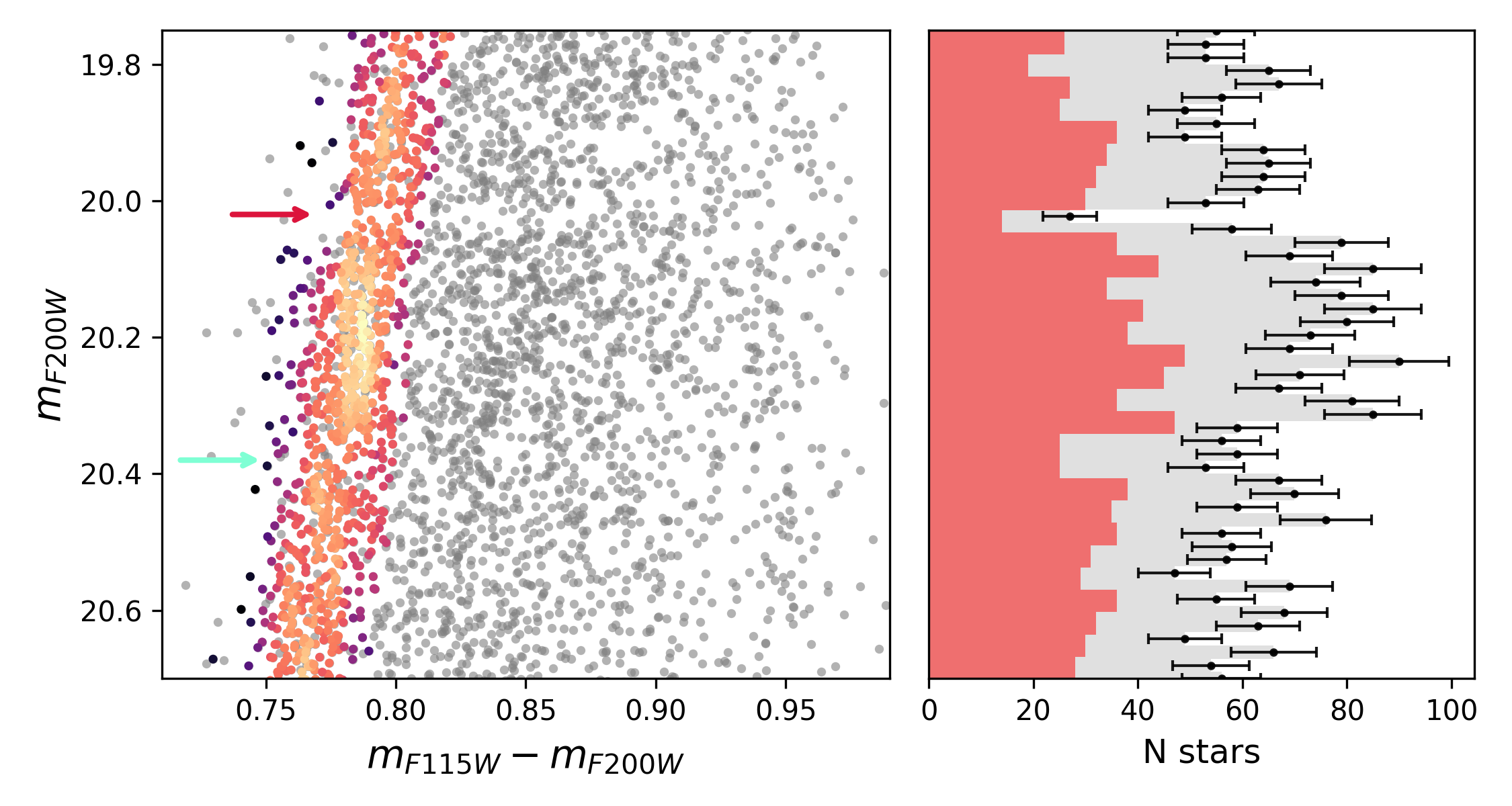}
\caption{\textit{Top panels:} Hess diagram of stars fainter than the MS knee in NGC\,104 (left) and corresponding luminosity function (right). \textit{Bottom panels:} CMD of NGC\,104. 1P stars are color-coded according to their local density in the CMD, while all other stars are shown in gray. The right panel compares the luminosity functions of all stars (gray histogram) and 1P stars (light-red histogram). In the left panels, the crimson arrows indicate the M-dwarf gap, whereas the aquamarine arrows mark the magnitude level at which a drop in the luminosity function is observed. }
    \label{fig:gap}
\end{figure}

\section{Other results and scientific potential of the Dataset}\label{sec:potential}

The high-precision astrometry and photometry provided by the GO-8960 program, combined with the archival data presented in this work, enable a wide range of investigations focused on GCs and their stellar populations. In the following, we outline the main research projects that will be carried out using this dataset.

\begin{itemize}

\item \textit{Detection and chemical characterization of multiple populations among M dwarfs.} 
Investigating the multiple-population phenomenon among low-mass stars is one of the primary goals of this project. In this paper, we have presented deep photometric diagrams for all clusters in our sample and highlighted the main features of their stellar populations. Building on these results, we will derive the fraction of stars belonging to 1P and 2P stars, as well as to the different 2P subpopulations, in close analogy with previous surveys based on \textit{HST} data \citep{milone2017a}.

The dataset is also well suited to constraining the chemical composition of multiple populations. In particular, it enables accurate estimates of oxygen abundances in late K and M dwarfs, as well as relative helium abundances for MS stars brighter than the MS knee. The synergy with archival \textit{HST} photometry, especially in the ultraviolet, will further allow precise measurements of relative nitrogen abundances among brighter MS stars.

\item \textit{The very low--mass regime and the stellar mass function.}
The infrared sensitivity and spatial resolution of JWST enable the detection of very low--mass stars, revealing previously unknown features such as gaps and discontinuities along the stellar sequences below $\sim 0.1\,M_\odot$, approaching the brown dwarf regime \citep{marino2024a}. The present dataset allows us to probe the late M-dwarf regime, which was largely inaccessible with previous facilities. An example is shown in Fig.~\ref{fig:coda}, where a sequence of cool dwarfs bends toward bluer colors at the faint end of the MS of NGC\,6656. These observations open a new window on the stellar mass function and provide stringent tests of stellar models in the very low--mass regime \citep[see][for models of very low-mass stars in NGC\,104]{gerasimov2024a, ventura2026a}.


\item \textit{Metallicity distribution of first-population stars.}
Studies of RGB stars have shown that 1P stars in most GCs exhibit extended sequences in ChM \citep{milone2015a, milone2018a, legnardi2022a}, reflecting small star-to-star metallicity variations \citep{marino2019a, marino2023a, legnardi2022a}. More recently, a similar feature has been identified among M dwarfs in NGC\,104 \citep{legnardi2024a}. The infrared dataset presented in this paper, when combined with optical \textit{HST} data, will allow us to constrain the metallicity distribution of low-mass stars with a precision of a few hundredths of a dex.
\footnote{{Synthetic spectra and spectroscopic analyses have demonstrated that the intrinsic extension of 1P stars in ChMs constructed from suitable combinations of {\it HST} and {\it JWST} filters is primarily driven by small star-to-star metallicity variations \citep{marino2019a, marino2019b, marino2023a, legnardi2022a}. Consequently, once the observed pseudo-colour distribution is corrected for photometric uncertainties, it can be directly translated into a metallicity distribution through comparison with synthetic stellar models. The accuracy of the inferred metallicities is therefore determined mainly by the photometric precision of the ChM, and can reach a few hundredths of a dex for high-quality {\it HST}+{\it JWST} photometry \citep[see, e.g.,][]{legnardi2022a, legnardi2024a}.}}
 providing insight into the chemical properties of the primordial gas from which 1P stars formed.

\item \textit{Mass functions of multiple populations.}
Determining the luminosity and mass functions of GC stars, including their distinct populations, is crucial for understanding cluster formation and evolution. In particular, comparing the MFs of 1P and 2P stars provides constraints on the possible dependence of the initial mass function (IMF) on the formation environment. If 2P stars formed under significantly denser conditions than 1P stars, differences in their present-day MFs may reveal whether the IMF is universal or varies with environment \citep{dondoglio2022a}.

\item \textit{Binary populations.}
The analysis of CMDs provides key constraints on the binary content of GCs, including the fraction of MS--MS systems, their mass-ratio distribution, and their radial distribution \citep{romani1991a, sollima2007a, milone2012a}. These properties are essential for understanding GC dynamical evolution and the formation of exotic stellar populations, such as blue straggler stars, cataclysmic variables, millisecond pulsars, and low-mass X-ray binaries.

We will exploit the CMDs presented in this work to refine measurements of the binary fraction, mass-ratio distribution, and radial segregation, extending previous studies \citep{milone2012a}, which were largely limited to cluster inner regions. We will also estimate the binary fraction for different stellar populations following the methodology introduced in our previous works \citep{milone2020a, milone2025b, bortolan2025a}. Measuring the binary fraction separately for 1P and 2P stars is particularly important, as 2P stars likely formed in denser environments where binaries are more efficiently disrupted. Differences in binary fraction and radial distribution between the two populations thus provide direct constraints on formation scenarios and dynamical evolution.

\item \textit{Internal kinematics of multiple populations.}
The study of the internal kinematics of multiple populations provides one of the few direct observational constraints on both their formation and subsequent dynamical evolution \citep{vesperini2021a}. The combination of the JWST dataset with complementary \textit{HST} and \textit{Gaia} data enables precise measurements of velocity dispersion, anisotropy, rotation, and the degree of energy equipartition for 1P and 2P stars. Since internal kinematics evolve through two-body relaxation and mixing processes, comparing the dynamical properties of different populations provides key insights into the dynamical age of clusters and the progressive mixing of initially distinct stellar populations \citep[e.g.][]{richer2013a, bellini2015a, cordoni2020a, cordoni2025a, ziliotto2023a, ziliotto2025a, ziliotto2026a}.

\end{itemize}

\subsection{The M-dwarf gap}
A careful inspection of the CMDs of NGC\,104 shown in Figs\,\ref{fig:CMDsSW} and \ref{fig:CMDsSWzoom} reveals a sharp discontinuity along the MS at nearly constant magnitude, $m_{\rm F200W}\simeq20.02$. This feature is clearly visible in the Hess diagram displayed in the upper-left panel of Fig.\,\ref{fig:gap} and corresponds to a pronounced decrease in the stellar number counts, as shown by the luminosity function in the upper-right panel.  Based on the isochrone from \cite{dotter2008a} with [Fe/H]=$-$0.75, [$\alpha$/Fe]=0.4 that betters fit the observed CMD by assuming ($m_M$)$_{0}$=13.27 and E(B$-$V)=0.01, we infer a mass for the  gap of 0.352 $m_{\odot}$.
Below the gap, the luminosity function rises again over an interval of about 0.35 mag in the F200W band (corresponding to $\sim$0.05 M$_{\odot}$), with the star counts increasing by approximately 25\%. 
 These features are well visible also when we consider 1P stars only, based on the location in the ChM ($\Delta_{F115W,F200W} \lesssim 0.03$ mag). 

Possibly, the M-dwarf gap is not a peculiarity of NGC\,104. As an example, a visual inspection of Fig.\,\ref{fig:ngc2808twopanel} suggests a similar feature in NGC\,2808 at $m_{\rm F200W} \sim 21.8$, corresponding to a mass of $\sim 0.35\,M_{\odot}$.

The observed feature closely resembles the Jao gap, a narrow underdensity identified in the CMD of nearby low-mass field stars by \citet{jao2018a}, as well as the discontinuity recently detected by \citet{marchuk2026a} in the $\sim$2,Gyr-old open cluster NGC\,2158. These CMD features have been interpreted as the observational signature of structural changes occurring near the fully convective boundary of low-mass stars. At masses of $M\sim0.3$--$0.4,M_{\odot}$, the redistribution of $^3$He through convective mixing modifies the nuclear energy generation rate and induces changes in the stellar structure \citep{dantona1982a}. Models further predict episodic mixing events, known as the \emph{convective kissing instability}, in which the convective core temporarily merges with the outer convective envelope \citep{andronov2004a,vansaders2012a}. The resulting thermal readjustments produce small variations in stellar radius and luminosity, giving rise to an underdensity in the CMD.

The detection of the M-dwarf gap in NGC\,104, together with the recent discovery of a similar feature in NGC\,6397 based on Euclid observations \citep{griggio2026a}, demonstrates that this phenomenon is not restricted to young and intermediate-age stellar populations but is also present in ancient ($\sim$12--13\,Gyr) GCs. This finding appears to be at odds with current theoretical predictions, which suggest that the feature should disappear at old ages \citep[see discussion in][]{marchuk2026a}.

The IR photometry used in this project can enable precise age determinations. Optical CMDs suffer from degeneracies between age, metallicity, distance, and reddening, whereas infrared CMDs offer a complementary approach. In this regime, the MS knee, produced by opacity effects in M dwarfs where cooler stars become bluer in infrared colors, provides a potential anchor point for age determinations, when combined with the age-sensitive MS turnoff \citep{bono2010a, correnti2016a}. However, its use {as a reference point} for high-precision age estimates requires careful modelling of the impact of multiple stellar populations on the knee morphology.

In contrast, the M-dwarf gap may provide an alternative and better-defined reference point, less affected by the complex morphology of the MS knee and therefore useful for improving age determinations.

\subsection{The Galactic Bulge}
Although the project is primarily focused on GCs, the dataset also provides deep photometry of proper-motion-selected field stars.

\begin{figure*}
    \centering
    \includegraphics[width=.48\linewidth]{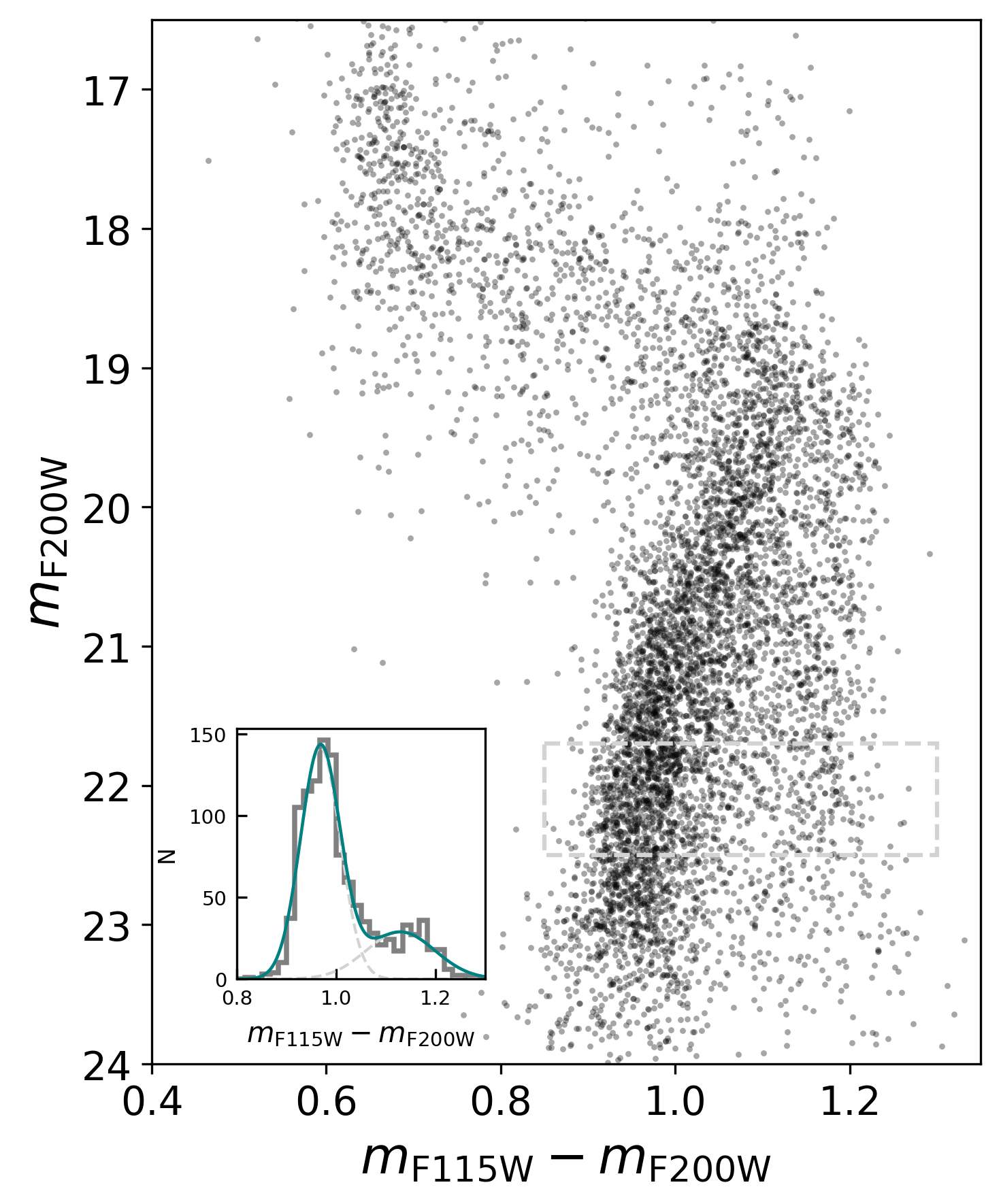}
    \includegraphics[width=.48\linewidth]{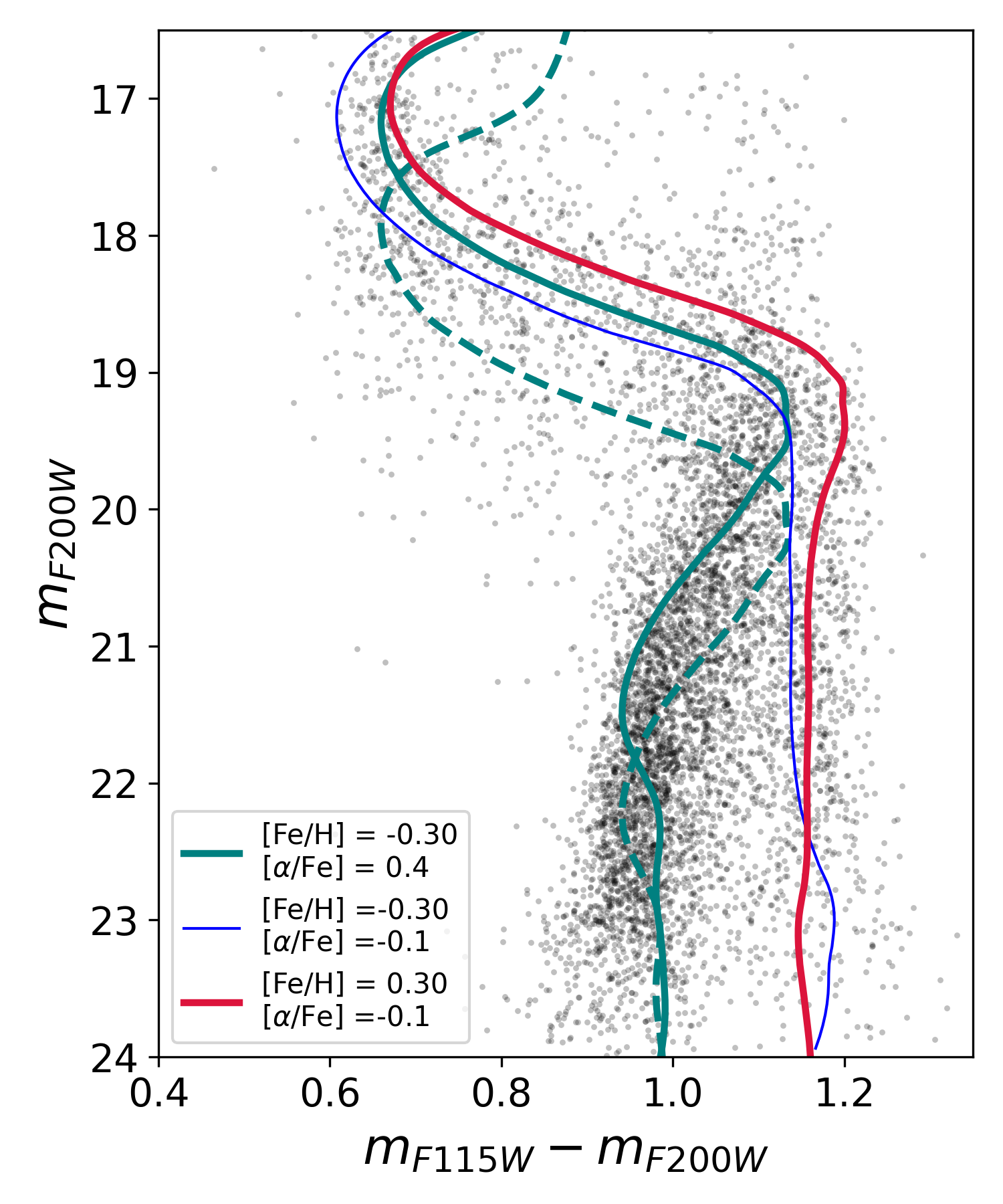}
\caption{ Proper-motion-selected CMDs of Bulge and field stars toward NGC\,6656, corrected for differential reddening. The inset shows the histogram distribution of the stars located withing the gray dashed rectangle. The teal line is the best-fit bi-Gaussian function and its two components are represented with gray dashed lines.  
The right-hand panel shows 12-Gyr isochrones from the BaSTI database, overplotted for different metallicities and $\alpha$-element abundances. We adopted a reddening of $E(B-V)=0.38$ mag. The solid isochrones were computed assuming a distance modulus of $(m-M)_0 = 14.15$ mag, whereas the dashed isochrone assumes $(m-M)_0 = 14.90$ mag.
}
    \label{fig:bulge}
\end{figure*}

As an example, it enables the construction of deep CMDs of stars along lines of sight toward the Galactic bulge, as illustrated in Fig.~\ref{fig:bulge} for the field of NGC\,6656.

The coexistence of GC and bulge stars within the same field of view offers a key advantage: the high-resolution reddening map derived from GC stars can be used to correct the photometry of bulge stars with unprecedented precision. This approach allows the detection of subtle features across the CMD \citep[e.g.][]{lagioia2014a}.

The CMD of field stars in the direction of NGC\,6656 reveals a variety of complex structures, including a bimodal color distribution below the MS knee, at $m_{\rm F200W} \gtrsim 19.5$. This fact is illustrated in the inset, where we show the histogram distribution of the $m_{\rm F115W}-m_{\rm F200W}$ color for the faint MS stars enclosed in the gray dashed box. The color distribution is fitted with a bi-Gaussian function.

A comparison with 12-Gyr isochrones from the BaSTI database \citep{pietrinferni2021a} is shown in the right-hand panel of Fig.~\ref{fig:bulge}. Above the MS knee, the CMD exhibits only a moderate sensitivity to metallicity. In particular, the two isochrones shown as solid blue and red lines, both with [$\alpha$/Fe]$=-0.1$ but differing in metallicity ([Fe/H]$=-0.3$ and [Fe/H]$=0.3$, respectively), are separated by a small amount compared to the observed spread in colour and magnitude.

The separation between isochrones with different [Fe/H] values is even smaller when considering {metal-poorer}, $\alpha$-enhanced models, such as the teal isochrone with [Fe/H]$=-0.3$ and [$\alpha$/Fe]$=0.4$. We also show the corresponding isochrone (dashed teal line) with identical chemical composition but shifted by a distance modulus larger by 0.75 mag. While the solid isochrones assume $(m-M)_0 = 14.15$ mag, the dashed teal isochrone adopts $(m-M)_0 = 14.90$ mag. This comparison suggests that a significant fraction of the observed spread in colour and magnitude among bright MS stars can be attributed to distance variations.

In contrast, below the MS knee, the $m_{\rm F115W}-m_{\rm F200W}$ colour becomes highly sensitive to variations in $\alpha$-element abundance. The observed split MS is therefore consistent with the presence of two stellar populations characterised mostly by different $\alpha$-element enhancements. In particular, based on the relative amplitudes of the best-fitting bi-Gaussian decomposition, we find that $\alpha$-rich stars account for $84 \pm 2\%$ of the total sample.

These results demonstrate that this CMD provides a powerful diagnostic for disentangling stellar populations with distinct chemical compositions along the faint MS of bulge stars and for constraining their $\alpha$-element distribution. Consequently, the present dataset allows not only the derivation of deep stellar mass functions in the bulge, but also the characterization of multiple MSs associated with chemically distinct populations.

\subsection{Extragalactic Sources}
In addition, the available multi-band photometry provides a valuable opportunity to investigate background galaxies. Deep imaging across multiple filters enables the detection and characterization of faint extragalactic sources located behind both the cluster and foreground field-star populations. The resulting spectral energy distributions (SEDs) can be used to derive robust photometric redshifts, thereby enabling statistical studies of galaxy number counts and spatial clustering, and ultimately providing constraints on models of galaxy evolution and large-scale structure formation.

\begin{figure*}
    \centering
    \includegraphics[width=.95\linewidth]{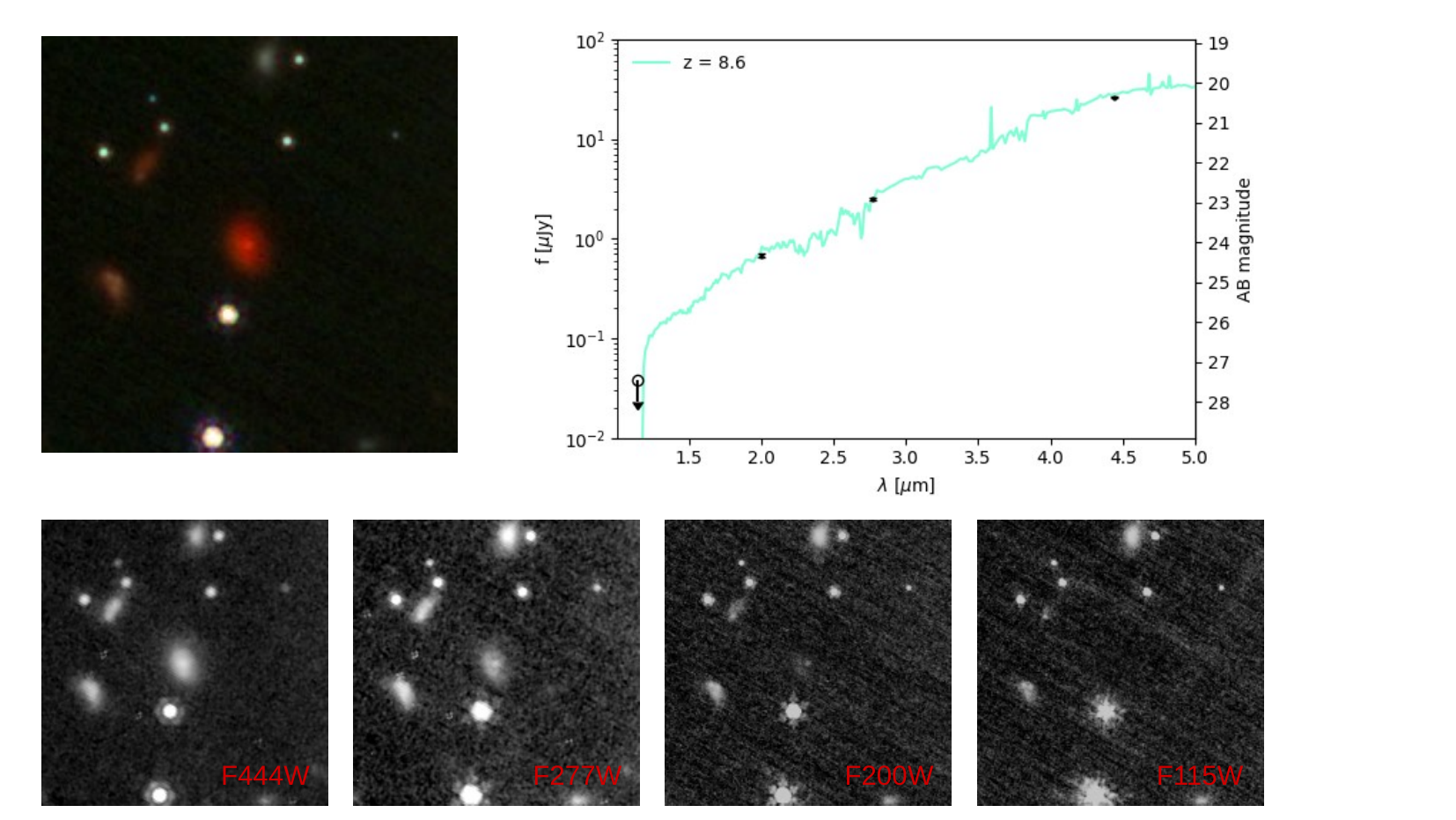}
\caption{Example of a background galaxy identified in the  field of NGC\,1851. The upper-left panel shows a three-colour composite image of a $6\times 6$ arcsec region centered at $\mathrm{RA}=05^{\mathrm h}13^{\mathrm m}45.48^{\mathrm s}$, $\mathrm{Dec}=-40^\circ\,04'\,04.5''$. The upper-right panel presents the observed SED of the central red galaxy together with the best-fitting template derived using the EAZY code. The bottom panels display the stacked images in the F444W, F277W, F200W, and F115W filters (from left to right).}
    \label{fig:gal1}
\end{figure*}

As an example, Fig.~\ref{fig:gal1} presents a three-colour image of a $6\times 6$ arcsec region centered at $\mathrm{RA}=05^{\mathrm h}13^{\mathrm m}45.48^{\mathrm s}$, $\mathrm{Dec}=-40^\circ\,04'\,04.5''$. We also show the stacked images in the F444W, F277W, F200W, and F115W bands, together with the SED of the central red galaxy, serendipitously discovered thanks to the depth of the present observations. The galaxy displays a regular morphology, characterized by a prominent central concentration and a possible disk-like component, indicative of a relatively evolved system. Although the non-detection in the F115W band could formally be interpreted as a Lyman-break dropout at $z\sim8.6$, its resolved morphology argues against such a high-redshift scenario. An alternative interpretation is that of a dusty and/or evolved galaxy at intermediate redshift ($z\sim3$--4), where a pronounced Balmer/4000\,\AA\ break, possibly combined with significant dust attenuation, gives rise to the observed red SED.

\begin{figure*}
    \centering
    \includegraphics[width=.95\linewidth]{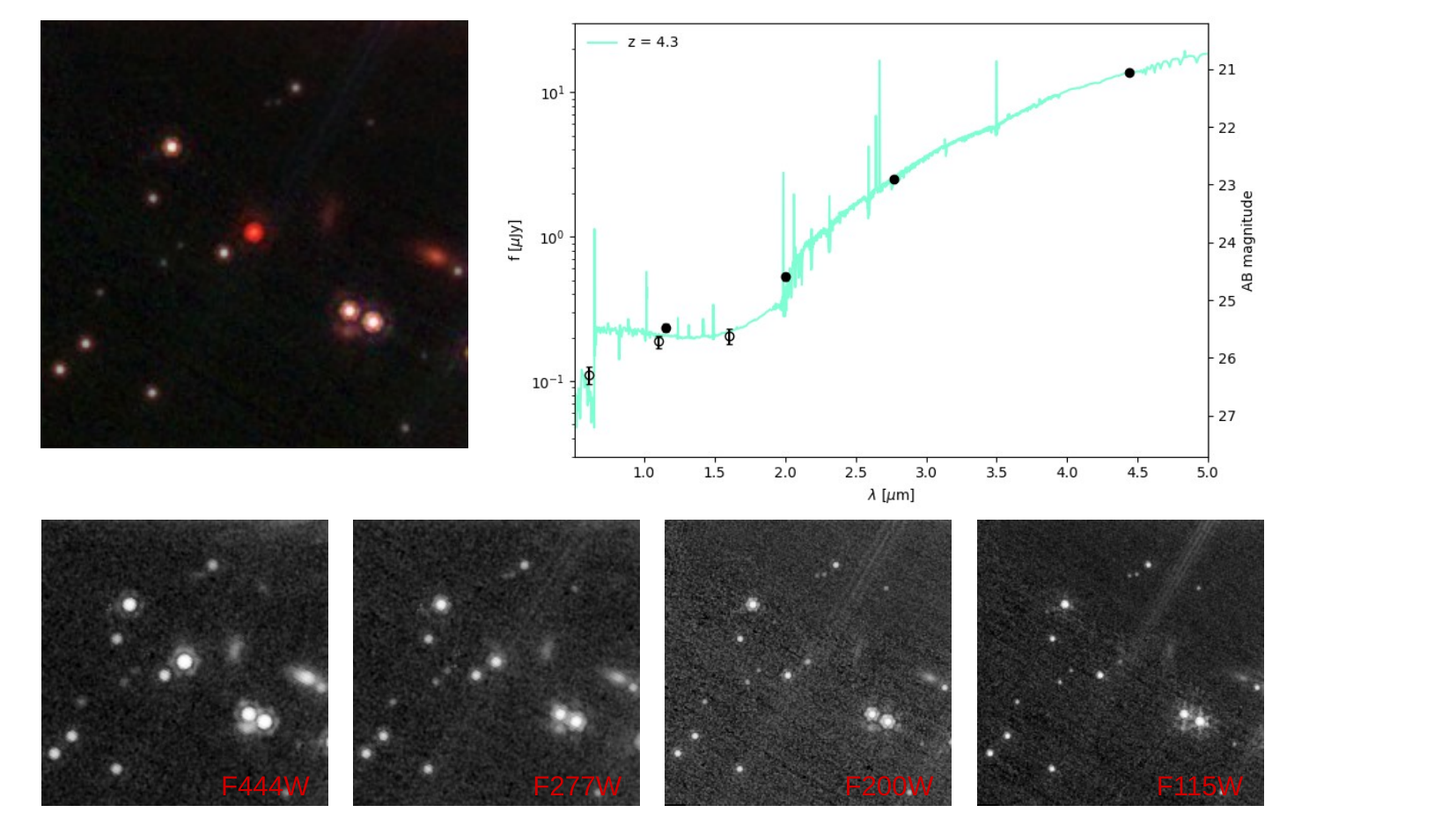}
\caption{Candidate LRD in the field of NGC\,1851. Same format as Fig.~\ref{fig:gal1}. The upper-left panel shows a three-colour composite image of a $6\times 6$ arcsec region centered at $\mathrm{RA}=05^{\mathrm h}13^{\mathrm m}54.14^{\mathrm s}$, $\mathrm{Dec}=-40^\circ\,04'\,17.5''$, and the upper-right panel the observed SED with the best-fitting EAZY template. Filled and open symbols correspond to JWST/NIRCam and HST photometry, respectively. The lower panels show the stacked images in the F444W, F277W, F200W, and F115W bands (left to right).}
    \label{fig:gal6}
\end{figure*}

A second example is shown in Fig.~\ref{fig:gal6}, which presents the same set of diagnostics for a compact source detected in the field of NGC\,1851 ($\mathrm{RA}=05^{\mathrm h}13^{\mathrm m}54.14^{\mathrm s}$, $\mathrm{Dec}=-40^\circ\,04'\,17.5''$). Its photometric properties are consistent with those of a LRD galaxy at a redshift of $z \simeq 4.3$.

Photometric redshifts were estimated using the EAZY code \citep[Easy and Accurate Redshifts from Yale;][]{brammer2008a}, which fits the observed SEDs with linear combinations of galaxy templates over a broad redshift range and returns both the best-fitting redshift and the corresponding redshift probability distribution function.



\section*{Acknowledgements}
This work is based on observations made with the NASA/ESA/CSA James Webb Space Telescope. The data were obtained from the Mikulski Archive for Space Telescopes at the Space Telescope Science Institute, which is operated by the Association of Universities for Research in Astronomy, Inc., under NASA contract NAS 5-03127 for JWST. These observations are associated with program GO-8960. E.\,P.\,L.\, acknowledges support by Special Project for High-End Foreign Experts ''Xingdian`` Funding from Yunnan Province and National Key R\&D Program of China Grant (No. 2024YFA1611601). HW acknowledges the support from the MUNI Award in Science and Humanities(MUNI/I/1762/2023).

\section*{Data Availability}
The data underlying this article will be shared on reasonable request to the corresponding author.

\section*{References}

\bibliographystyle{mnras}
\bibliography{ms}

\bsp	
\label{lastpage}
\end{document}